\documentclass[reprint,aps,prd,nofootinbib,a4paper,10pt,twocolumn,superscriptaddress,floatfix]{revtex4-2}
\usepackage[utf8]{inputenc}
\usepackage{textgreek}
\usepackage[section]{placeins}
\usepackage{hyperref} 
\usepackage{changes}
\usepackage{graphicx,subcaption}
\usepackage[titletoc]{appendix}
\usepackage{xcolor}
\usepackage{dsfont}
\usepackage{bm}
\usepackage{mathtools} 
\usepackage{amsfonts}
\usepackage{multirow}
\usepackage{enumerate}
\usepackage{comment}
\hypersetup{
     colorlinks   = true,
     citecolor    = cyan,
     linkcolor = pink
     }

\definecolor{MyB}{rgb}{0.1,0.1,1.0}

\definecolor{MyA}{rgb}{1.0,0.1,1.0}

\definecolor{MyE}{rgb}{0.0,0.65,0.0}

\begin{document}

\title{Wave Optics Effects from Gravitational Wave Propagation Through Dark Matter Halos}

\author{Annamalai P. Shanmugaraj}
\email{annamalaips@cp3.sdu.dk, annamalai.pshanmugaraj@gmail.com}
\affiliation{CP$^3$-Origins, University of Southern Denmark, Campusvej 55, DK-5230 Odense M, Denmark}

\author{Roland Haas}
\email{rhaas@mail.ubc.ca}
\affiliation{Department of Physics \& Astronomy, Vancouver Campus,
325 - 6224 Agricultural Road,
Vancouver BC, V6T 1Z1, Canada}

\author{Erik Schnetter}
\email{eschnetter@perimeterinstitute.ca}
\affiliation{Perimeter Institute for Theoretical Physics, Waterloo, ON, Canada}
\affiliation{Department of Physics and Astronomy, University of Waterloo, Waterloo, ON, Canada}
\affiliation{Center for Computation \& Technology, Louisiana State University, Baton Rouge, LA, USA}

\author{Sofie Marie Koksbang}
\email{koksbang@cp3.sdu.dk}
\affiliation{CP$^3$-Origins, University of Southern Denmark, Campusvej 55, DK-5230 Odense M, Denmark}

\begin{abstract}
Gravitational wave (GW) propagation is usually studied under the geometric optics approximation. However, when GWs propagate through structures of sizes similar to their wavelength, this approximation breaks down.  To study this situation, we present a scheme for numerically evolving linearised plane GWs through stationary, spherical astrophysical structures in both weak and strong gravity regimes. Our simulations evolve the full Einstein equations (with all 10 components) for Gaussian, NFW and Burkert potentials. Our simulations show that the scattering of the GWs depends not only on the mass of the lens but also strongly on the gravitational potential distribution of the lens. Among our most important results, we find that the GWs do not propagate along null geodesics when propagating through the Gaussian density, neither in the strong nor weak gravity setting. Another significant finding is that in the strong gravity case of the Burkert potential, the convexity of the wave is flipped when leaving the structure.
\newline\indent
We compare our results with the linearized scalar wave predictions and find that the difference between these and the exact GW modes are of order one when the wave is inside the central potential. However, the difference reduces to only a few percent when the wave has passed through the structure. Although these effects are small, future GW detectors and Pulsar Timing Arrays (PTAs) could be sensitive to these signals which could thus potentially help in constraining the structure of dark matter spikes or halos.
\end{abstract}
\maketitle

\section{Introduction}
The direct detection of gravitational waves (GWs) by the LIGO–Virgo–KAGRA collaboration \cite{LVK1, LVK2, LVK3, LVK4, ligo2026gwtc, LVK5, LVK7, LVK8} has opened a new observational window on the Universe, enabling the exploration of strong field predictions of general relativity and a large array of other endeavours within fundamental physics such as constraining the graviton mass, detecting possible large extra dimensions, exploring dark matter properties and providing new insights to the Hubble tension (see e.g. \cite{roadmap, review_GW,Cai:2026swf} for reviews). In most observational and theoretical analyses, GW propagation is treated within the framework of geometric optics where it is assumed that the wavelength of the GW is much smaller than the characteristic scale of any gravitational lens it encounters. Under this approximation, the effect of lensing reduces to a frequency-independent magnification and a time delay. However, in situations where the Schwarzschild radius of the lens is comparable to or smaller than the wave length of the GW, wave optics becomes important and the geometric optics approximation breaks down \cite{Takahashi_2003, Schwarzschild_3, Schwarzschild_4,99review}. In the wave optics regime, diffraction and interference modify the GW signal in a frequency-dependent manner, producing amplitude modulations and phase shifts.
\newline\indent
Theoretical work on wave optics for GWs has primarily relied on analytic approximations such as employing series expansions/perturbative methods \cite{approximation1, Nakamura1998GWLensingPointMass,Cusin_2020,Guo_2020, Tambalo_2023, Yarimoto2025BornApproximation, VillarrubiaRojo2025GLoW,amoruso2026gravitationalwavelensingraysdisorderedsystem}, using simple lens models such as point lenses \cite{pointmass1, wu2025weaklensingapproximationwaveoptics, Nakamura1998GWLensingPointMass}, and by evaluating diffraction integrals \cite{Tambalo_2023,Guo_2020,guo2024testingexactdiffractionformula,99review} for scalar GWs and thin lenses, traditionally restricting to the small-angle regime (see e.g. the introduction of \cite{li2025rigorouscalculationscalarscattering} for a review on shortcomings of analytical approaches in the literature). Such studies demonstrate that wave optics signatures carry information about the mass, structure, and spatial distribution of the lensing object \cite{leung2023wavemechanicsinterferencedecoherence,Takahashi_2003,_al_kan_2023,Tambalo_2023}, and can be utilized to probe dark matter and small-scale structures in the Universe \cite{Jung_2019,Diego_2019,Yeung_2023,Guo_2022,Fairbairn_2023,Savastano_2023,urrutia2025darktimbregravitationalwaves}. Although analytical methods offer valuable insights, they are thus often restricted to idealized geometries and cannot easily accommodate complex or time-dependent lens configurations. But realistic astrophysical environments are characterized by complex, inhomogeneous matter distributions that cannot readily be captured by the idealized lens models usually considered in analytical studies. The general shapes, substructures, and clustering of matter can lead to rich and distinctive wave-optics effects such as diffraction, interference, and mode conversion, that are too complicated to study analytically. Numerical approaches are essential for accurately modelling these effects when considering gravitational wave propagation in realistic cosmological settings and thus for interpreting future observations. Thus, to fully understand to what extent wave optics affects our interpretation of GW signals in terms of e.g. inferring lens properties or cosmological parameters \cite{chen2026prospect,ahmed2026gravitational}, we should ideally move beyond the simplified analytical treatments, and study more complex systems with numerical relativity. 
However, simulations face significant challenges of their own: The need to resolve both the wavelength of the GW and the macroscopic scales of the lens, the computational cost of high-resolution wave propagation, and the difficulties of handling arbitrary lens geometries. Significant progress has been made on the numerical front for standard astrophysical configurations. For instance, \cite{He_2021} performed numerical propagation of GWs through a potential well, although without considering curvature corrections. Another notable work in this regard is \cite{yeung2024wolensingpythonpackagecomputing}, where the authors developed a Python package that offers solutions for GW lensing computations within the full wave-optics regime. But this work uses two-dimensional diffraction integrals with small angle approximation and hence does not capture effects such as polarization mixing, mode coupling due to curvature terms, wave scattering due to curvature, tail formation and Lorenz gauge preservation.
\newline\indent
As gravitational wave detectors continue to improve in sensitivity, the lensing of gravitational waves by intervening matter is expected to become increasingly relevant \cite{Chen_2026, Meena2020GWLensingProspects, Wierda2021ForecastingGWLensing, VillarrubiaRojo2025WaveOptics,Li_2018,Wang_1996,ding2015strongly,hannuksela2020localizing,sereno2011cosmography,tambalo2023gravitational,cao2019direct,xu2022please}. The current ground-based observatories LIGO, Virgo and KAGRA are sensitive to the frequency range $10$Hz - $10$kHz. For these wave lengths, wave optics effects occur when the waves propagate through structures with mass around or below $10^4\ M_\odot$. Astrophysical objects with masses in this range include stars and low mass compact object remnants like white dwarfs, neutron stars, stellar black holes and primordial black holes. Future
space-based detectors such as LISA \cite{Colpi2024LISA} and DECIGO \cite{Kawamura2020DECIGO} are sensitive to frequencies of $0.1$--$100\,\mathrm{mHz}$ and $1\,\mathrm{mHz}$--$100\,\mathrm{Hz}$,
respectively, corresponding to lens masses of $M_L \sim 10^6$--$10^9\,M_\odot$ for LISA and $10^3$--$10^8\,M_\odot$ for
DECIGO. Astrophysical structures in this mass range include dark matter subhalos and larger compact structures \cite{universe9050200}. Pulsar Timing Arrays (PTA) are sensitive to ultra low frequency GWs of nano-Hz or micro-Hz which have very high wavelength that are affected by wave optics due to cosmological structures and dark matter halos.
\newline\newline
In this work, we present a numerical framework for studying GW propagation in the wave optics regime. Our approach solves the full Einstein equation for the linearized tensor perturbations corresponding to the GW. In contrast to the previous work like \cite{He_2022}, where the authors propagate GWs through a black hole, we propagate GWs through a Gaussian density distribution to mimic propagation through a dark matter halo structure. Additionally, we consider the importance of the halo shape by comparing results obtained by considering GW propagation through Gaussian, Navarro–Frenk–White (NFW) \cite{Navarro_1996} and Burkert \cite{Burkert_1995} density distributions. We use the \texttt{CarpetX} framework of the \texttt{Einstein Toolkit} \cite{schnetter2010carpetx,Löffler_2012} for the evolution with Adaptive Mesh Refinement (AMR) capabilities. This enables our simulations to accurately model interference and lensing effects for a wide range of astrophysical scenarios. Our simulations can be used to make realistic models of GW propagation and hence e.g. more realistic training data for Machine Learning efforts. The results from these simulations can also be used for consistency checks of simplified analytical methods so that these can be tested in more complex situations than what is usually done, where comparison is made with e.g. point lenses.
\newline\indent
The paper is organized as follows: In section \ref{sec:theoretical_setup} we consider a background spacetime with metric perturbations and derive the evolution equation for GWs propagating on this ``background''. In section \ref{sec:numerical setup} we describe our numerical setup and numerical methods involved in solving this set of equations including setting up initial conditions. In section \ref{sec:numerical results} we report our findings for GW propagation through a Gaussian density distribution in both the weak and strong gravity setting while section \ref{sec:results_nfw_bk} reports the findings for GW propagation through dark matter halo profiles, i.e. for NFW and Burkert densities. Section \ref{sec:summary_conclusions} is dedicated to conclusions and a discussion of possible future extensions.
\newline\indent
Throughout this paper, we adopt geometrized units with $G=1=c$ where space and time have the same units and is expressed as an arbitrary reference mass $M$. Since the equations are scale-free, the results can be converted to physical units for any choice of $M$.

\section{Theoretical setup}\label{sec:theoretical_setup}
Similar to light propagation studies, most gravitational lensing studies apply the geometric optics approximation. In this approximation, one assumes that the wavelength of the light or GW is much smaller than the length scale over which the amplitude, polarization and radius of curvature change \cite{misner1973gravitation}. However, when the wavelength is comparable to the radius of curvature of the background, wave optics effects become prominent and the geometric optics approximation fails. We here study the appearance of such wave-optics phenomena by numerically propagating gravitational waves through steep curvature gradients. For describing the spacetime around an astrophysical structure such as a dark matter halo, we assume that the background metric through which the GW propagates has the line element
\begin{equation}
    g^B_{\mu \nu} = \text{diag}[ \ -\exp(2\phi)dt^2 + \exp(-2\phi) (dx^2 +dy^2 +dz^2) \ ],
    \label{metric_exp}
\end{equation}
where $\phi$ is a gravitational potential obtained by solving the Poisson equation sourced by the central density of the halo. The metric of this form is more appropriate for capturing non-linear curvature terms compared to the usual linearly perturbed perturbed metric of the form (considering only scalar perturbations for now)
\begin{equation}
    g_{\mu \nu} = \text{diag}[ \ -(1 + 2\phi)dt^2 + (1 - 2\phi) (dx^2 +dy^2 +dz^2) \ ]
\label{metric}
\end{equation}
and is e.g. used in the weak-field approximation of the relativistic N-body simulation \texttt{gevolution} \cite{Adamek_2017}.
\newline\indent
Assuming the line element of equation \eqref{metric_exp}, Einstein's equation for the linearized trace-reversed metric perturbation satisfying the Lorenz gauge condition $\nabla_{\mu} \bar h^{\mu \nu} = 0$ is \cite{thorne1989gravitational}
\begin{equation}
\begin{split}
    \nabla^{\alpha} \nabla_{\alpha} \bar h_{\mu \nu} + 2 R^B_{\alpha \mu \beta \nu} \bar h^{\alpha \beta} = - 16 \pi \ \bigg( \mathcal{T}_{\mu \nu} - \frac{1}{2} \bar{h}_{\alpha \beta} T_B^{\alpha \beta} g^B_{\mu \nu} \\ -\frac{1}{2} \bar{h} T^B_{\mu \nu} + \frac{1}{4} \bar{h} \ T_B \ g^B_{\mu \nu} \bigg) 
    \label{gweqn}
\end{split}
\end{equation}
where $\bar h_{\mu \nu} = h_{\mu \nu} - \frac{1}{2} h g^B_{\mu \nu}$ is the trace-reversed metric perturbation, $\mathcal{T}_{\mu \nu}$ is the stress-energy-momentum tensor, $\bar h \equiv \bar{h}_{\alpha \beta} g_B^{\alpha \beta} $ and $T_B = T_B^{\alpha \beta} g^{B}_{\alpha \beta}$.
\newline
\indent
Expanding the first term in the Left-hand-side of equation \eqref{gweqn} yields
\begin{equation}
\begin{split}
    \bar{h}_{\mu \nu ; \alpha}^{\ ; \alpha} =\; & g_B^{\alpha \beta} \bigg[ 
\partial_\beta \partial_\alpha \bar{h}_{\mu \nu} 
- \partial_\beta \left( \Gamma^\lambda_{\alpha \mu} \bar{h}_{\lambda \nu} \right) 
- \partial_\beta \left( \Gamma^\lambda_{\alpha \nu} \bar{h}_{\mu \lambda} \right) \\
& - \Gamma^\lambda_{\beta \alpha} \partial_\lambda \bar{h}_{\mu \nu} 
+ \Gamma^\lambda_{\beta \alpha} \Gamma^\sigma_{\lambda \mu} \bar{h}_{\sigma \nu}
+ \Gamma^\lambda_{\beta \alpha} \Gamma^\sigma_{\lambda \nu} \bar{h}_{\mu \sigma} \\
& - \Gamma^\lambda_{\beta \mu} \partial_\alpha \bar{h}_{\lambda \nu} 
+ \Gamma^\lambda_{\beta \mu} \Gamma^\sigma_{\alpha \lambda} \bar{h}_{\sigma \nu}
+ \Gamma^\lambda_{\beta \mu} \Gamma^\sigma_{\alpha \nu} \bar{h}_{\lambda \sigma} \\
& - \Gamma^\lambda_{\beta \nu} \partial_\alpha \bar{h}_{\mu \lambda} 
+ \Gamma^\lambda_{\beta \nu} \Gamma^\sigma_{\alpha \mu} \bar{h}_{\sigma \lambda}
+ \Gamma^\lambda_{\beta \nu} \Gamma^\sigma_{\alpha \lambda} \bar{h}_{\mu \sigma} 
\bigg],
\end{split}
\end{equation}
where the Christoffel symbols are computed with respect to the background metric $g^B_{\mu \nu}$.
\newline\indent
Similarly, the second term of equation \eqref{gweqn} can be expanded to (ignoring the overall factor of 2)
\begin{equation}
\begin{split}
R^B_{\alpha\mu\beta\nu} \bar h^{\alpha\beta} &= g^B_{\alpha \delta} \ (
\partial_{\beta}\Gamma^{\delta}{}_{\nu\mu}
- \partial_{\nu}\Gamma^{\delta}{}_{\beta\mu}
\\
&+ \Gamma^{\delta}{}_{\beta\lambda}\Gamma^{\lambda}{}_{\nu\mu}
- \Gamma^{\delta}{}_{\nu\lambda}\Gamma^{\lambda}{}_{\beta\mu} \ ) \ g_B^{\rho\alpha} \ g_B^{\sigma\beta} \ \bar h_{\rho\sigma}
\end{split}
\end{equation}
In equation \eqref{gweqn}, we will assume that the contribution from the GWs to the stress-energy-momentum tensor are much smaller than the background contributions and may be ignored. The evolution equations in terms of the trace-reversed metric perturbation thus become
\begin{equation}
\begin{split}
g_B^{\alpha \beta} \bigg[ 
\partial_\beta \partial_\alpha \bar{h}_{\mu \nu} 
- \partial_\beta \left( \Gamma^\lambda_{\alpha \mu} \bar{h}_{\lambda \nu} \right) 
- \partial_\beta \left( \Gamma^\lambda_{\alpha \nu} \bar{h}_{\mu \lambda} \right) \\
- \Gamma^\lambda_{\beta \alpha} \partial_\lambda \bar{h}_{\mu \nu} 
+ \Gamma^\lambda_{\beta \alpha} \Gamma^\sigma_{\lambda \mu} \bar{h}_{\sigma \nu}
+ \Gamma^\lambda_{\beta \alpha} \Gamma^\sigma_{\lambda \nu} \bar{h}_{\mu \sigma} \\
- \Gamma^\lambda_{\beta \mu} \partial_\alpha \bar{h}_{\lambda \nu} 
+ \Gamma^\lambda_{\beta \mu} \Gamma^\sigma_{\alpha \lambda} \bar{h}_{\sigma \nu}
+ \Gamma^\lambda_{\beta \mu} \Gamma^\sigma_{\alpha \nu} \bar{h}_{\lambda \sigma} \\
- \Gamma^\lambda_{\beta \nu} \partial_\alpha \bar{h}_{\mu \lambda} 
+ \Gamma^\lambda_{\beta \nu} \Gamma^\sigma_{\alpha \mu} \bar{h}_{\sigma \lambda}
+ \Gamma^\lambda_{\beta \nu} \Gamma^\sigma_{\alpha \lambda} \bar{h}_{\mu \sigma} 
\bigg] \\
+ \  2 \ g^B_{\alpha \delta} \ (
\partial_{\beta}\Gamma^{\delta}{}_{\nu\mu}
- \partial_{\nu}\Gamma^{\delta}{}_{\beta\mu}
+ \Gamma^{\delta}{}_{\beta\lambda}\Gamma^{\lambda}{}_{\nu\mu} \\
- \Gamma^{\delta}{}_{\nu\lambda}\Gamma^{\lambda}{}_{\beta\mu} \ ) \ g_B^{\rho\alpha} \ g_B^{\sigma\beta} \ \bar h_{\rho\sigma} = 8 \pi \ \bigg( \bar{h}_{\alpha \beta} T_B^{\alpha \beta} g^B_{\mu \nu} \\ + \ \bar{h} T^B_{\mu \nu} - \frac{1}{2} \bar{h} \ T_B \ g^B_{\mu \nu} \bigg) .
\end{split}
\label{evol}
\end{equation}
The corresponding Lorenz gauge condition reads
\begin{equation}
\mathcal{C}_{\nu} :=\nabla^{\mu} \bar h_{\mu \nu} = g_B^{\nu \rho} ( \ \partial_{\rho}{ \bar h_{\mu \nu} } - \Gamma^{\lambda}_{\mu \rho} \bar h_{\lambda \nu} - \Gamma^{\lambda}_{\nu \rho} \bar h_{\mu \lambda} \ ),
\label{cv}
\end{equation}
where we defined the quantity $\mathcal{C}_{\nu}$ for later convenience. In addition to the evolution equations \eqref{evol}, we also evolve the linearly approximated simplified (scalar wave) equation defined by $g_
B^{\alpha \beta}
\partial_\beta \partial_\alpha h_{\mu \nu} = 0$ which does not include curvature terms (Christoffel symbols and its derivatives ), i.e.,
\begin{equation}
    \left[ \exp(2 \phi)^{-1}\partial_{tt} - \exp(- 2 \phi)^{-1}( \partial_{xx} + \partial_{yy} + \partial_{zz}) \right] h = 0.
    \label{simplified_evol}
\end{equation}
Since this is a scalar wave evolution with no coupling between different components of $h_{\mu \nu}$, it is sufficient to evolve a single component which is denoted as $h$ in the above equation. Evolving \eqref{simplified_evol} and comparing the results with \eqref{evol} will allow us to distinguish the contributions from the curvature terms.
\newline\indent
For a final comparison, we will also evolve the density-free equations where we set $T_{\mu \nu} =0$ in \eqref{evol} for the Gaussian profile to distinguish the contributions from matter-field coupling from the direct curvature contributions.

\section{Numerical setup}\label{sec:numerical setup}
In this section, we describe our numerical setup for evolving the evolution equations derived in the previous section. We will run six simulations that propagate initially plane-parallel gravitational waves through three different spherical potential fields motivated by models of dark matter spikes and halos.
\newline\indent
In subsection \ref{subsec:1+3d} we describe our code architecture and numerical methods, in subsection \ref{subsec:densities} we discuss the three density profiles we consider, and in subsection \ref{subsec:initialconditions} we describe our initial conditions. In subsections \ref{subsec:AMR} and \ref{subsec:constraint_damp} we present the numerical schemes adopted for constraint damping and Adaptive Mesh Refinement (AMR).

\subsection{Numerical evolution in 3+1D}\label{subsec:1+3d}
The linearized gravitational wave equations in curved spacetime are a set of ten second order linear hyperbolic partial differential equations (PDEs) in ten variables. We turn this into 20 first order equations by introducing the auxiliary fields $H_{\alpha\beta} \equiv \partial_t \bar h_{\alpha\beta}$. We use \texttt{Cadabra} \cite{peeters2018introducingcadabrasymboliccomputer,Peeters2018,Peeters_2007}, an open-source computer algebra software package to expand the tensor equations \eqref{evol}. We then use \texttt{Sympy-bridge} in \texttt{Cadabra} to convert the equations into expressions compatible with \texttt{C++}. The equations are then evolved in time using a fourth order Runge Kutta time integrator. To compute the gradients and divergences appearing in the PDEs, we use the Method of Lines \cite{scheisser1991numerical} to discretize in all but one dimension. This leads to a system of ordinary differential equations (ODEs) where we can use a fourth order central finite difference stencil to compute the derivatives. We use \texttt{Einstein Toolkit}'s \texttt{CarpetX} framework to evolve this set of twenty coupled space and time-discretized ODEs. We make use of Adapative-Mesh-Refinement (AMR) capabilities of CarpetX to resolve the regions that have steep curvature gradients. 
\newline\newline
As we are only interested in the simulation domain's interior, we use symmetric, reflecting boundary conditions in all directions. We stop the simulation when the wave reaches close to the end of the simulation box to prevent reflected waves from affecting our results. To be conservative of our computational resources, we propagate plane wave that are constant in $x-y$ axis in a rectangular simulation domain.

\subsection{Considered densities and corresponding potentials}\label{subsec:densities}
We consider three different spherical mass density profiles for our simulations, namely a Gaussian profile, the Navarro–Frenk–White (NFW) profile \cite{Navarro_1996} and Burkert density profile \cite{Burkert_1995}. The Gaussian profile mimics the usual spherical mass distribution whereas the NFW and Burkert profiles are common models of the shape of cold dark matter haloes and dwarf galaxies, respectively. Specifically, we consider the profiles defined as
\newline
\begin{equation}
    \rho(r)= \begin{cases}
    \frac{M}{ \lambda^3 (2\pi)^{3/2}  } \cdot \exp( -\frac{r^2}{2 \lambda^2}), & \text{Gaussian} \\[1em]
    
    \frac{\rho_{0}}{\frac{r}{R_s}\left( 1 + \frac{r}{R_s} \right)^2}, & \text{NFW profile } \\[1em]
    
    \frac{\rho_{0}^{\text{bk}} R_s^3}{(r + R_s)\left( r^2 + R_s^2 \right)}, & \text{Burkert profile},
    \end{cases}
\end{equation}
\newline
where $\lambda$ determines the spread of the Gaussian density, $\rho_0 \equiv M/4 \pi R_s^3$, and $R_s$ (which we will refer to as the scale radius) sets the size of the structures.
\newline
\newline
We obtain the gravitational potential by solving the Poisson equation $\nabla^2 \phi(r) = 4\pi \rho(r)$. Specifically, the corresponding expressions for $\phi$ are
\begin{widetext}
\begin{equation}
\phi(r) = \begin{cases}
-\dfrac{M_{\text{gauss}}}{r} \cdot \operatorname{erf}\!\left(\dfrac{r}{\sqrt{2}\,\lambda}\right), & \text{Gaussian} \\[1em]
-\dfrac{M_{\text{NFW}}}{r} \ln\!\left(1 + \dfrac{r}{R_s}\right), & \text{NFW profile \cite{GalRotpy}} \\[1em]
\pi \rho_0^{\text{bk}} R_s^2 \Bigg\{ \Big(1 - \dfrac{R_s}{r}\Big) \ln\!\Big(1 + \Big(\dfrac{r}{R_s}\Big)^2\Big)
+ 2\Big(1 + \dfrac{R_s}{r}\Big)\Big[\tan^{-1}\!\Big(\dfrac{r}{R_s}\Big) - \ln\!\Big(1 + \dfrac{r}{R_s}\Big)\Big] - \pi \Bigg\}, & \text{Burkert profile\cite{GalRotpy}}
\end{cases}
\end{equation}
\end{widetext}
where we defined $M_{\text{gauss}} :=   \phi_{\text{max}}\lambda \sqrt\frac{\pi}{2}$, $M_{\text{NFW}} := R_s \phi_{\text{max}}$ and $\rho_0^{\text{bk}} := \dfrac{\phi_{\text{max}}}{ \pi^2 R_s^2}$. Here, $\phi_{\text{max}}$ is the maximum value $\phi$ takes, and setting it to a specific value allows us to compare simulations among these different profiles. In equation \eqref{metric}, setting $\phi=0$, corresponds to flat spacetime and $|\phi| = -1/2$ results in formation of a horizon-like surface similar to a black hole. Thus, a weak gravity regime is obtained by setting $|\phi|$ slightly above zero while the strong gravity regime is probed with $|\phi|$ close to 1/2. For our simulations, we probe both situations by considering first $\phi_{\text{max}} = -0.1$ and then $\phi_{\text{max}} = -0.49$.
\newline
\newline \indent
The NFW density diverges as $r \rightarrow0$ while the potential $\phi$ remains finite. For this profile, the singular density region is handled implicitly by the numerical discretisation. In particular, we first compute the potential $\phi(r)$ from the analytical expression and instead of computing $\rho(r)$ directly from its (divergent) analytical form, we use the numerical derivative expression $\dfrac{\nabla^2 \phi}{4 \pi}$ from the Poisson equation. Provided the grid spacing is coarse enough, the discretised $\nabla^2 \phi$ remains finite at every sampled radius. With this approach, the convergence of the numerical simulation degrades near the origin due to the assumed implicit smoothness of the finite-difference stencil. Since $\rho(r)$ is obtained by differentiating the known analytical potential $\phi(r)$ pointwise, the Laplacian at each grid point $r_i$ depends only on the local stencil values $\phi_{i-1}, \phi_i, \phi_{i+1}$ and not on the solution elsewhere. Thus, the reduced accuracy near the centre remains localised and does not propagate outward through the grid. Any quantity derived from the density at radii away from this core region remains unaffected. 

For the Gaussian profile, to demonstrate how the density coupling affects the wave propagation, we will solve the full Einstein equation as well as the equations with the right hand side set to zero, i.e. assuming $T_{\mu \nu} = 0$. We will refer to the resulting equations as the density-free equations. This comparison is not shown for the other two profiles (NFW and Burkert) because their density contrasts are highly localised and/or have very small amplitude. (see figures \ref{fig:densities_weak}, \ref{fig:densities_strong} in Appendix \ref{app:dens_negative_energy}) and the density-free results are indistinguishable from the results obtained by solving the full Einstein equation.
\newline
\newline
For the stress-energy-momentum tensor $T_{\mu \nu}$, we assume the matter to be pressure-less (mimicking cold dark matter) and hence $T_{\mu \nu}^B = \text{diag}[\rho(r)\exp(2 \phi),0,0,0]$.

\subsection{Initial conditions}\label{subsec:initialconditions}
We consider circularly polarized monochromatic waves propagating along the z-axis as our initial conditions. These waves can e.g. represent GWs originating from binary compact object coalescence or stochastic GWs permeating through space. We initialize them at a specific distance from the central density and let them propagate through the curved spacetime near the centre of our simulation box. The initial strain of $h_+ = h_{xx} = -h_{yy}$ and $h_\times = h_{xy} = h_{yx}$ are set as
\begin{equation}\label{ic_planewave}
    \begin{split}
        h_+ & = \mathcal{A} \cos[ 2 \pi (k_zz-\omega t)] \\
        h_\times & = \mathcal{A} \sin[ 2 \pi (k_zz-\omega t)]. \\
    \end{split}
\end{equation}
Here, $\mathcal{A}$ is the amplitude of the wave and $\omega = \sqrt{k_x^2+k_y^2+k_z^2} = |k_z|$.
\newline
\newline
For a full evolution of Einstein's equation \eqref{gweqn} in curved spacetime, we need initial conditions for all the ten components of $\bar h_{\mu \nu}$ and another ten initial conditions for its time derivatives satisfying the constraints in equation \eqref{cv}. Specifying only the above plane wave initial condition in equation \ref{ic_planewave} along with its derivatives results in under-determined initial conditions.
Although we could use equation \eqref{cv} to determine the other components, this is non-trivial since it requires careful consideration not to introduce spurious gauge modes. To avoid this, we initialize the wave in a confined region in flat Minkowski background similar to the approach in \cite{Shanmugaraj_2025}. Specifically, we introduce a quintic spline at a finite distance from the central density and use it to smoothly force the gravitational potential, $\phi$, and its derivatives $\phi'$ and $\phi''$ to zero. Our potential is thus given by
\begin{equation}
    \phi(r)= 
        \begin{cases}
    \phi(r), & \text{if } r\leq r_1\\
    \text{Spline approximation},              & \text{if } r_1 \leq r \leq r_2 \\
    0, & \text{if } r \geq r_2,
\end{cases}
\end{equation}
where $r = \sqrt{x^2 + y^2 +z^2}$ and $r_1$, $r_2$ are set by hand. In table \ref{tab:spline} we list the values of $r_1$ and $r_2$ used in our simulations. In figure \ref{fig:d_phi} we present the derivatives of $\phi$ for each potential configuration for the case of strong gravity and we note that the behaviour of the derivatives remains similar for the weak gravity case.

Introducing the above described spline setup means that we are forcing $\phi$ and its derivatives to vanish at finite $r$, thus rendering the central overdensity gravitationally ``invisible'' at distances $r\geq r_2$. This requires the introduction of a region with small negative density, here well below the numerical precision of the simulation at $\sim 10^{-5}$ (see fig. \ref{negative_dens_plots} in appendix \ref{app:dens_negative_energy}). Although the negative density is not physical, we introduce it here as a purely {\em numerical} approximation and in appendix \ref{subsec:nullgeodesics} (fig. \ref{fig:geodesics1} and \ref{fig:geodesics2}) we show that the deviation of null geodesics due to the spline is similar to the non-spline case and that the spline only affects the sharpness of the focus and focusing distance of the GWs. Due to this shift in focusing distance and sharpness, the results presented here are only affected quantitatively, while the observed wave optics phenomena are qualitatively valid.
\newline
\newline
We initialize the wave at a distance slightly greater than $r_2$ where $\phi = 0$. More precisely, we initialize the wave as a wave packet by introducing the wave amplitude
\begin{equation}
    \mathcal{A_{\text{pw}}} = \left(  \frac{1}{1 + e^{-2q(z-l_1)}} + \frac{1}{1 + e^{-2q(l_2 - z)}} -1  \right),
\end{equation}
where $l_1$ and $l_2$ are the distances inside which the wave fronts are confined at initial time. The parameter $q$ determines how smoothly the wave fronts reach their peak amplitude. For very small $q$, the wave packet would look like a Gaussian pulse. For larger values of $q$, we instead have a square wave function. 
\newline
\newline\indent
Finally, we specify the wave frequency $k_z$ to be close to the width of $\phi(r)$ since this should lead to wave optics effects. For the Gaussian potential, we set $\lambda = 1.0$ and with that, we define $k_z = \dfrac{2}{\text{FWHM}(\phi)} \simeq 0.4 \ M^{-1}$ (we chose twice the size of the FWHM-Full Width Half Maximum to make the simulation domain shorter). For the case of the NFW and Burkert potentials, we ensure that these potentials have the same FWHM as the Gaussian potential. This corresponds to setting $R_s^{NFW} \simeq 0.98 M $ and $R_s^{bk} \simeq 5.74 M $. Since the FWHM of all three potentials are equal, the GW that propagates through them are also given the same initial wavelength.
\newline\newline
Since we propagate the waves along the z-axis, we set up a rectangular-cuboid simulation domain with the longer side being along the z-axis. This setup minimizes the computational cost of our simulations. We have tabulated the simulation parameters in detail in table \ref{table:sim_parameters}.

\begin{table}
    \centering
    \begin{tabular}{|c|c|c|c|c|c|c|}\hline
         &  \multicolumn{6}{|c|}{Simulations}\\\hline
 & \multicolumn{2}{|c|}{Gaussian}& \multicolumn{2}{|c|}{NFW}& \multicolumn{2}{|c|}{Burkert}\\\hline
 3D grid& \multicolumn{2}{|c|}{$128^2_{xy} \times 512_z$}& \multicolumn{2}{|c|}{$128^2_{xy} \times 512_z$}& \multicolumn{2}{|c|}{$128^2_{xy} \times 512_z$}\\\hline
 CFL factor& \multicolumn{2}{|c|}{$0.5$}& \multicolumn{2}{|c|}{$0.5$}& \multicolumn{2}{|c|}{$0.5$}\\\hline
 FWHM of $\phi$& \multicolumn{2}{|c|}{$\simeq 2.473 \ M$}& \multicolumn{2}{|c|}{$\simeq 2.473 \ M$}& \multicolumn{2}{|c|}{$\simeq 2.473 \ M$}\\\hline
 Scale radius& \multicolumn{2}{|c|}{$\lambda = 1$}& \multicolumn{2}{|c|}{$R_s \simeq 0.984$}& \multicolumn{2}{|c|}{$R_s \simeq 5.742$}\\\hline
 $\lambda_{\text{GW}}$& \multicolumn{2}{|c|}{$\simeq 2.473 \ M$}& \multicolumn{2}{|c|}{$\simeq 2.473 \ M$}& \multicolumn{2}{|c|}{$\simeq 2.473 \ M$}\\\hline
 $k_z ^{\text{GW}}$& \multicolumn{2}{|c|}{$\simeq 0.404 \ M^{-1}$}& \multicolumn{2}{|c|}{$\simeq 0.404 \ M^{-1}$}& \multicolumn{2}{|c|}{$\simeq 0.404 \ M^{-1}$}\\\hline
    \end{tabular}
\caption{The parameters specifying the six simulations presented in this work. We use the same parameters for strong and weak gravity regime simulations.}

\label{table:sim_parameters}
\end{table}

In appendix \ref{app:dens_negative_energy}, plots of derivatives of potential $\phi$ (Fig. \ref{fig:d_phi}) are presented along with a table \ref{tab:spline} that lists the spline distance $r_1$ and $r_2$ chosen for each potential configuration.

\subsection{Constraint Damping}\label{subsec:constraint_damp}
If the Lorenz gauge conditions are satisfied on the initial Cauchy surface, the constraint violation should in principle be satisfied throughout the evolution \cite[Chapter VI, pp. 150--160]{ChoquetBruhat2008}. However, when numerically evolving Einstein's field equation, equation \eqref{gweqn}, the numerical precision of the computations lead to growth in the constraint violation over time even when they are perfectly satisfied initially. This eventually leads to development of unwanted spurious gauge modes. To avoid this, we damp the constraint violations appearing due to this numerical error. We follow the formalism developed by \cite{gundlach2005constraint} and adopt an approach similar to \cite{barack2005perturbations}. We add a term $- \kappa (t_\alpha Z_\beta + t_\beta Z_\alpha)$ to \eqref{evol} where $Z_\alpha \equiv \mathcal{C}_\alpha = \nabla^\mu \bar h_{\mu \nu}$. Here, $\kappa$ is a positive constant and $t_\alpha$ is a future pointing time-like vector field. We now choose a simple form of $\kappa$ and $t_\alpha$ that makes the addition of this extra term directly proportional to \eqref{evol}. Hence, we set $\kappa = 1$ and $t_\alpha = -\delta^t_\alpha$. In total, we add  $\Box_{tt} = 2Z_t$, $\Box_{tx} = Z_x$, $\Box_{ty} = Z_y$ and $\Box_{tz} = Z_z$ to the right-hand-side of the evolution equations.
\newline
\newline
In appendix \ref{app:convergence}, we show that constraints converge to zero with higher simulation resolution, supporting that the constraint violations that arise during the evolutions are numerical artifacts due to the numerical discretisation error and thus do not represent a physical Lorenz gauge violation. In figures \ref{fig:gauss_wg_CV},\ref{fig:gauss_sg_CV},\ref{fig:NFW_wg_CV},\ref{fig:NFW_sg_CV},\ref{fig:bk_wg_CV} and \ref{fig:bk_sg_CV}, we show that the maximum constraint violations are well below the order of magnitude of the GW signal and most of the violation happens due to the wave reflection at the boundary.

\subsection{Adaptive Mesh Refinement (AMR)}\label{subsec:AMR}
To resolve the features in the waves better, we set up two AMR grids on top of the base grid with one refinement level that has twice the resolution of the base grid. To better resolve $\phi$ and its derivatives in the simulation centre, we set up a stationary box-like AMR cuboid with a specific radius to resolve the steep curvature gradients. We set up another AMR grid on top of the wave packet with a size slightly greater than the wavepacket. We then move this AMR grid along with the wave as it propagates. At each iteration the AMR grid is moved by calculating the energy weighted centroid along the z-axis given by
\begin{equation}
    z_{\text{wave-centre}}(t) = \frac{\sum \ z \cdot E(z,t)}{ \sum E(z,t) }
\end{equation}
where $E(z,t) = \dfrac{1}{2} u_t^2 + \dfrac{1}{2} \exp(4\phi) (\nabla u)^2$ is the energy of the wave packet along z-axis.
\newline
\newline
We emphasize that our simulations evolve all ten components of the GW tensor together with their associated first and second derivatives. To do this, the simulations are required to store the background metric and its derivatives in memory. As a result, the memory requirements are substantial, and the computations also demand sufficient memory bandwidth. An alternative strategy would be to recompute these quantities each time they appear in the evolution equations, but this would further increase the computational cost and make the simulations significantly slower. Consequently, performing simulations without adaptive mesh refinement (AMR) becomes extremely expensive and, beyond a certain point, computationally impractical. For this type of simulations, adaptive mesh refinement (AMR) is therefore essential. High resolution is required only in the vicinity of the central potential and in the regions occupied by the propagating wave, whereas the rest of the computational domain can be evolved at comparatively low resolution. This substantially reduces the memory requirements and computational cost, while preserving the accuracy needed in the physically relevant regions.

\section{Numerical Results for GW propagation through a Gaussian density distribution}\label{sec:numerical results}
In this section, we present the simulation results from propagating GWs through a Gaussian density by solving the full Einstein equation, $G_{\mu \nu}=8 \pi T_{\mu\nu}$. In addition, we compare these results with the linearly approximated simplified (scalar wave) equation (eq. \ref{simplified_evol}). We also perform two simulations (corresponding to strong and weak gravity cases) with $T_{\mu \nu} = 0$ and compare it with the simulation evolving the full Einstein equation. This comparison enables us to distinguish the contributions from the $T_{\mu \nu}$-part of the Einstein equation. In this section we thus present results from evolving a total of three sets of equations: the full Einstein equation, a simplified (scalar wave) equation (eq. \eqref{simplified_evol}) and the density-free equation with $T_{\mu \nu}=0$.
\newline\indent
In the presented figures, we overlay the geodesics on top of our simulation results. For details on how the geodesics were calculated, we refer the reader to appendix \ref{subsec:nullgeodesics}.

\subsection{Gaussian potential: weak gravity regime}\label{subsec:gauss_wg}
\begin{figure*}
\includegraphics[width=16cm]{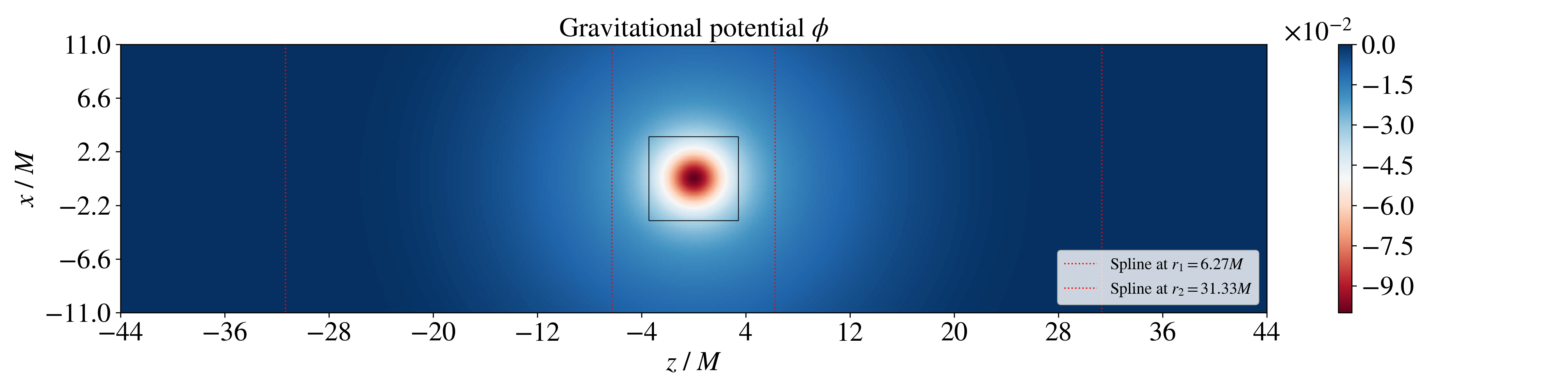}
\includegraphics [width=16cm]{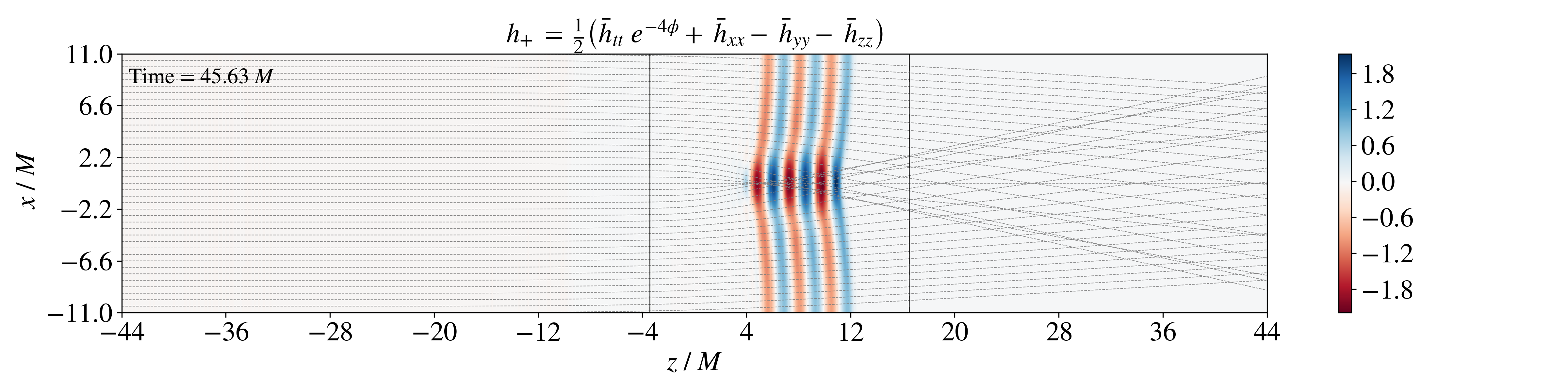}
\includegraphics [width=16cm]{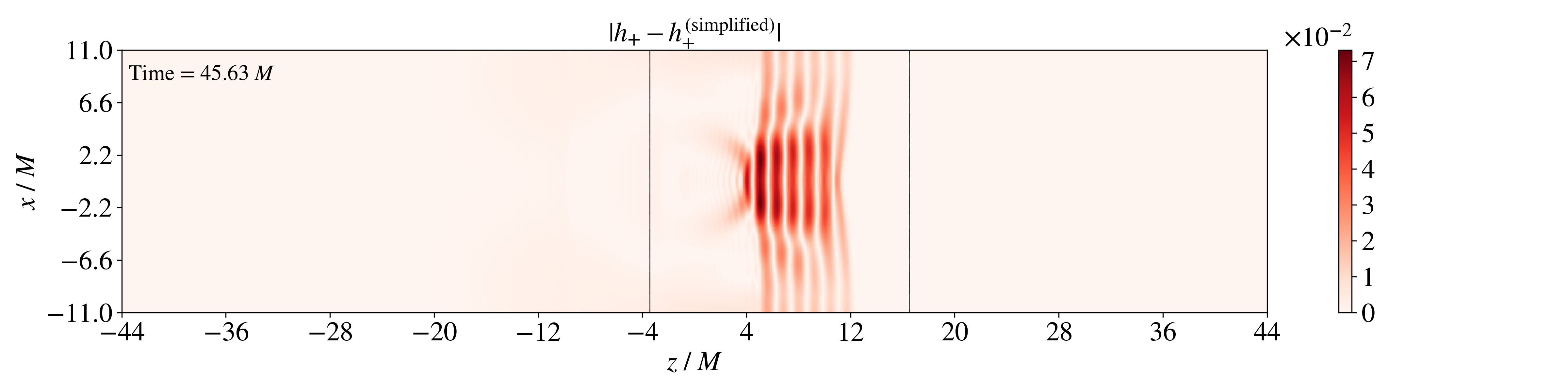}
\includegraphics [width=16cm]{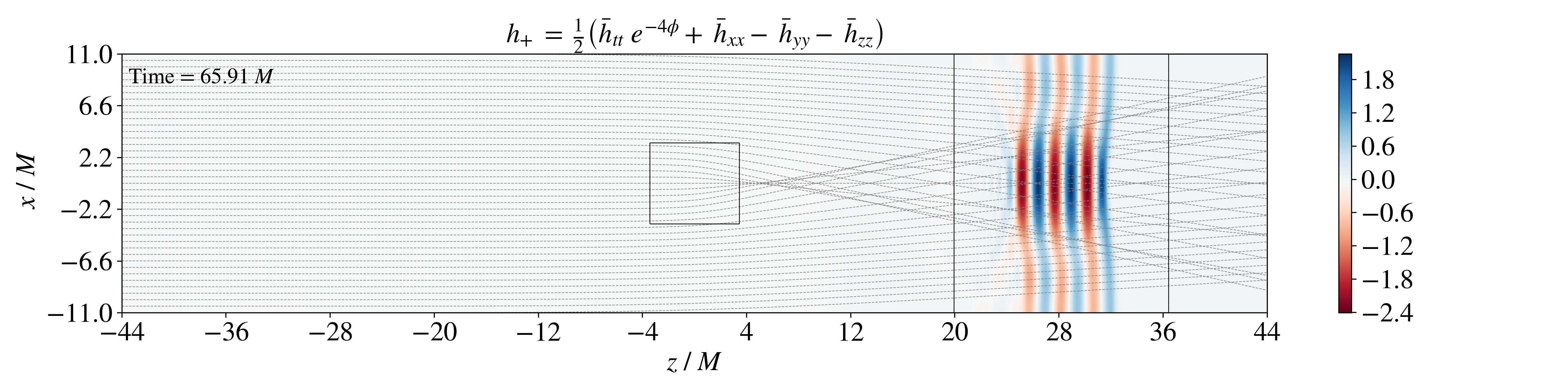}
\includegraphics [width=16cm]{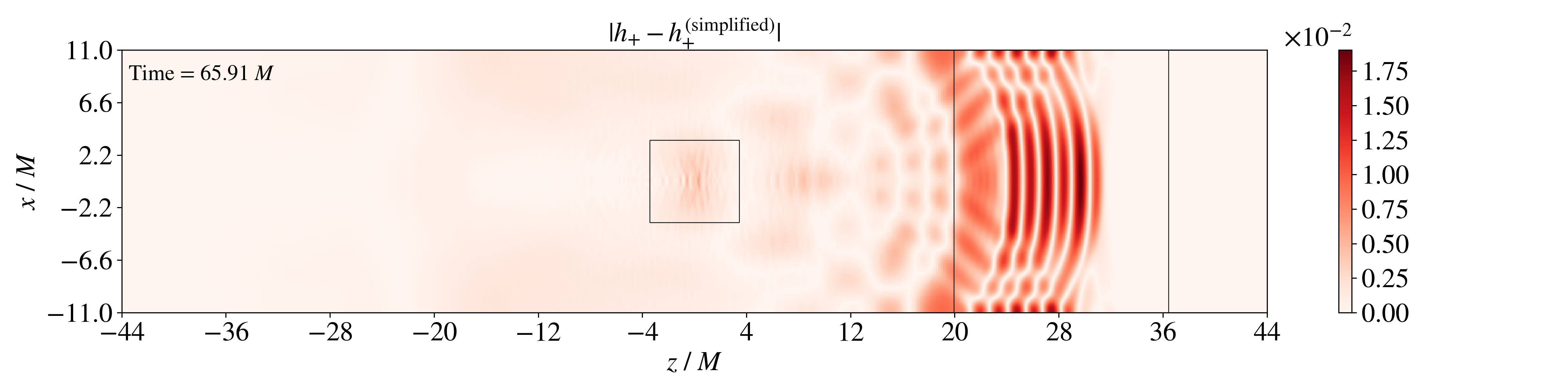}

\caption{From top to bottom: The top figure shows the gravitational potential $\phi$ along $xz$-plane for the Gaussian potential in the weak gravity regime. The second figure shows the 2D slice of $h_+$ after the wave has traversed the central density at $t=45.63 \ M$. The third figure shows the absolute difference between $h_+$ and $|h_+^{\text{simplified}}|$ which are evolved independently with the same initial conditions. The fourth and fifth figures show the wave after it has completely exited the curvature at a later time $t=65.91 \ M$. The figure for $h_+$ has an overlay of null geodesics for the potential including the spline. The black outlined boxes inside the simulation domain is where the AMR patch exists and has twice the resolution of base grid. }\label{fig:gauss_evol_wg}
\end{figure*}

\begin{figure*}
\centering
    \begin{subfigure}[b]{0.475\textwidth}
    \centering
    \includegraphics[width=8cm]{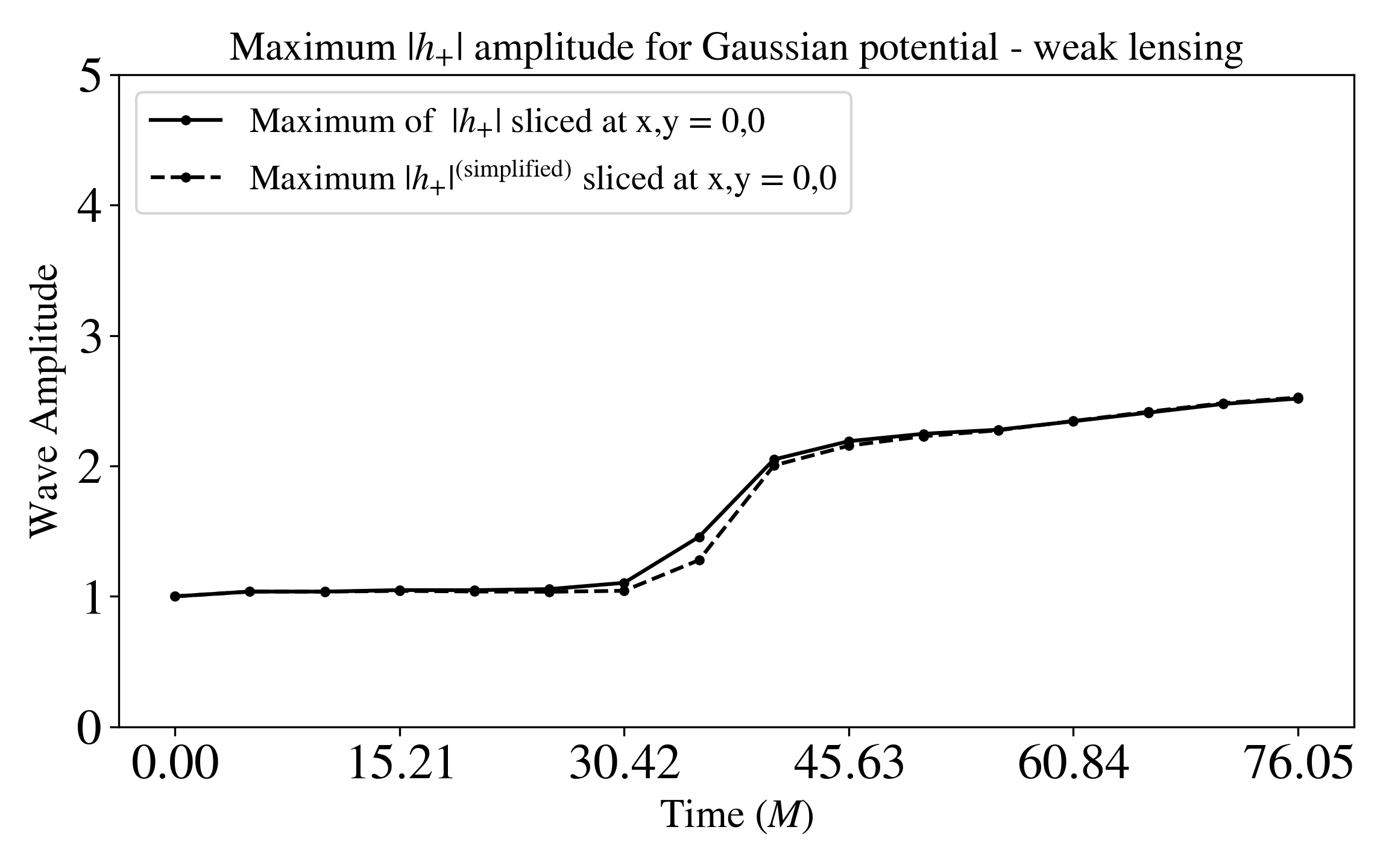}
    \caption{\small Maximum wave amplitude of $h_+$ and $|h_+^{\text{simplified}}|$ - Gaussian potential, weak gravity case}    
    \label{subfig:gauss_wg_amp}
    \end{subfigure}
        \hfill
    \begin{subfigure}[b]{0.475\textwidth}
    \centering
    \includegraphics[width=8cm]{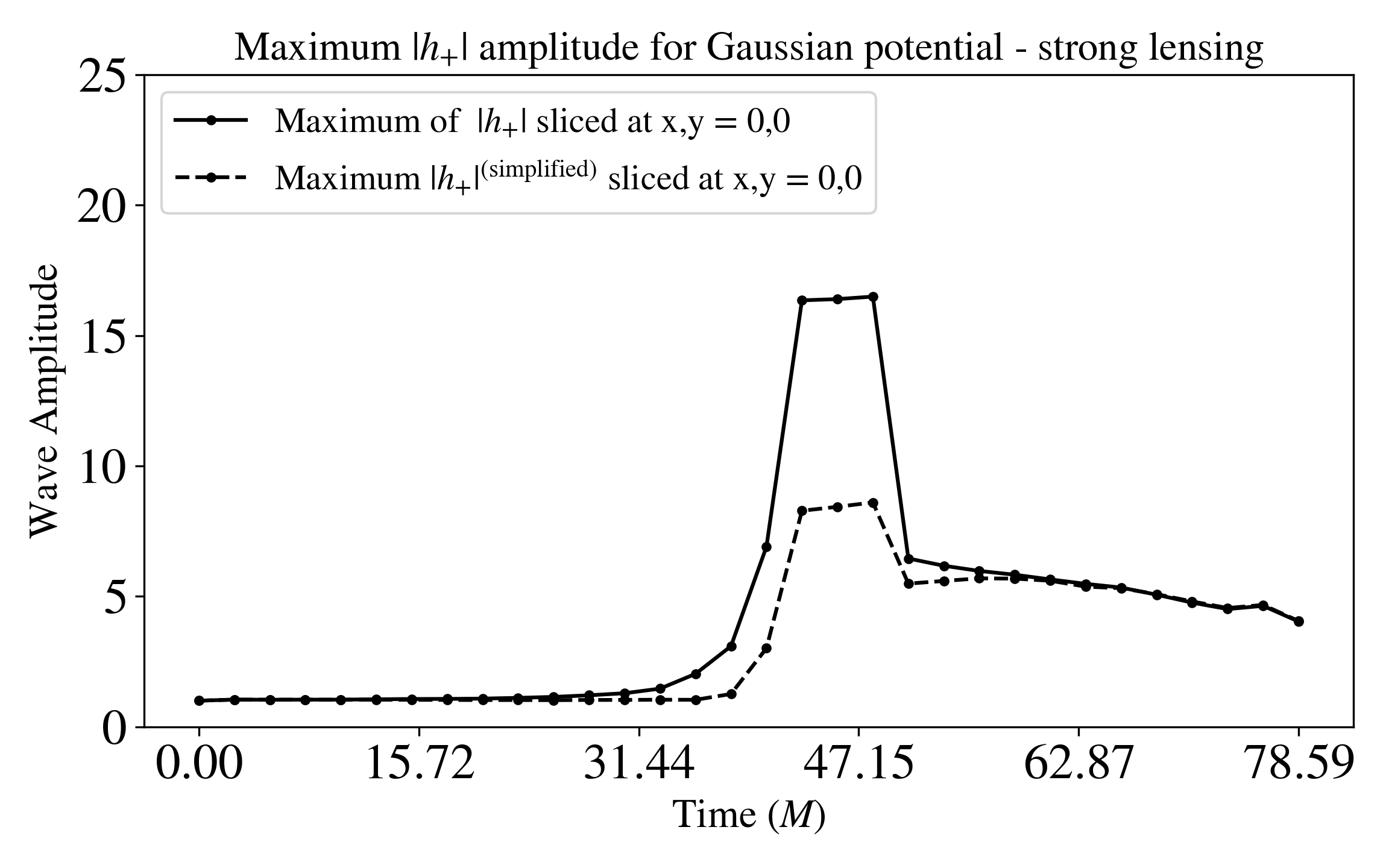}
    \caption{\small Maximum wave amplitude of $h_+$ and $|h_+^{\text{simplified}}|$ - Gaussian potential, strong gravity case}    
    \label{subfig:gauss_sg_amp}
    \end{subfigure}
        \vskip\baselineskip
    \begin{subfigure}[b]{0.475\textwidth}
    \centering
    \includegraphics[width=8cm]{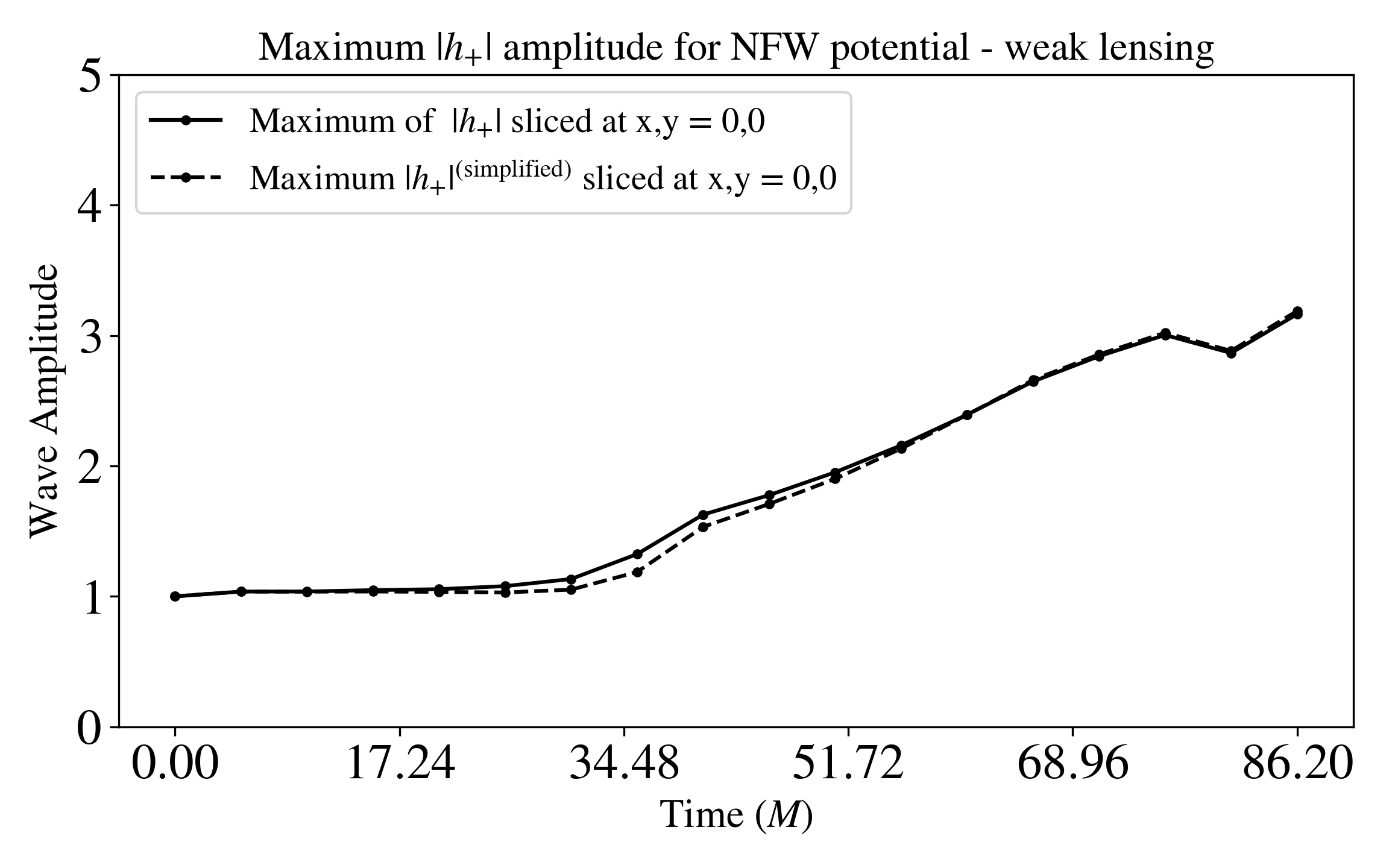}
    \caption{\small Maximum wave amplitude of $h_+$ and $|h_+^{\text{simplified}}|$ - NFW potential, weak gravity case}
    \label{subfig:nfw_wg_amp}
    \end{subfigure}
        \hfill
    \begin{subfigure}[b]{0.475\textwidth}
    \centering
    \includegraphics[width=8cm]{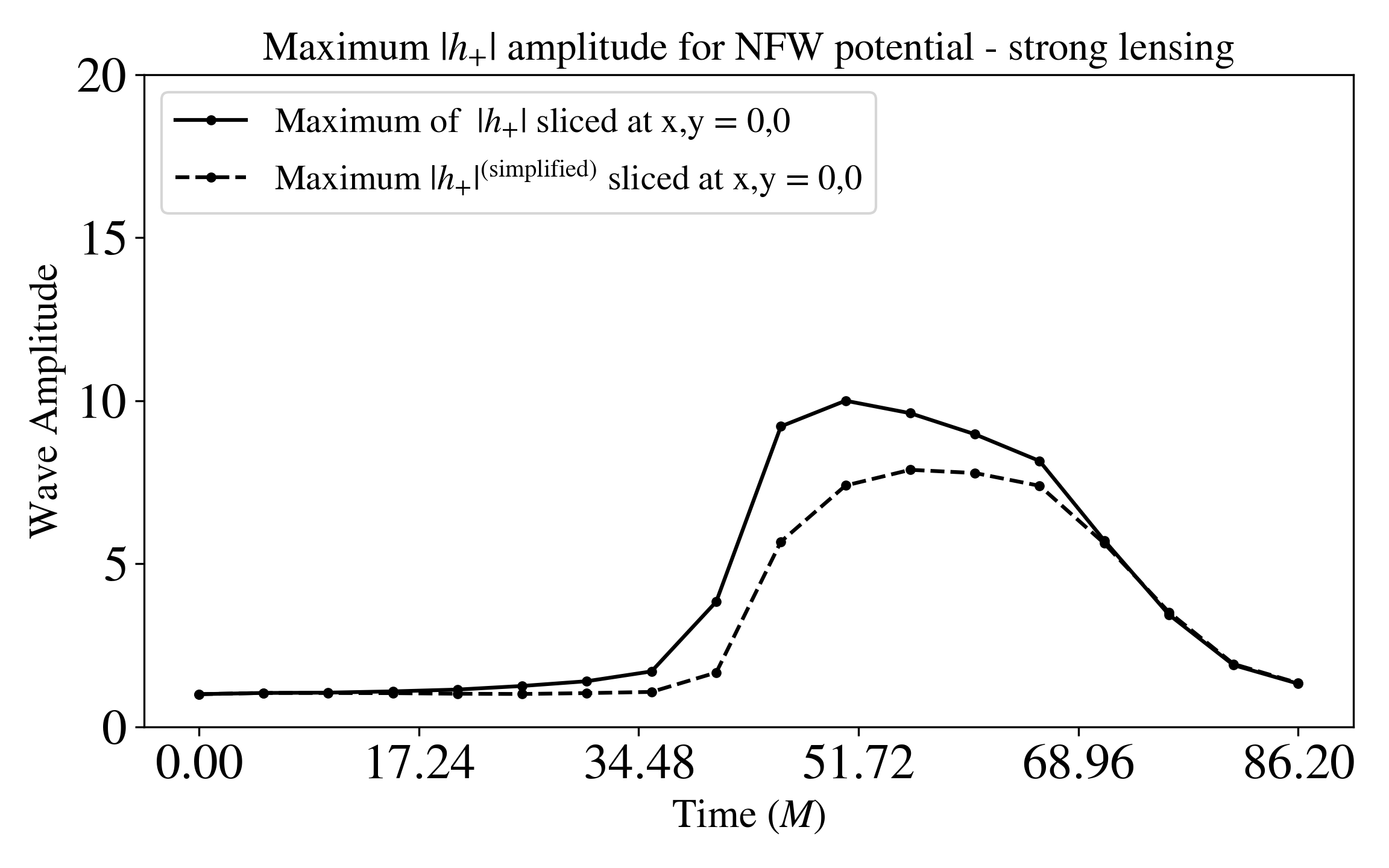}
    \caption{\small Maximum wave amplitude of $h_+$ and $|h_+^{\text{simplified}}|$ - NFW potential, strong gravity case} 
    \label{subfig:nfw_sg_amp}
    \end{subfigure}
        \vskip\baselineskip
    \begin{subfigure}[b]{0.475\textwidth}
    \centering
    \includegraphics[width=8cm]{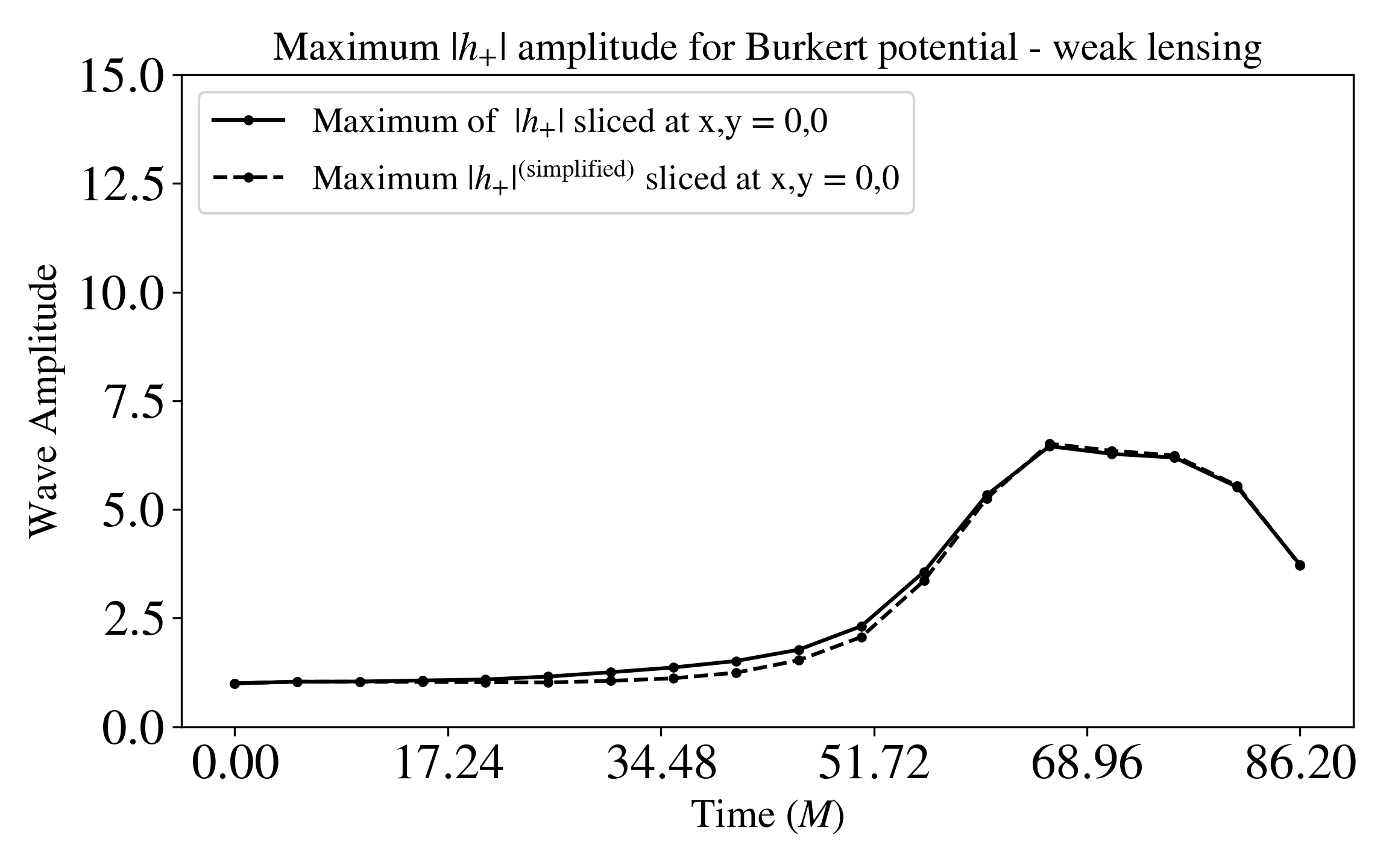}
    \caption{\small Maximum wave amplitude of $h_+$ and $|h_+^{\text{simplified}}|$ - Burkert potential, weak gravity case}    
    \label{subfig:bk_wg_amp}
    \end{subfigure}
        \hfill
    \begin{subfigure}[b]{0.475\textwidth}
    \centering
    \includegraphics[width=8cm]{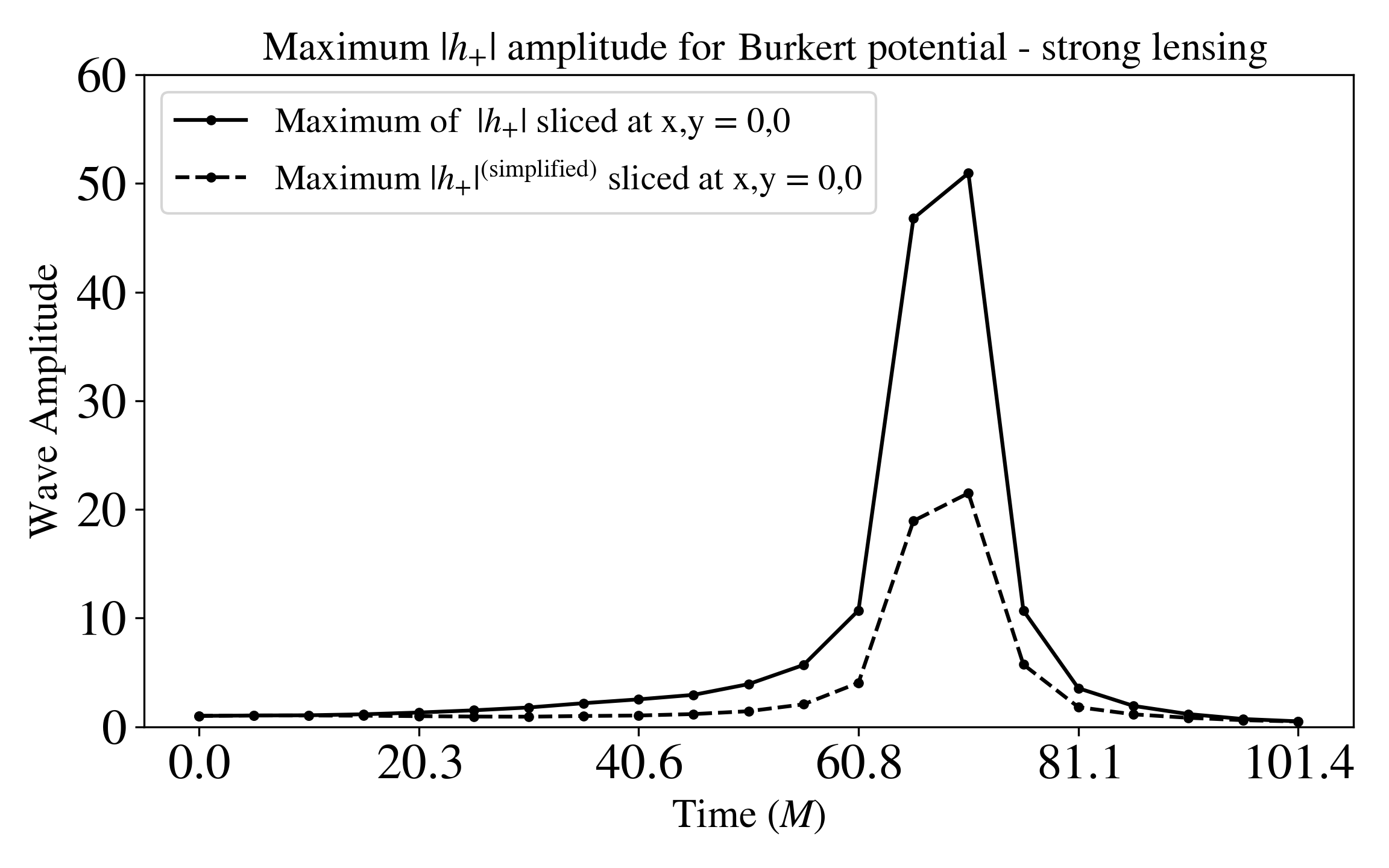}
    \caption{\small Maximum wave amplitude of $h_+$ and $|h_+^{\text{simplified}}|$ - Burkert potential, strong gravity case}    
    \label{subfig:bk_sg_amp}
    \end{subfigure}
\caption{The figure shows the evolution of the maximum amplitude of the $h_+$ and $|h_+^{\text{simplified}}|$ waves, sampled at $(x,y)=(0,0)$, for the Gaussian, NFW, and Burkert potentials in both the weak and strong gravity regimes. In all cases, the maximum amplitude increases during the evolution as a result of gravitational focusing. Among the weak gravity configurations, the Gaussian potential produces the smallest amplification. This is consistent with the corresponding null geodesic structure shown in Fig. \ref{subfig:geodesic_WG_gauss}, where the region over which the ray bundle intersects is the most extended, leading to weaker concentration of the wave energy. In contrast, the Burkert potential in the strong gravity regime yields the largest amplification. As seen in Fig. \ref{subfig:geodesic_SG_bk}, the ray bundles intersect within a much smaller region, resulting in stronger focusing and hence a larger peak amplitude.}\label{fig:amplitudes}
\end{figure*}

\begin{figure*}
\includegraphics[width=16cm]{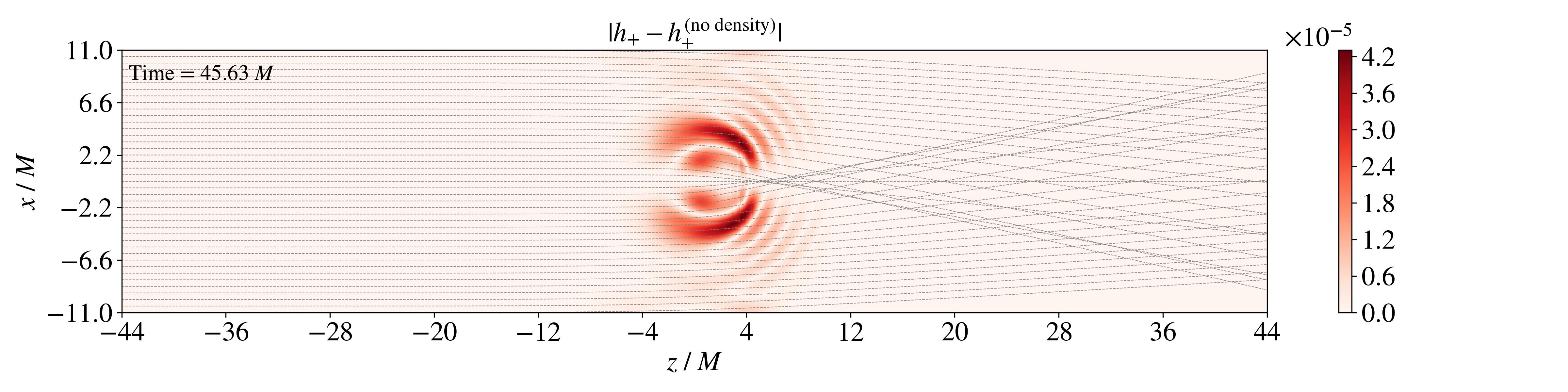}
\includegraphics [width=16cm]{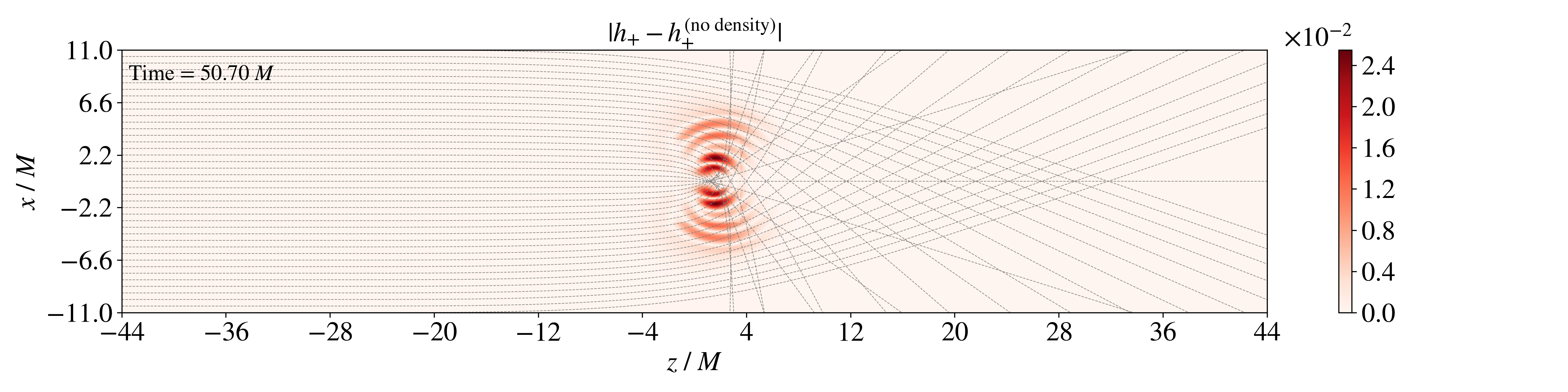}

\caption{The top and bottom panels of the figure show the difference between the two evolutions considered, namely the full Einstein-equation evolution and the density-free evolution with $T_{\mu\nu}=0$, for the Gaussian density in the weak and strong gravity regimes, respectively. The plots are for a time stamp where the GW is traversing the density distribution. The plots also include an overlay of the corresponding null geodesics shown as gray, dashed lines, for both gravity regimes. We find that the difference between the two solutions is extremely small (of order $10^{-5}$) in the weak gravity case, whereas in the strong gravity case it reaches the level of a few percent of the overall GW signal.}\label{fig:no_dens_comparison}
\end{figure*}

In this subsection, we present the results for GWs propagating through a Gaussian potential in the weak gravity regime. Figure \ref{fig:gauss_evol_wg} shows a two-dimensional slice of the evolution of the $h_+ = h_{xx}$ component as the wave propagates from left to right through the density distribution. The figure also shows an overlay of null geodesics for the spacetime, by which we can asses if the wave roughly travels along null geodesics. We present the results in terms of the observable $h_+$ rather than the trace-reversed metric perturbation $\bar{h}_{\mu\nu}$, since these two are related by
$h_{\mu \nu} = \bar{h}_{\mu \nu} - \frac{1}{2}\bar{h}\, g^B_{\mu \nu}.$
\newline
\newline
The incident plane wave is initially located in a flat Minkowski region, where $\phi = 0$, on the negative $z$-axis and $\partial_t h_+$ is set such that the wave propagates from negative to positive $z$. The wave's initial peak amplitude is normalized to unity. Figure \ref{fig:gauss_evol_wg} displays the wave profile at two representative times: shortly after the wave has passed through the central density at $t = 45.63\,M$, and after it has completely exited the curved region, at $t = 65.91\,M$. At $t = 45.63\,M$, the portion of the wave near $x=0$ lags behind the wavefront near the outer boundaries. This is the well-known Shapiro time delay \cite{shapiro1964fourth}. Simultaneously, the amplitude in the central region is enhanced due to gravitational focusing. In the language of wave optics, this enhancement arises from constructive interference of the wave with itself.
\newline\indent
The figure also includes the quantity $|h_+ - h_+^{\text{simplified}}|$, where $h_+^{\text{simplified}}$ denotes the solution of the simplified (scalar wave) equation defined in equation \ref{simplified_evol}. This comparison allows us to isolate the effects of the additional geometric terms involving the Christoffel symbols and the Riemann curvature tensor that appear in the full Einstein equations. The difference is of order $10^{-2}$, and the corresponding plot shows that a small tail-like structure that appears behind the wavefronts near $z=4 \ M$ when we solve the full equations.
\newline\indent
At the later time $t = 65.91\,M$, the plot of $|h_+ - h_+^{\text{simplified}}|$ now shows a characteristic pattern with a maximum amplitude of order $10^{-2}$. Here, we note that the interference pattern in the longitudinal direction is a numerical artifact of the reflecting boundary conditions. We do not expect that this pattern represents a physical effect. 
\newline\indent
Figure \ref{subfig:gauss_wg_amp} shows the evolution of the maximum amplitude of $h_+$ and $|h_+^{\text{simplified}}|$, tracked along $(x,y)=(0,0)$. As expected from focusing, the amplitude increases as the wave traverses the curved region. The corresponding null geodesic overlay in Fig. \ref{fig:gauss_evol_wg} shows that the strongest convergence of the ray bundle occurs at a distance of approximately $z = 40\,M$ from the peak of the potential. A noticeable difference develops between $h_+$ and $h_+^{\text{simplified}}$ when the wave passes through the central density. However, their final amplitudes are similar, differing only at the level of $10^{-2}$, consistent with the behaviour seen in Fig. \ref{fig:gauss_evol_wg}.
\newline\indent
In figure \ref{fig:no_dens_comparison} (top figure), we show the difference between the simulations evolving the full Einstein equation versus the density-free equations (indicated as $h_+^{\text{(no density)}}$ in the figure). From this comparison, we see that a very weak amplitude of GW is scattered in almost all directions by the density field. Again, we included an overlay of null-geodesic trajectories on top of the two-dimensional slice. We see that parts of the wave fronts is moving in directions that are clearly not orthogonal to the null-geodesics.

\subsection{Gaussian potential: strong gravity regime}\label{subsec:gauss_sg}
\begin{figure*}
\includegraphics[width=16cm]{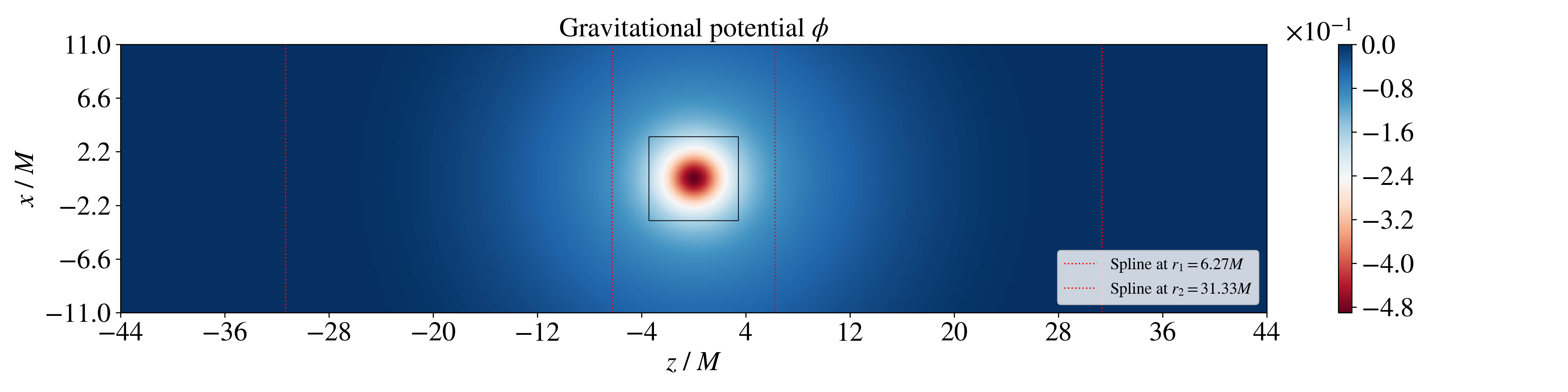}
\includegraphics [width=16cm]{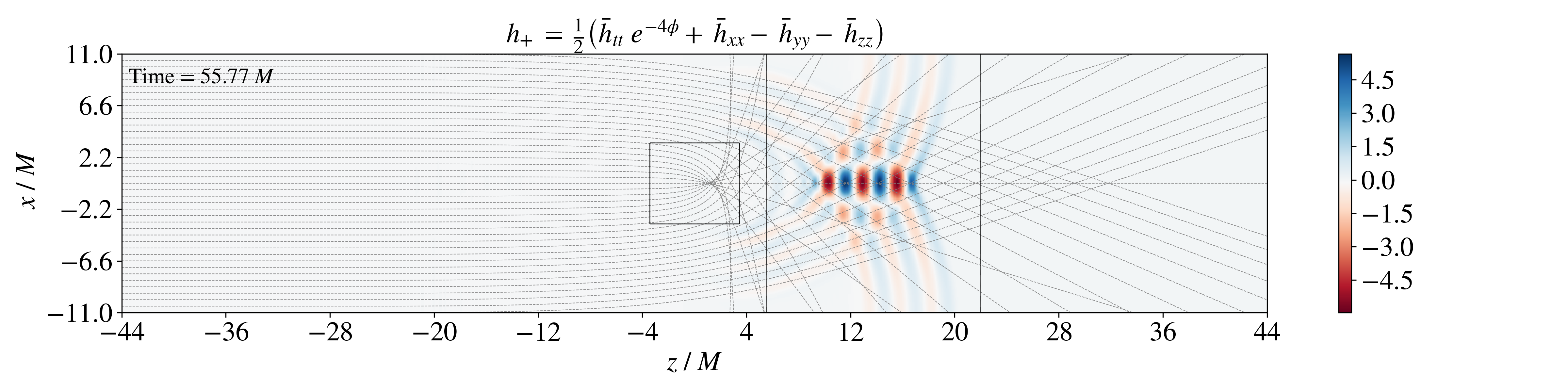}
\includegraphics [width=16cm]{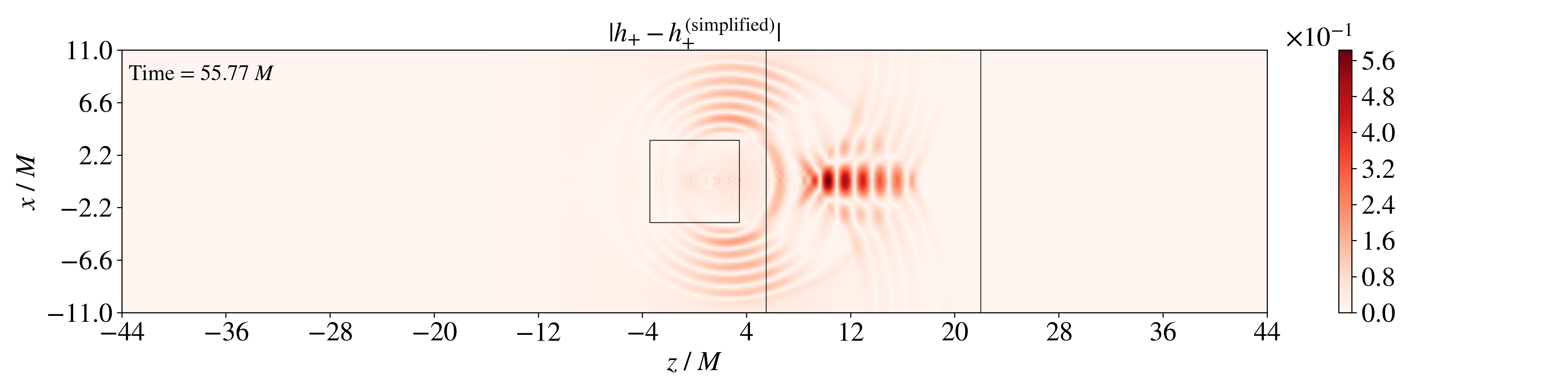}
\includegraphics [width=16cm]{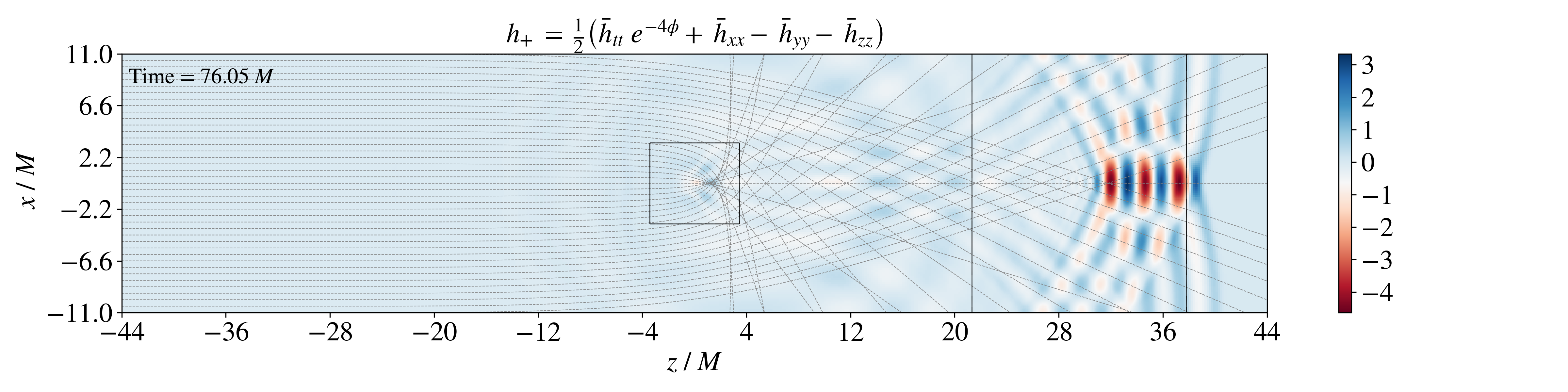}
\includegraphics [width=16cm]{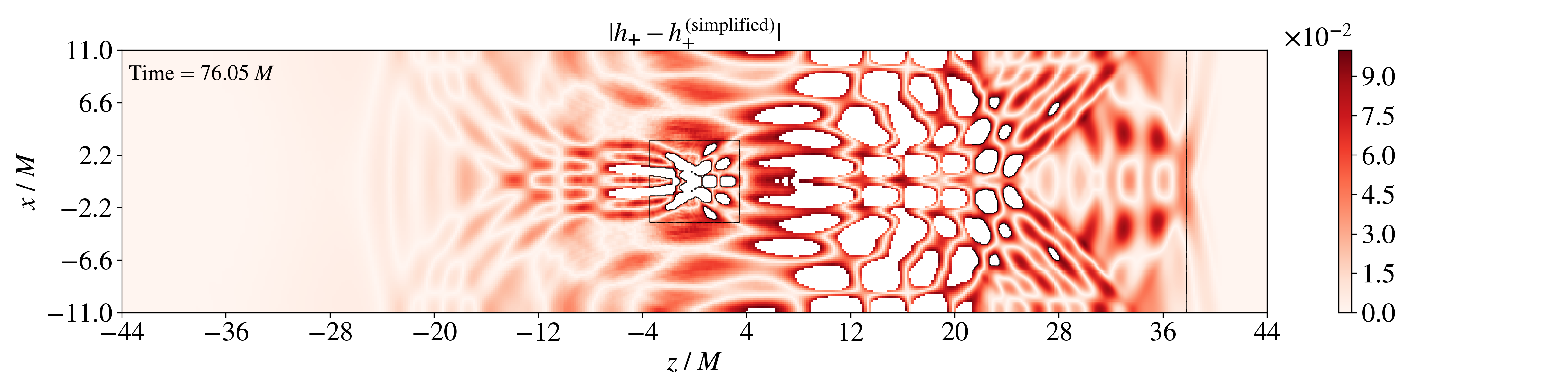}

\caption{From top to bottom: The top figure shows the distribution of gravitational potential $\phi$ along the $xz$-plane for the Gaussian potential in the strong gravity regime. The second figure shows the 2D slice of $h_+$ after the wave has traversed the central density at $t=55.77 \ M$. The third figure shows the absolute difference between $h_+$ and $|h_+^{\text{simplified}}|$ which are evolved independently with the same initial conditions. The fourth and fifth figures show the wave after it had completely exited the curved region at the later time $t=76.05 \ M$. The figure for $h_+$ has an overlay of null geodesics for the potential (including the spline). White regions indicate values that fall outside the range of the colour bar on the right. The black outlined box within the simulation domain mark the AMR patch, where the resolution is twice that of the base grid.}\label{fig:gauss_evol_sg}
\end{figure*}

In this subsection, we present the results for GW propagation through a Gaussian potential in the strong gravity regime, corresponding to $\phi_{\mathrm{max}} = -0.49$. Figure \ref{fig:gauss_evol_sg} shows a two-dimensional slice of the evolution of the $h_+$ component as the wave propagates from left to right through the lens. The figure also shows an overlay of null geodesics for the potential configuration.
\newline\indent
As before, the incident plane wave is initially located in a flat Minkowski region on the negative $z$-axis and propagates toward positive $z$. Its initial peak amplitude is normalized to unity. Figure \ref{fig:gauss_evol_sg} displays the wave profile at two representative times: shortly after the wave has passed through the central density, at $t = 55.77\,M$, and after it has completely exited the curved region, at $t = 76.05\,M$. At $t = 55.77\,M$, the portion of the wave near $x=0$ exhibits a pronounced Shapiro time delay. In addition, the wave amplitude in the central region is significantly enhanced by strong gravitational focusing. The figure also shows that the wave develops a distinct ``tail'' produced by the large curvature near the lens/density distribution. This appears because waves originating from the upper part of the $x$-axis are deflected downward, while those from the lower part are deflected upward. This feature is therefore much more pronounced than in the weak gravity case. The corresponding plot of $|h_+ - h_+^{\text{simplified}}|$ shows differences of order $10^{-1}$. The figure also shows that, in the region surrounding the density peak, part of the wave is scattered in multiple directions with prominent amplitude in the direction orthogonal to the initial direction of propagation. Parts of these scattered waves do not seem to propagate orthogonally to any null geodesics. This is also seen in Fig. \ref{fig:no_dens_comparison} (bottom figure), where we plot the difference between simulations solving the full Einstein equation vs. the density-free equation (represented as $h_+^{\text{(no density)}}$). Here, we see that an amplitude of a few percent of the total GW signal is scattered in almost all directions, similar to the weak gravity case but with much stronger amplitude. This behaviour is not predicted by the null geodesic description of wave propagation. We therefore find that the wave dynamics exhibit a few percentage deviation from the null geodesic approximation when the wave propagates through the density distribution. This indicates that, for sensitive GW observations, the effects associated with the full Einstein equations are relevant and should be incorporated into accurate signal modelling. Nonetheless, the density-free approximation captures most of the wave dynamics.  
\newline\indent
At the later time $t = 76.05\,M$, after the wave has exited the curved region, its amplitude is slightly reduced compared to the earlier snapshot. This shows that wave amplification does not necessarily persist at all stages of lensing in the wave-optics regime. In the $|h_+ - h_+^{\text{simplified}}|$ plot, the portion of the wave in the range $28 \leq z \leq 40$ exhibits differences with a maximum amplitude of order $10^{-2}$. Thus, despite the strong gravity setting, the final magnitude of the difference is comparable to that found in the weak gravity case. This is, however, only true along the principal axis.
\newline\indent
As for the weak gravity simulation, the imposed reflection symmetry about the $x$-axis causes part of the waves seen at $t = 55.77\,M$ to be reflected back into the computational domain from both of the $x$-axis boundaries. The resulting interference pattern across the domain is therefore a numerical artifact of the reflecting boundary conditions rather than a physical effect.
\newline\indent
Figure \ref{subfig:gauss_sg_amp} shows the evolution of the maximum amplitude of $h_+$ and $|h_+^{\text{simplified}}|$ tracked along the principal axis (the $z$-axis). By comparing with Fig. \ref{fig:gauss_evol_sg}, we see that the strongest amplification occurs in the region where the null geodesics converge and overlap significantly. Once the wave has passed through this region of overlapping geodesics, the amplitude decreases. Although a substantial deviation develops between $h_+$ and $h_+^{\text{simplified}}$ when the wave is in the central density, their final amplitudes are again similar when the wave has left the density, and the amplitudes differ only at the level of $10^{-2}$, consistent with the behaviour shown in Fig. \ref{fig:gauss_evol_sg}.

\section{Numerical results for GW propagation through dark matter halos: NFW and Burkert profiles}\label{sec:results_nfw_bk}
In this section, we present the results from propagating GWs through NFW and Burkert gravitational potentials. For each potential, we overlay the corresponding null geodesics on the simulation results (see appendix \ref{subsec:nullgeodesics} for details on null geodesic computations). In addition, we compare the results obtained by solving the full Einstein equation with 
(scalar wave)equation given in Eq. \ref{simplified_evol}.

\subsection{NFW profile: weak gravity regime}\label{subsec:nfw_wg}
\begin{figure*}
\includegraphics[width=16cm]{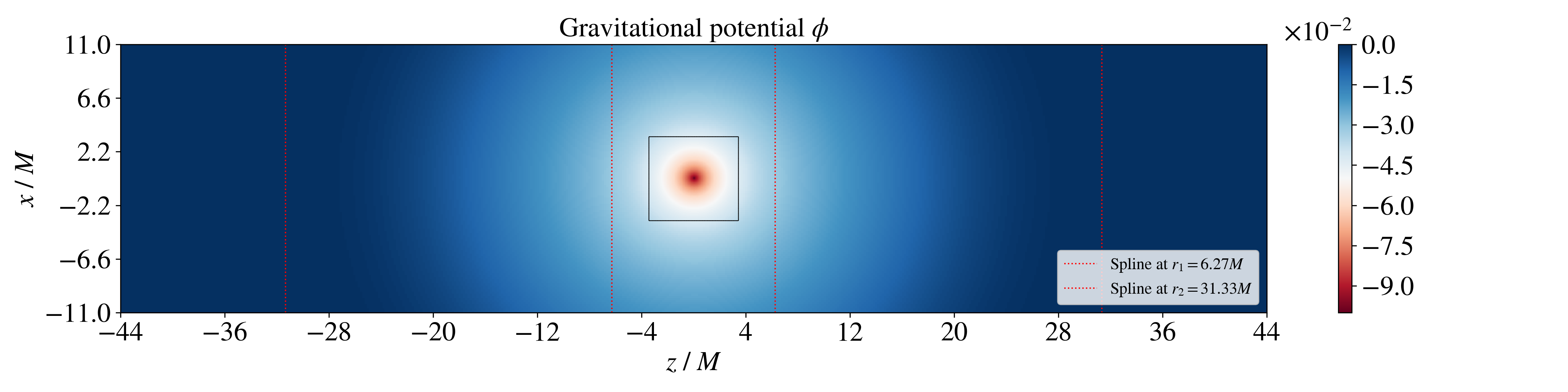}
\includegraphics [width=16cm]{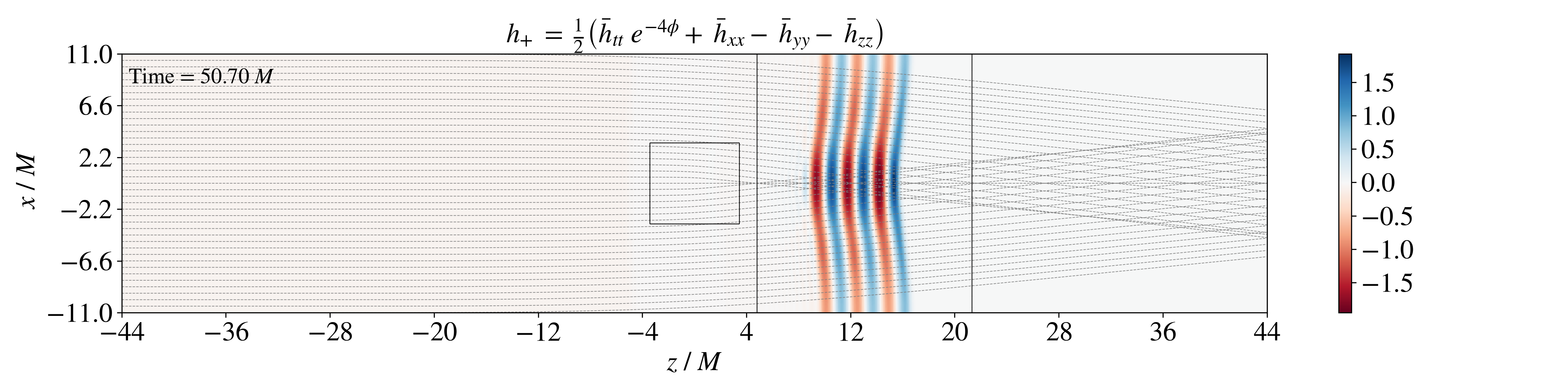}
\includegraphics [width=16cm]{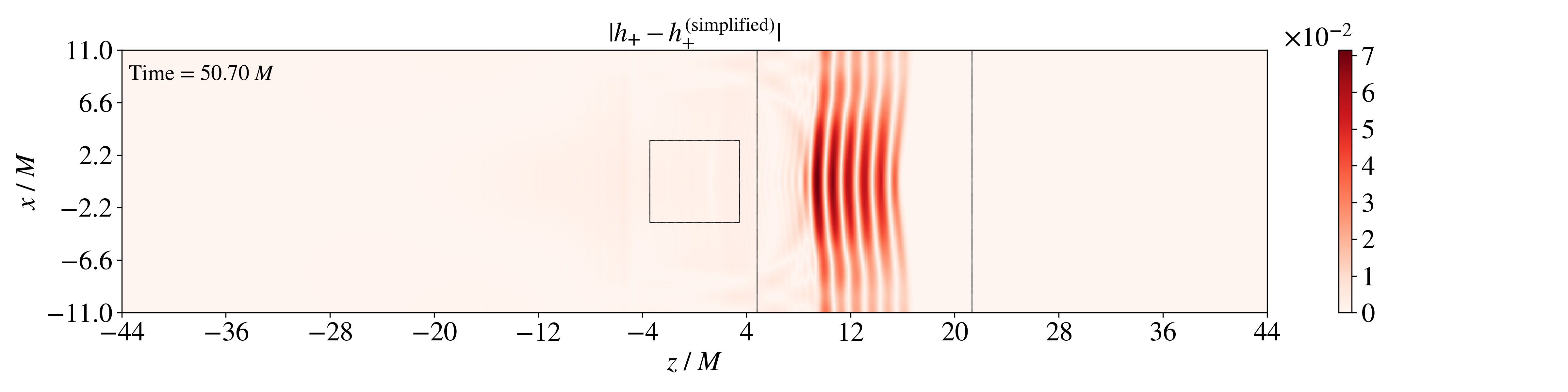}
\includegraphics [width=16cm]{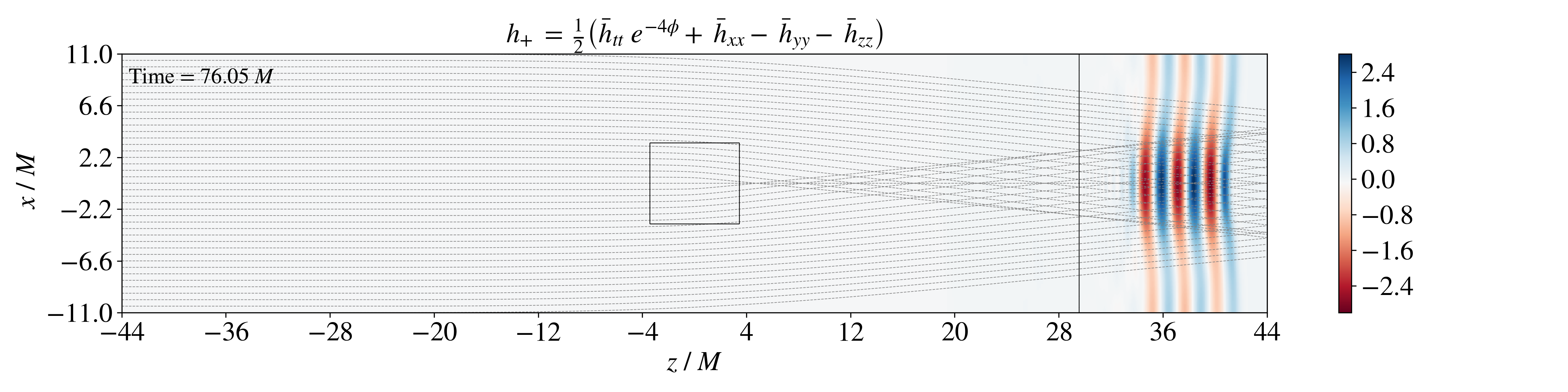}
\includegraphics [width=16cm]{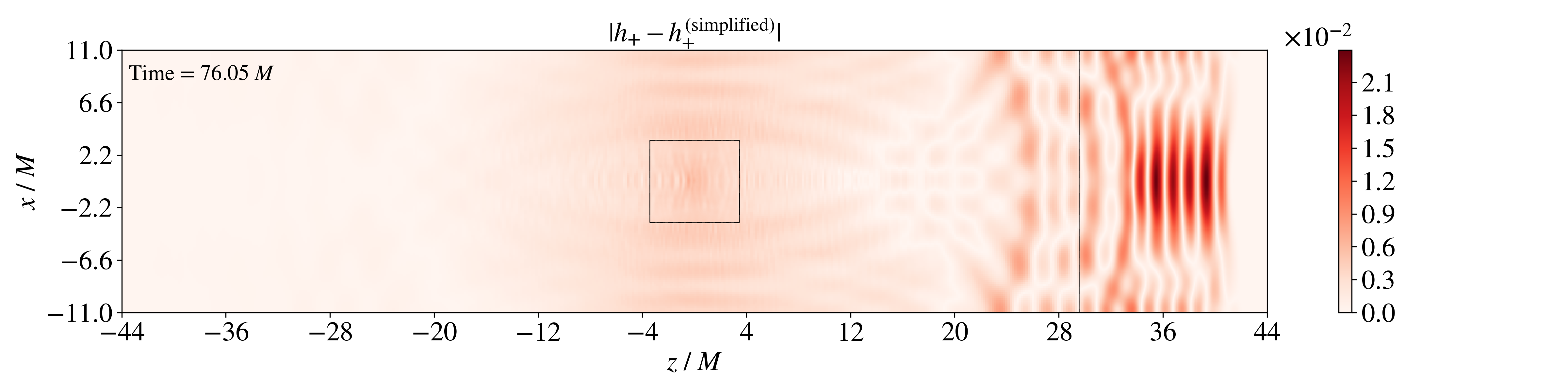}

\caption{From top to bottom: The top figure shows the gravitational potential $\phi$ along $xz$-plane for the NFW potential in the weak gravity regime. The second figure shows the 2D slice of $h_+$ after the wave has traversed the central density at time $t=50.70 \ M$. The third figure shows the absolute difference between $h_+$ and $|h_+^{\text{simplified}}|$ which are evolved independently with the same initial conditions. The fourth and fifth figures show the wave after it had completely exited the curved region at the later time $t=76.05 \ M$. The plot of $h_+$ has an overlay of null geodesics of the spacetime. The black outlined boxes inside the simulation domain is where the AMR patch exists and has twice the resolution of base grid. }\label{fig:nfw_evol_wg}
\end{figure*}

In this subsection, we present the results for GW propagation through the NFW potential in the weak gravity regime. Figure \ref{fig:nfw_evol_wg} shows a two-dimensional slice of the evolution of the $h_+ = h_{xx}$ component as the wave propagates from left to right through the density distribution.
\newline\indent
As in the Gaussian case, the incident plane wave is initially located in a flat Minkowski region on the negative $z$-axis and propagates towards positive $z$. Figure \ref{fig:nfw_evol_wg} displays the wave profile at two representative times: shortly after the wave has passed through the central potential region, at $t=50.70\,M$, and after it has completely exited the curved region, at $t=76.05\,M$. At $t=50.70\,M$, the portion of the wave near $x=0$ is delayed relative to the outer parts of the wavefront, indicating the expected Shapiro time delay. Additionally, the amplitude in the central region is enhanced due to gravitational focusing. In the plot of $|h_+ - h_+^{\text{simplified}}|$, the difference is of order $10^{-2}$. At the later time $t=76.05\,M$, after the wave has exited the curved region, its amplitude has increased further. The overall wave pattern is qualitatively similar to that found in the weak gravity Gaussian case.
\newline\indent
In the plot of $|h_+ - h_+^{\text{simplified}}|$ a more pronounced interference pattern is visible than in the Gaussian case, although the magnitude of the difference remains at the same order. As before, reflections occur on the boundaries of the simulation along the $x$-axis, causing part of the wave to re-enter the computational domain in the longitudinal direction. The resulting interference pattern is therefore a numerical artifact of the boundary conditions rather than a physical effect.
\newline\indent
Figure \ref{subfig:nfw_wg_amp} shows the evolution of the maximum amplitude of the waves $h_+$ and $|h_+^{\text{simplified}}|$, tracked along the principal axis. The corresponding null geodesics overlayed in Fig. \ref{fig:nfw_evol_wg} show that the strongest convergence of the ray bundle occurs at $z \gtrsim 44\,M$. This explains why the wave amplitude continues to increase even after the wave has passed through the central potential region. The geodesic overlay further suggests that the wave propagation roughly follows the expected null geodesic structure. Although a small difference develops between $h_+$ and $h_+^{\text{simplified}}$ during the evolution, their final amplitudes remain very similar and differ only at the level of $10^{-2}$ as shown in Fig. \ref{fig:nfw_evol_wg}.

\subsection{NFW potential: strong gravity regime}\label{subsec:nfw_sg}
\begin{figure*}
\includegraphics[width=16cm]{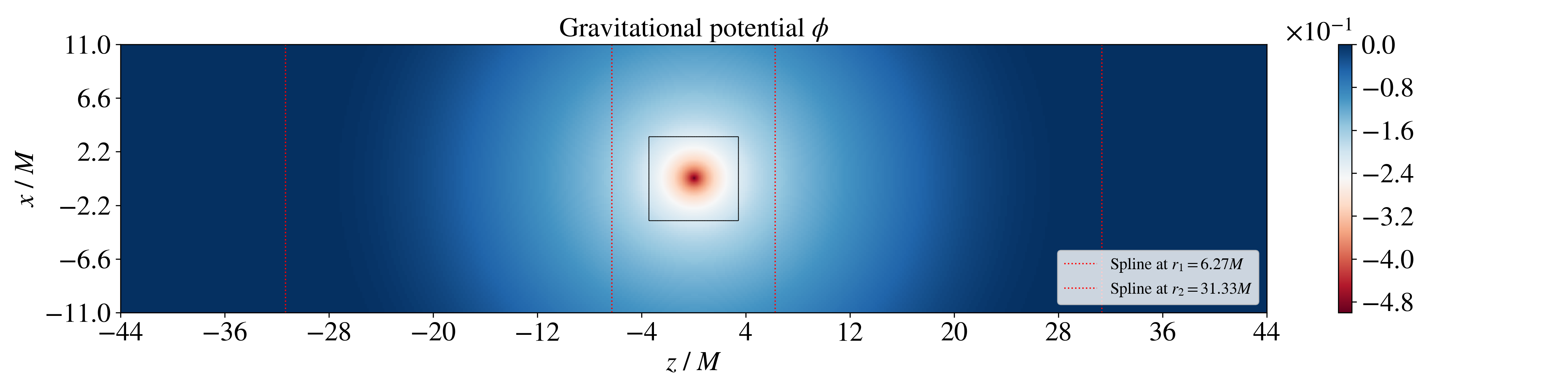}
\includegraphics [width=16cm]{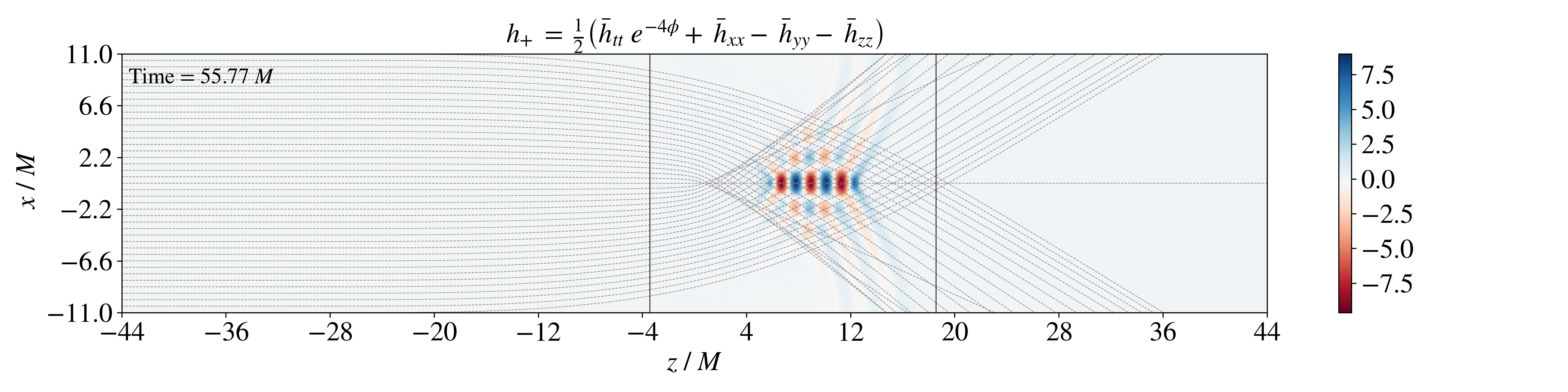}
\includegraphics [width=16cm]{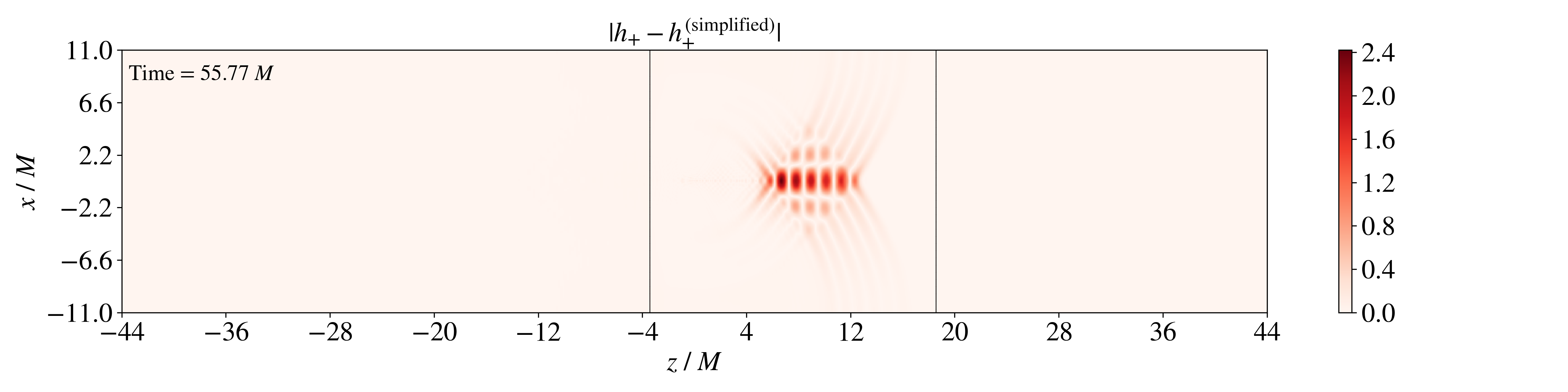}
\includegraphics [width=16cm]{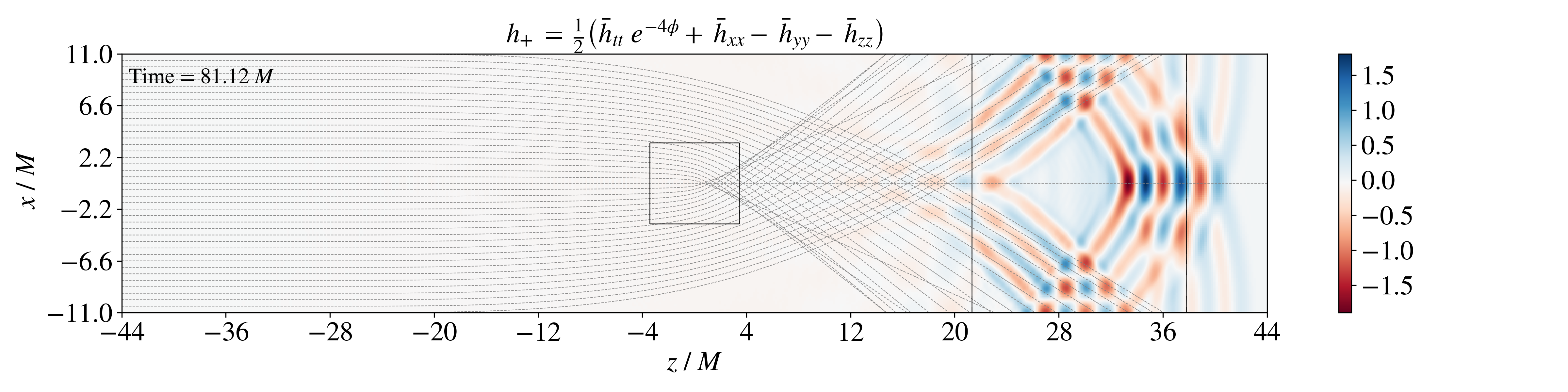}
\includegraphics [width=16cm]{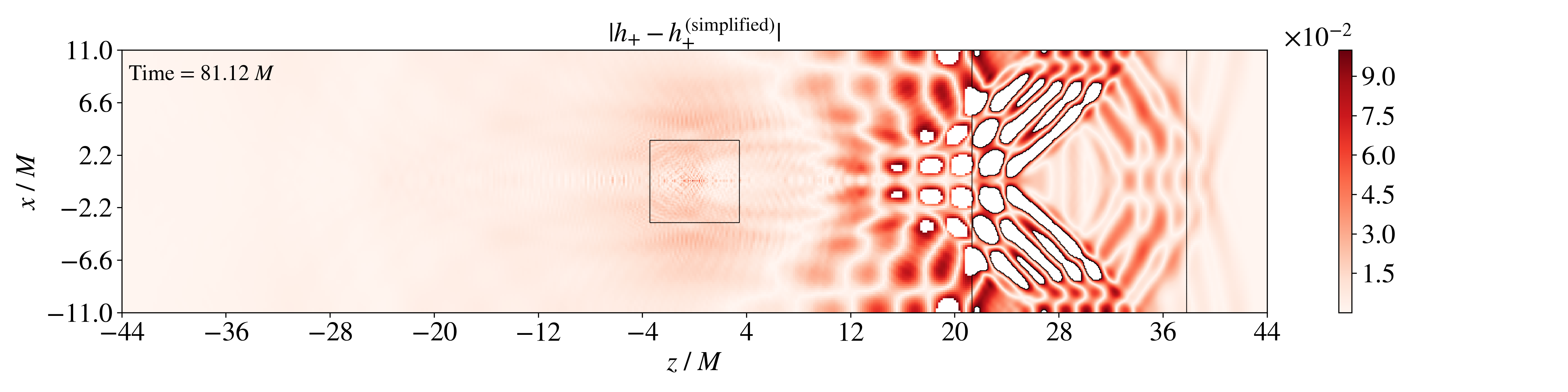}

\caption{From top to bottom: The top figure shows the gravitational potential $\phi$ along $xz$-plane for the NFW potential in the strong gravity regime. The second figure shows the 2D slice of $h_+$ after the wave has traversed the central density at $t=55.77 \ M$. The third figure shows the absolute difference between $h_+$ and $|h_+^{\text{simplified}}|$ which are evolved independently with the same initial conditions. The fourth and fifth figures show the wave after it had completely exited the curvature at the later time $t=81.12 \ M$. White regions indicate values that fall outside the range of the colour bar on the right. The black outlined box within the simulation domain mark the AMR patch, where the resolution is twice that of the base grid. }\label{fig:nfw_evol_sg}
\end{figure*}

In this subsection, we present the results for GWs propagating through the NFW potential in the strong gravity regime, corresponding to $\phi_{\mathrm{max}} = -0.49$. Figure \ref{fig:nfw_evol_sg} shows a two-dimensional slice of the evolution of the $h_+$ component as the wave propagates from left to right through the density distribution.
\newline\indent
As in the earlier simulations, the incident plane wave is initially located in a flat Minkowski region on the negative $z$-axis and propagates toward positive $z$. Figure \ref{fig:nfw_evol_sg} displays the wave profile at two representative times: shortly after the wave has passed through the central potential region, at $t=55.77\,M$, and after it has completely exited the curved region, at $t=81.12\,M$.
\newline\indent
At $t=55.77\,M$, the waves in the central region exhibit a substantial increase in amplitude due to strong gravitational focusing. In addition, the wave develops a pronounced tail structure, caused by the strong curvature which mixes the upper and lower parts of the wavefront. Unlike the Gaussian case, the wave is not scattered by the density inhomogeneity with the NFW profile. This is due to the strong localisation of the dark matter density inhomogeneity. The corresponding plot of $|h_+ - h_+^{\text{simplified}}|$ shows that the difference is of the same order of magnitude as the GW signal itself, indicating that the (scalar wave)wave equation fails to capture important features of the evolution in this regime. This is again similar to the Gaussian case, although with a distinctly different interference pattern.
\newline\indent
At the later time $t=81.12\,M$, after the wave has exited the curved region, its amplitude has decreased significantly relative to the earlier snapshot. Thus, the amplification does not persist throughout the evolution. Due to the strong bending induced by the curvature, parts of the wave are reflected from the $x$-axis boundaries and re-enter the computational domain. In the plot of $|h_+ - h_+^{\text{simplified}}|$, the portion of the wave in the range $32 \leq z \leq 40$ shows a maximum difference of order $10^{-2}$. Thus, despite the strong gravity setting, the final deviation in this region is comparable in magnitude to that found in the weak gravity case, which is also $10^{-2}$.
\newline\indent
Figure \ref{subfig:nfw_sg_amp} shows the evolution of the maximum amplitude of the waves $h_+$ and $|h_+^{\text{simplified}}|$, tracked along the principal axis. The strongest amplification occurs in the region where the null geodesics exhibit the greatest overlap (seen in the overlay in Fig. \ref{fig:nfw_evol_sg}), namely for $0<z<20$. After the wave has traversed this region, the amplitude decreases. This behaviour is consistent with the interpretation that the wavefront propagates along directions determined by the null geodesic structure. Although a substantial difference develops between $h_+$ and $h_+^{\text{simplified}}$ while the wave passes through the central potential region, their final amplitudes again become similar and differ only at the level of $10^{-2}$, as seen in Fig. \ref{fig:nfw_evol_sg}.
\newline\newline
Overall, the results obtained with the Gaussian and NFW profiles are quantitatively very similar, but qualitatively they differ in their exact interference patterns.

\subsection{Burkert potential: weak gravity regime}\label{subsec:bk_wg}
\begin{figure*}
\includegraphics[width=16cm]{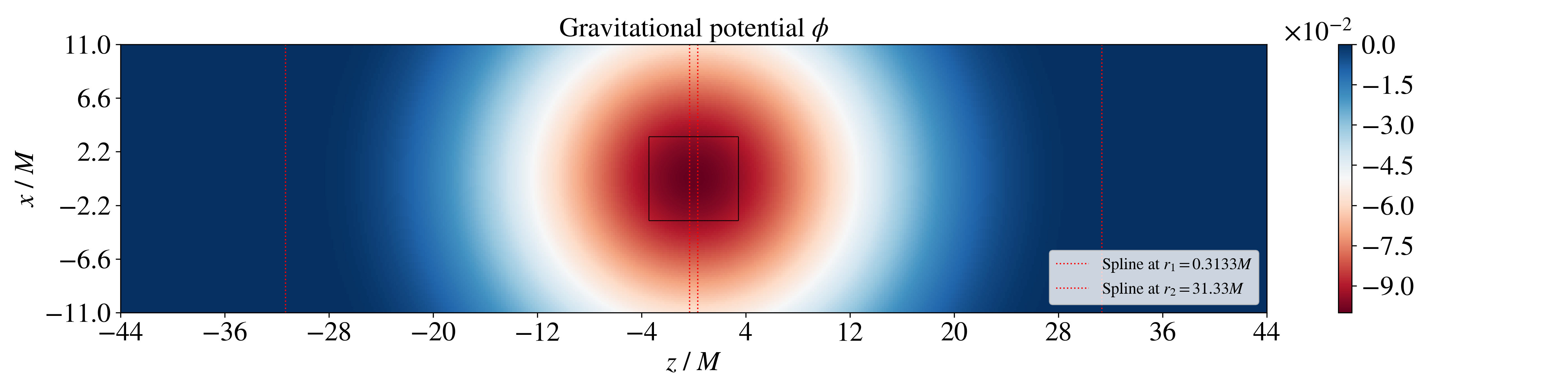}
\includegraphics [width=16cm]{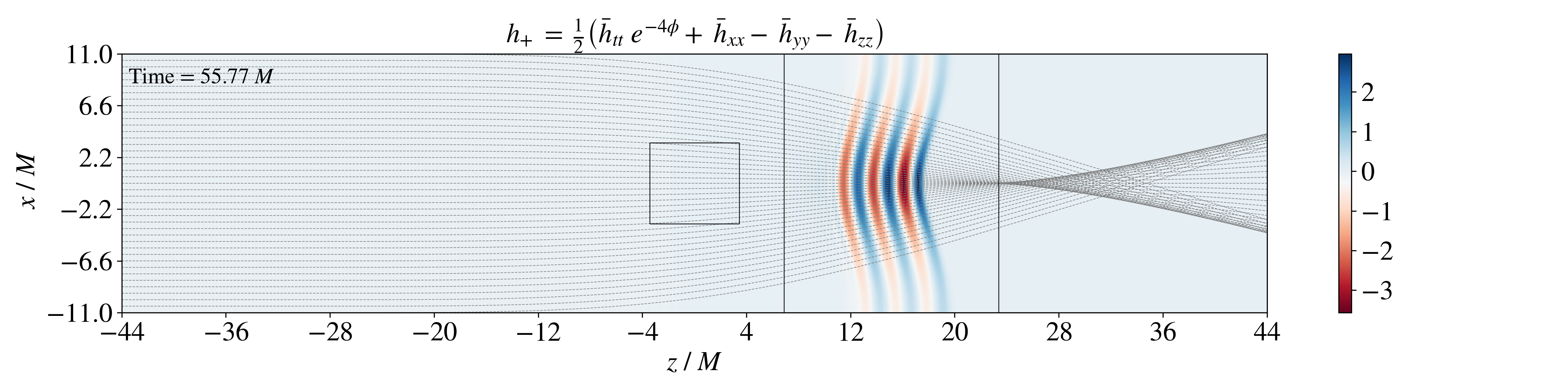}
\includegraphics [width=16cm]{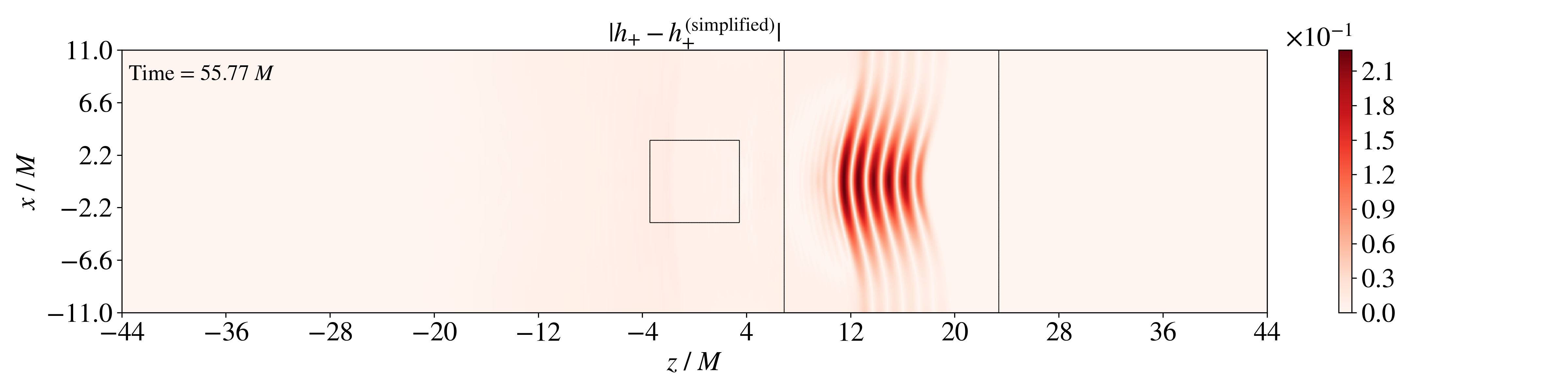}
\includegraphics [width=16cm]{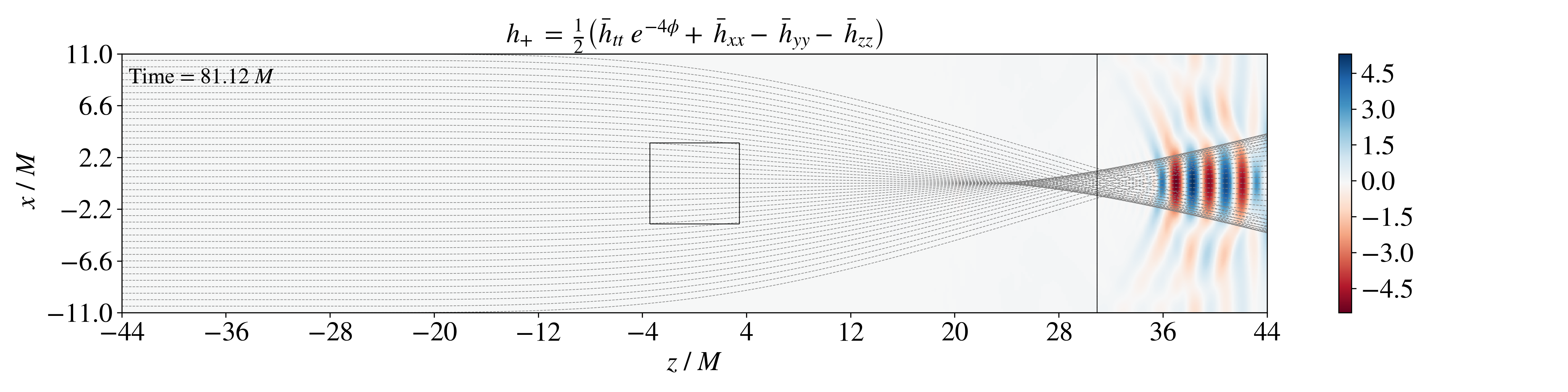}
\includegraphics [width=16cm]{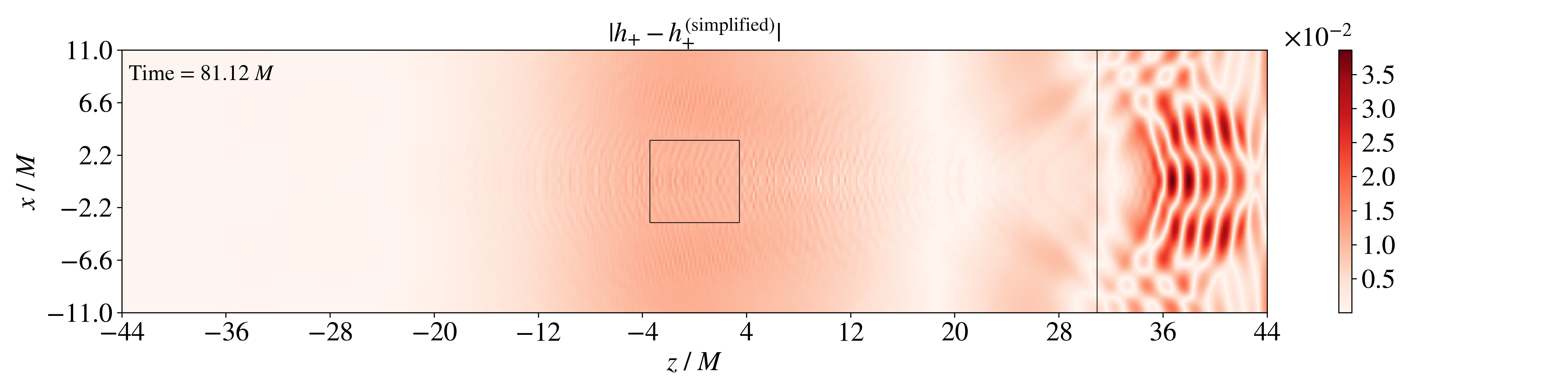}

\caption{From top to bottom: The top figure shows the gravitational potential $\phi$ along $xz$-plane for the Burkert potential in the weak gravity regime. The second figure shows the 2D slice of $h_+$ after the wave has traversed the central density at $t=55.77 \ M$. The third figure shows the absolute difference between $h_+$ and $|h_+^{\text{simplified}}|$ which are evolved independently with the same initial conditions. The fourth and fifth figures show the wave after it had completely exited the curvature at the later time $t=81.12 \ M$. The figure for $h_+$ has an overlay of null geodesics of the considered spacetime. The black outlined boxes inside the simulation domain are where the AMR patch exists and has twice the resolution of base grid. }\label{fig:bk_evol_wg}
\end{figure*}

In this subsection, we present the results for GW propagation through the Burkert potential in the weak gravity regime. Figure \ref{fig:bk_evol_wg} shows the gravitational potential together with a two-dimensional slice of the evolution of the $h_+$ component as the wave propagates from left to right through the lens/density distribution. Note that because of the spline implementation (see table \ref{tab:spline}), the Burkert profile differs noticeably from the Gaussian and NFW potentials. In particular, the spline is implemented so as to avoid steep gradients in the derivatives of $\phi$, thereby producing a smoother potential profile. For the Gaussian and NFW profiles, the spline starts well outside the core of the potential, so only the weak outer region of $\phi$ is modified. For the Burkert profile, whose scale radius $R_s$ is larger, $\phi$ and its derivatives are still large at this radius and forcing them to zero between $r_1 = 6.27 \ M$ and $r_2 = 31.33\ M$ would introduce rapid variations in $\phi'$ and $\phi''$. We therefore setup $r_1$ inside the core potential, so that the transition is spread over a larger range and the derivatives of $\phi$ remain smooth. Another consequence of such large $R_s$ value is that this results in a small amplitude of the central overdensity.
\newline\newline
As in the previous cases, the incident plane wave is initially located in a flat Minkowski region on the negative $z$-axis and propagates towards positive $z$. Figure \ref{fig:bk_evol_wg} displays the wave profile at two representative times: shortly after the wave has passed through the central potential region, at $t=55.77\,M$, and after it has completely exited the curved region, at $t=81.12\,M$. At $t=55.77\,M$, the portion of the wave near $x=0$ is delayed relative to the outer parts of the wavefront, while the amplitude in the central region is enhanced, as in the other weak gravity simulations. In the plot of $|h_+ - h_+^{\text{simplified}}|$, the difference is of order $10^{-1}$. At the later time $t=81.12\,M$, after the wave has exited the curved region, its amplitude has increased further. The plot of $|h_+ - h_+^{\text{simplified}}|$ exhibits a clear interference pattern, more pronounced than in the Gaussian and NFW cases, although the overall magnitude of the difference remains of the same order. As in the previous simulations, reflections from the $x$ axis boundaries appear, causing part of the wave to re-enter the computational domain in the $x$-direction. Again, the resulting interference pattern is a numerical artifact of the boundary conditions rather than a physical effect.
\newline\indent
Figure \ref{subfig:bk_wg_amp} shows the evolution of the maximum amplitude of the waves $h_+$ and $|h_+^{\text{simplified}}|$ tracked along the principal axis. The corresponding null geodesic overlay in Fig. \ref{fig:bk_evol_wg} indicates that the strongest convergence of the ray bundle occurs at $z \simeq 30\,M$ from the peak of the potential. This is precisely the region where the wave amplitude reaches its maximum. Beyond this point, the null geodesics begin to diverge from each other, and the wave amplitude correspondingly decreases. This suggests that wave propagation direction is in agreement with the null geodesics. Although a small difference develops between $h_+$ and $h_+^{\text{simplified}}$ while the wave traverses the central region, their final amplitudes remain very similar and differ only at the level of $10^{-2}$, as shown in Fig. \ref{fig:bk_evol_wg}.

\begin{figure*}
\includegraphics [width=15cm]{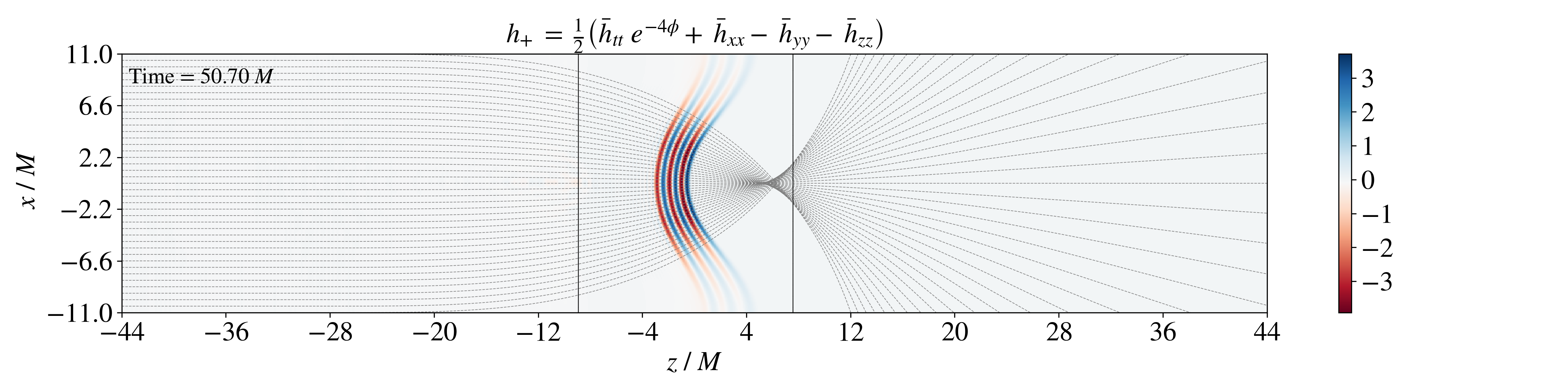}
\includegraphics [width=15cm]{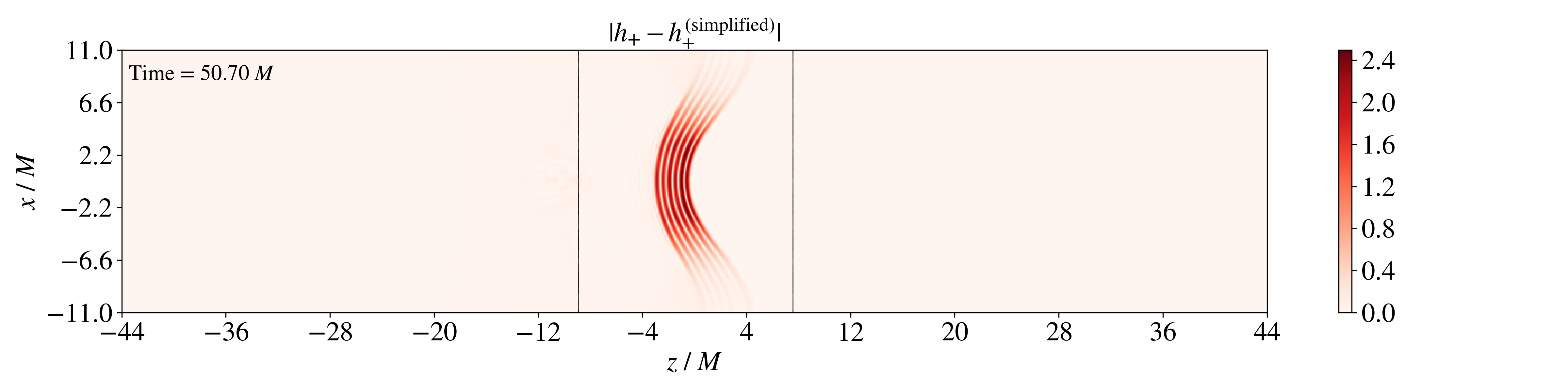}
\includegraphics [width=15cm]{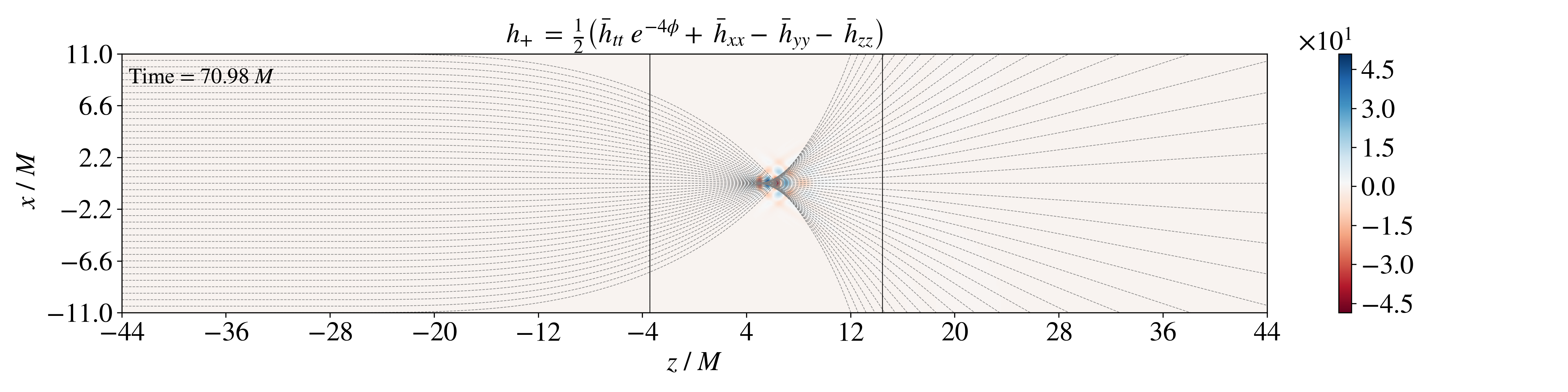}
\includegraphics [width=15cm]{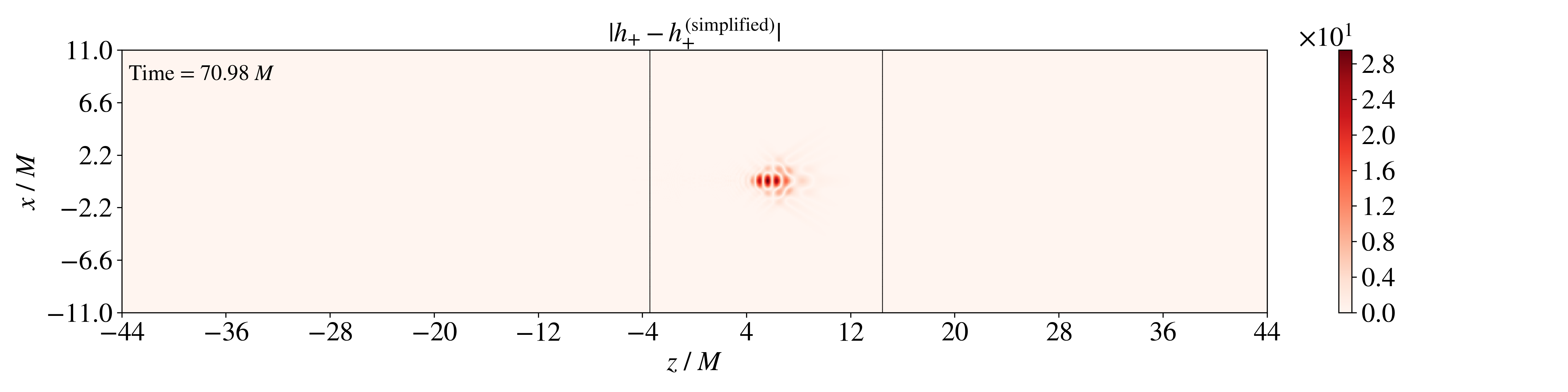}
\includegraphics [width=15cm]{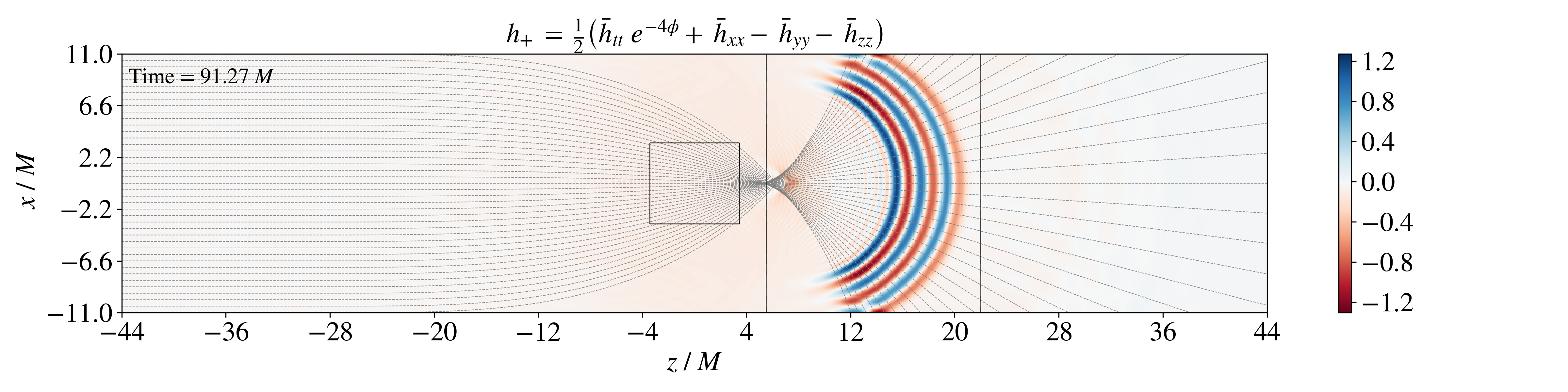}
\includegraphics [width=15cm]{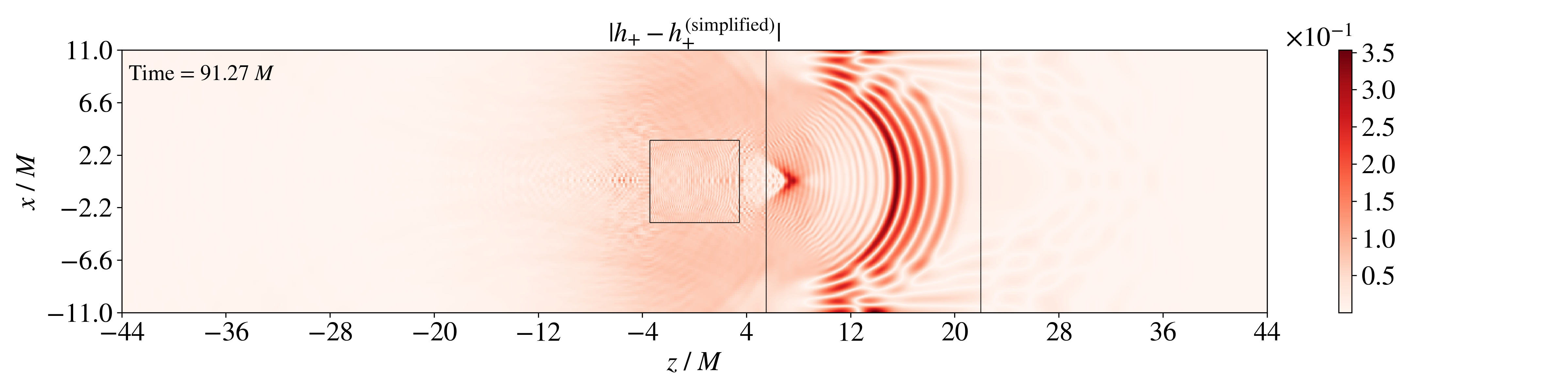}

\caption{2D slices of $h_+$ for waves propagating through the Burkert potential in strong gravity regime. For each time instance, we show two plots - one that shows the evolution of full Einstein equation and another that shows the absolute difference between $h_+$ and $|h_+^{\text{simplified}}|$. From top to bottom: The top two plots shows the $h_+$ after the wave has traversed the central density at $t=50.7 \ M$. The second and third plots at $t=70.98 \ M$ show the instance when waves are intensely focused and localised within a small region. The two last plots show the wave scattering after it has traversed the peak of the potential at the later time $t=91.27 \ M$, respectively. The plot of $h_+$ has an overlay of null geodesics of the corresponding spacetime. The black outlined boxes inside the simulation domain are where the AMR patch exists and has twice the resolution of the base grid. }\label{fig:bk_evol_sg}
\end{figure*}

\subsection{Burkert potential: strong gravity regime}\label{subsec:bk_sg}
In this subsection, we present the results for GWs propagating through the Burkert potential in the strong gravity regime, corresponding to $\phi_{\mathrm{max}}=-0.49$. Figure \ref{fig:bk_evol_sg} shows the two-dimensional evolution of the $h_+$ component as the wave propagates from left to right through the lens. The results are presented in terms of the observable $h_+$. As in the earlier simulations, the incident plane wave is initially located in a flat Minkowski region on the negative $z$-axis and propagates toward positive $z$. Figure \ref{fig:bk_evol_sg} shows the evolution at three representative times. The first snapshot, at $t=50.70\,M$, corresponds to the stage when the wave is about to enter the peak-potential region. The second, at $t=70.98\,M$, is taken shortly after the wave has propagated through the central density. The final snapshot, at $t=91.27\,M$, shows the later stage of the evolution, where the wave has exited the curved region. In the two first snapshots, the difference $|h_+ - h_+^{\text{simplified}}|$ is significantly stronger than in the earlier simulations.
\newline\indent
In the second snapshot ($t=70.98\,M$) the wave exhibits strong focusing (convergence of null geodesics), and the maximum interference occurs in a localized region approximately within $4<z<10$. This is why the amplitude is so strongly enhanced in this region. In contrast to the Gaussian case and similar to the NFW case, we do not observe orthogonal scattering due to the density.
At the later time, $t=91.27\,M$, the wave is scattered over a broad angular range, extending from negative to positive $x$-directions and almost resembling an approximately spherical wave source. In particular, we note that the wave front has now flipped its convexity. This did not happen in the simulations with the Gaussian or NFW profiles. Furthermore, a small part of the wave remains within the central potential region. At this stage, the difference $|h_+ - h_+^{\text{simplified}}|$ decreases to order $10^{-1}$. We do not show the subsequent evolution, since reflections from the $x$-boundaries begin to dominate the dynamics and the simulations therefore no longer provide physically useful information.
\newline \newline \indent
Figure \ref{subfig:bk_sg_amp} shows the evolution of the maximum amplitude of the waves $h_+$ and $|h_+^{\text{simplified}}|$, tracked along the principal axis. The strongest amplification occurs in the region where the null geodesics exhibit the greatest overlap, namely in the region $0<z<4$ (see the overlay in Fig. \ref{fig:bk_evol_sg}). After the wave has traversed this region, the amplitude decreases (corresponding to the null geodesics diverging from each other). All these wave propagation effects are in line with null geodesics predictions. Although a substantial difference develops between $h_+$ and $h_+^{\text{simplified}}$ while the wave passes through the central density, their final amplitudes become closer again and differ at the level of $10^{-1}$ when the wave has exited the curved region.

\section{Summary and Conclusions}\label{sec:summary_conclusions}
We have investigated gravitational wave (GW) lensing in the wave optics regime by numerically propagating plane GW packets through spherically symmetric spacetime inhomogeneities representing e.g. dark matter halos. Such analyses are of particular interest for future GW observations, including those by pulsar timing arrays (PTAs), since lensing-induced distortions of the waveform in the wave optics regime may provide an additional probe of the underlying matter distribution.
\newline\indent
A central goal of this work was to go beyond the usual linearized treatment of wave propagation and instead evolve the full Einstein equations. Overall, the six resulting simulations show that the qualitative behaviour of GW propagation is controlled jointly by the lens profile and the strength of the gravitational potential. In the weak gravity regime, all three considered density profiles (Gaussian, NFW, and Burkert) exhibit the expected signatures of Shapiro time delay and moderate focusing, with the locations of maximum amplification broadly consistent with the corresponding null geodesic convergence regions. In this regime, the (scalar wave)evolution equation corresponding to the scalar GW equations often studied in the literature reproduces the main features of the full solution reasonably well, and the final discrepancies remain small, only of a few percent.
\newline\indent
The situation changes significantly in the strong gravity regime. Here, the wave dynamics becomes more sensitive to the full geometric structure of the spacetime, and the propagation can no longer be described adequately by the (scalar wave)scalar GW evolution alone. In particular, we find that the wave may develop pronounced tails, multidirectional scattering, and stronger deviations from the null geodesic expectation when it traverses the central high curvature region. The comparison between the full Einstein equation evolution and the corresponding density-free evolution indicates that the presence of the density source induces corrections at the level of a few percent in the propagated GW signal. This is consistent with recent analytical work \cite{Atli:2025ilq} which showed that finite-wavelength propagation through matter distributions can lead to deviations from the pure geometric-optics limit. For sufficiently precise GW observations, such effects may therefore become relevant for accurate signal modelling and could, in principle, provide additional information about e.g. dark matter halo distributions.
\newline\indent
Our simulations show that the structure and shape of the gravitational potential plays an important role in determining the lensing of the GWs. The wave amplification is not controlled solely by the total mass of the lens, but also by the shape of the potential and, more specifically, by how sharply it focuses the null-ray bundles. This is evident from the comparison among the Gaussian, NFW, and Burkert profiles. In particular, the strong gravity Burkert case exhibits the most pronounced amplification, with the peak amplitude increasing by nearly a factor of $50$ relative to the incident wave. However, while the full and simplified (scalar wave) evolutions can differ substantially when the wave is inside the central curvature region, the discrepancy typically decreases after the wave exits the lens, often settling to the level of a few percent. The only exception we find is simulations with the Burkert profile in the strong-gravity limit. This suggests that the curvature-dependent terms in the full Einstein equation have their strongest impact while the wave is traversing the high curvature region, rather than producing a large cumulative effect far from the lens similar to Integrated Sachs-Wolfe effect \cite{Sachs1967}. To confirm this, it would nonetheless be useful to extend our simulations to situations where several consecutive structures are traversed in order to better understand to what extent wave-optics effects accumulate over long distances.
\newline\indent
A further finding is that the simplified (scalar wave) evolution equation often studied in the literature is reliable only in restricted regimes. In the weak gravity simulations, it reproduces the main features of the full solution reasonably well, at least qualitatively and often semi-quantitatively. In the strong gravity cases, however, its limitations become evident. In particular, it does not capture the full multidirectional scattering pattern that appears when the wave interacts with the central density, nor does it fully reproduce the large transient amplification observed in the strongest focusing configurations. Furthermore, for the Burkert profile in the strong gravity regime, the discrepancy remains large even after the wave has exited the high-curvature region: the (scalar wave) evolution differs from the full Einstein equation result by more than $35\%$ after the waves have completely exited the curved region. The simplified (scalar wave) description should therefore be used with caution in regimes where the curvature is strong and the wavelength is not negligible compared with the characteristic lensing scale.
\newline\newline
For the Gaussian density, we studied three sets of equations: the usual full Einstein equation, the simplified (scalar wave) equation and a set of density-free equations where we set $T_{\mu\nu}=0$. In the latter case, the GW evolution is treated as propagation through curved spacetime but where the matter content inducing the curvature itself is ignored. This comparison for instance shows that once the density source is included, a weak but non-negligible component of the GW is scattered into multiple directions, with an amplitude reaching a few percent of the overall signal in the strong gravity case. This behaviour is qualitatively similar to that seen in the weak gravity case, although its magnitude is larger in the strong gravity case. Such multidirectional scattering is not predicted by the null geodesic description, which only tracks the propagation of rays in the geometric-optics limit and therefore cannot account for finite-wavelength effects of this kind. The comparison therefore indicates that, when the wave propagates through a finite sized density distribution, the full dynamics can deviate from the null geodesic approximation at the level of a few percent. Although small, this deviation is systematic and may become relevant in precision modelling of lensed GW signals. We did not observe such an effect for GWs propagating through NFW and Burkert profiles (results not shown). We expect this is simply because the considered FWHM of $\phi(r)$ for these profiles leads to density fields that are too localised and/or weak to produce any significant deviations as those seen in the Gaussian case.
\newline\indent
From an observational perspective, our results suggest that lensing by intermediate astronomical structures may imprint subtle but potentially measurable signatures on GW signals. Such effects could otherwise be misidentified as evidence for modified gravity or other beyond General Relativity phenomena if they are not modelled carefully. More broadly, the present analysis demonstrates that full wave simulations in inhomogeneous spacetimes can provide a useful framework for disentangling propagation effects from source physics and from genuine departures from general relativity.
\newline \indent
Finally, several directions remain open for future investigation. In the present work, we have focused primarily on the $h_+$ component and have not studied in detail the behaviour of the other components of $h_{\mu\nu}$ during propagation through curved spacetime. It would be particularly interesting to investigate polarization mixing and possible birefringence-type effects in more general settings. Additional extensions include improved numerical infrastructure with AMR and better absorbing boundaries, and a more systematic analysis of how the lens profile, wave frequency, and source-lens-observer geometry jointly determine the observable waveform distortions.
\newline \indent
In summary, our results show that GW propagation in curved spacetimes is well approximated by the geometric optics picture only in sufficiently weak gravitational fields. In stronger curvature regimes, finite-wavelength effects, density induced corrections, and the detailed structure of the lens can produce non-trivial modifications of the waveform that are not captured by (scalar wave)evolution equations alone. Accurate modelling of such phenomena therefore requires the full Einstein equation and will be increasingly important as GW observations enter a regime of greater precision and sensitivity. In addition, when training machine learning models to make inferences based on GW data including lensing data, there is a risk of obtaining biased results if the training is done on simple, analytical models (as in e.g. \cite{liu2025identifying}) instead of more realistic simulations (see e.g. \cite{koloniari2025robustness} for some of the many considerations necessary for good machine learning training).

\begin{acknowledgments}
This project is funded by VILLUM FONDEN, grant VIL53032 (PI: SMK). Research at Perimeter is supported by the Government of Canada through the Department of Innovation, Science, and Economic Development, and by the Province of Ontario through the Ministry of Colleges and Universities. Part of the computational work was performed using Perimeter's HPC system``Symmetry" and DeiC Large Memory HPC System (Hippo) managed by the eScience Center at the University of Southern Denmark.
\newline\indent
APS thanks Mandana Amiri and Mark Halpern of the Physics and Astronomy department, University of British Colombia (UBC) for hosting him during his research stay at UBC where part of this work was carried out. The authors thank the anonymous referee for comments that significantly improved the manuscript and in particular for identifying an error in one of the original setups.
\newline\newline
{\bf Author contribution statement}: APS performed the analytical computations. APS performed the numerical work under the guidance of SMK, RH and ES. Analysis of the numerical work was led by APS but all authors contributed. Conclusions were drawn by APS and SMK who jointly wrote the manuscript with comments from ES and RH.
\end{acknowledgments}

\appendix

\section{Null geodesics}\label{subsec:nullgeodesics}
\begin{figure}[htbp]
\centering
   \begin{subfigure}[b]{0.475\textwidth}
   \centering
   \includegraphics[width=8cm]{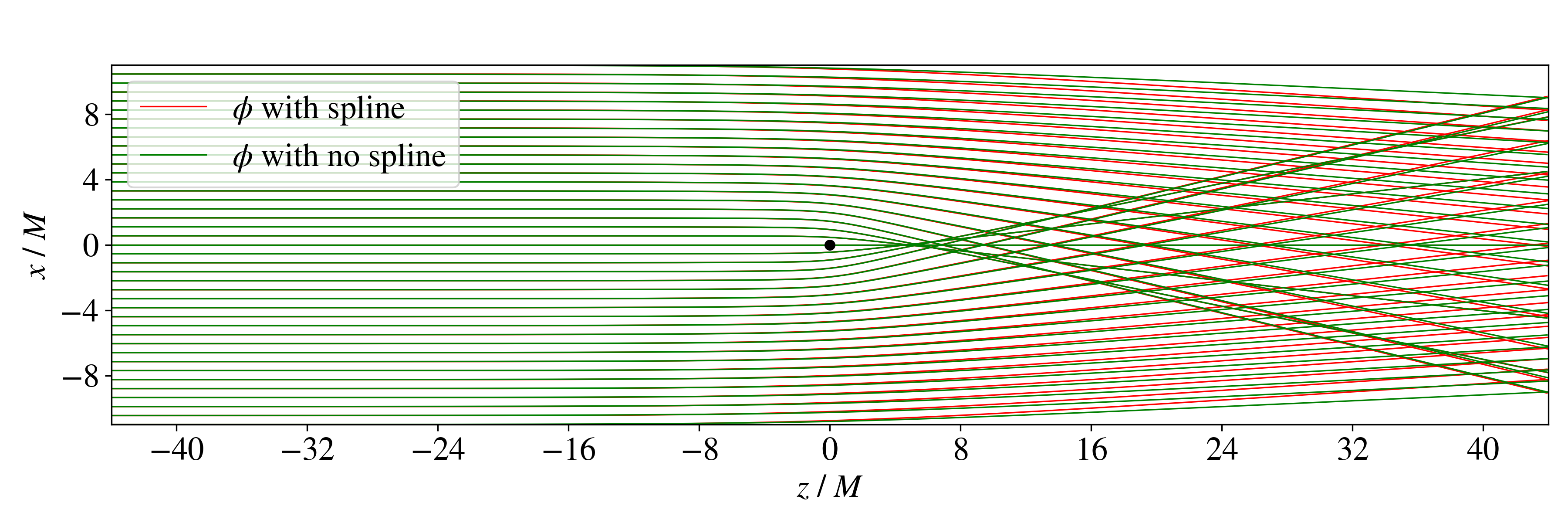}
   \caption{\small Null geodesics for Gaussian potential - weak gravity case}    
   \label{subfig:geodesic_WG_gauss}
   \end{subfigure}
       \hfill
   \begin{subfigure}[b]{0.475\textwidth}
   \centering
   \includegraphics[width=8cm]{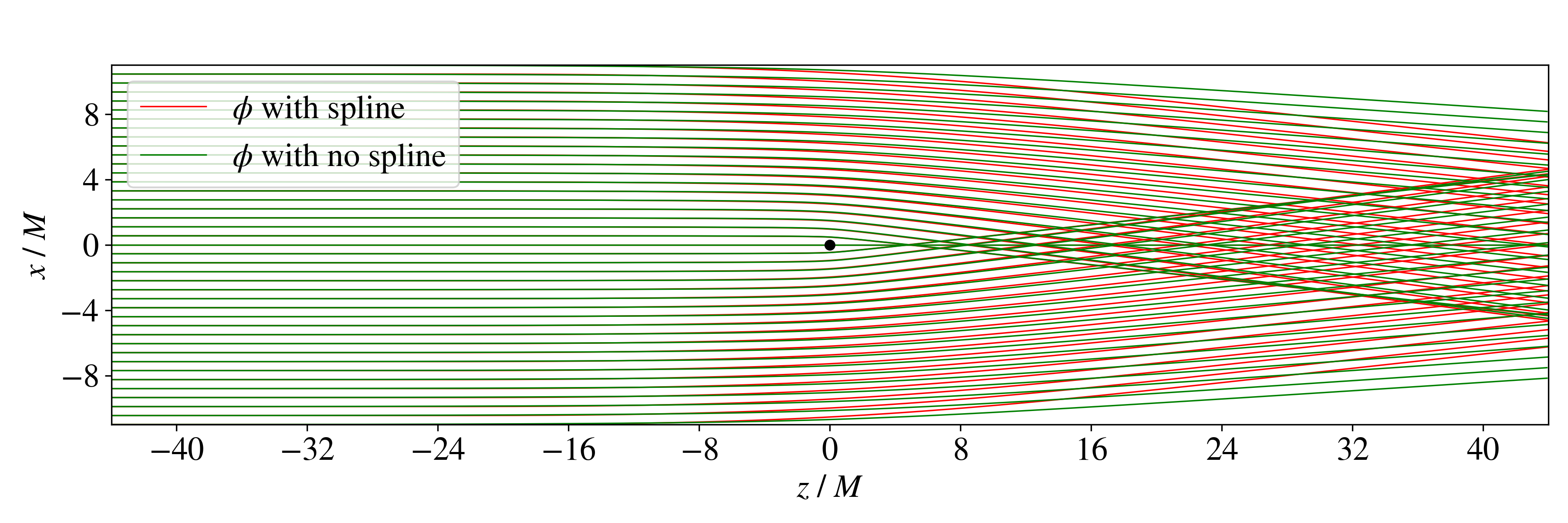}
   \caption{\small Null geodesics for NFW potential - weak gravity case}    
   \label{subfig:geodesic_WG_nfw}
   \end{subfigure}
       \hfill
   \begin{subfigure}[b]{0.475\textwidth}
   \centering
   \includegraphics[width=8cm]{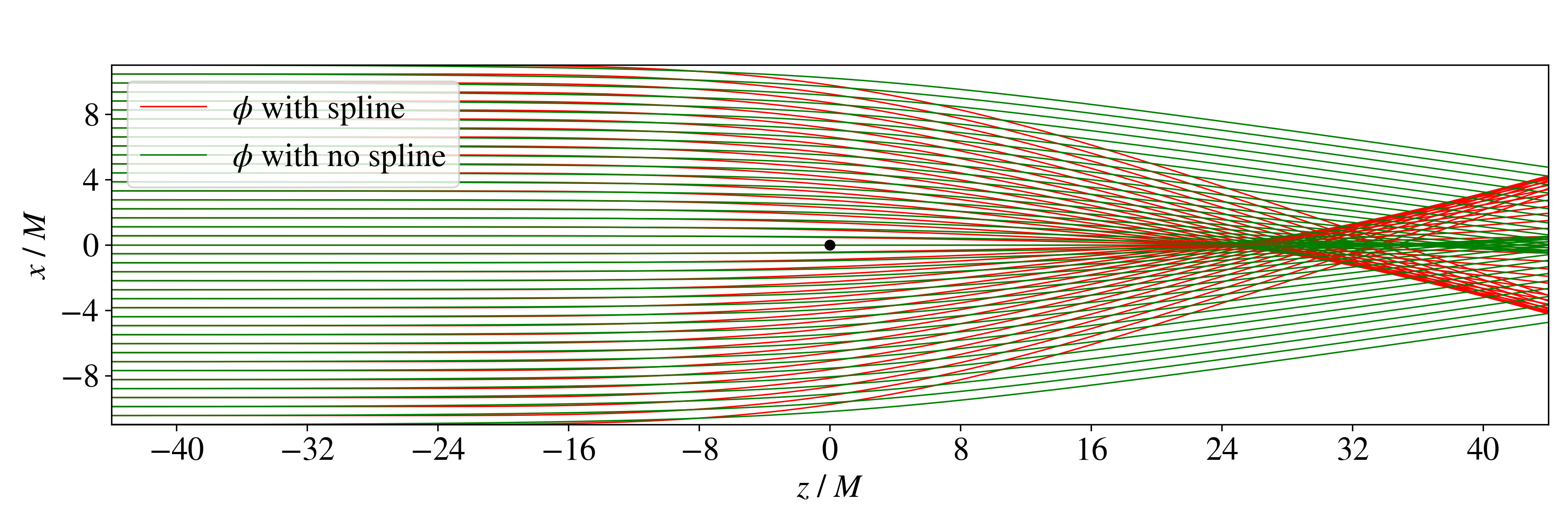}
   \caption{\small Null geodesics for Burkert potential - weak gravity case}    
   \label{subfig:geodesic_WG_bk}
   \end{subfigure}

\caption{Null geodesic trajectories of a plane wavefront for three gravitational potentials (the Gaussian, NFW, and Burkert profiles) with $\phi_{\text{max}} = -0.1$ corresponding to the weak gravity case. The black dot denotes the peak curvature point. We compare the geodesics of the spacetime with and without the splines, showing that they are qualitatively similar and differ only by focusing distance and sharpness of focus.}
\label{fig:geodesics1}
\end{figure}

\begin{figure}
\centering
   \begin{subfigure}[b]{0.475\textwidth}
   \centering
   \includegraphics[width=8cm]{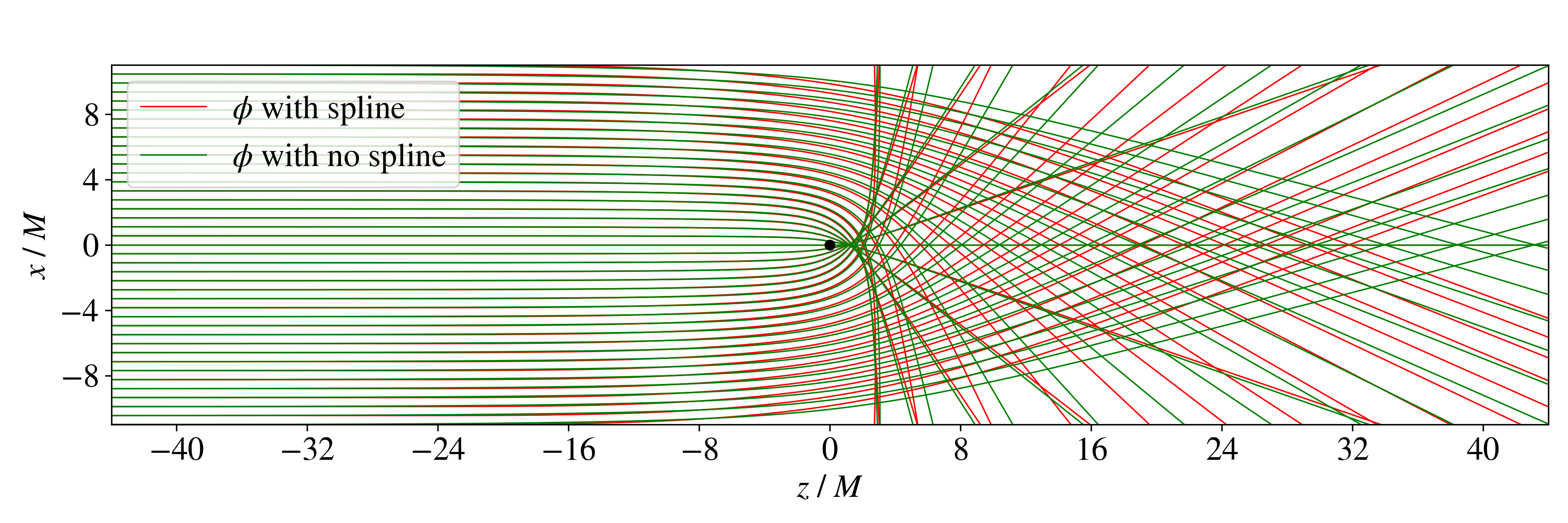}
   \caption{\small Null geodesics for Gaussian potential - strong gravity case}    
   \label{subfig:geodesic_SG_gauss}
   \end{subfigure}
       \hfill
   \begin{subfigure}[b]{0.475\textwidth}
   \centering
   \includegraphics[width=8cm]{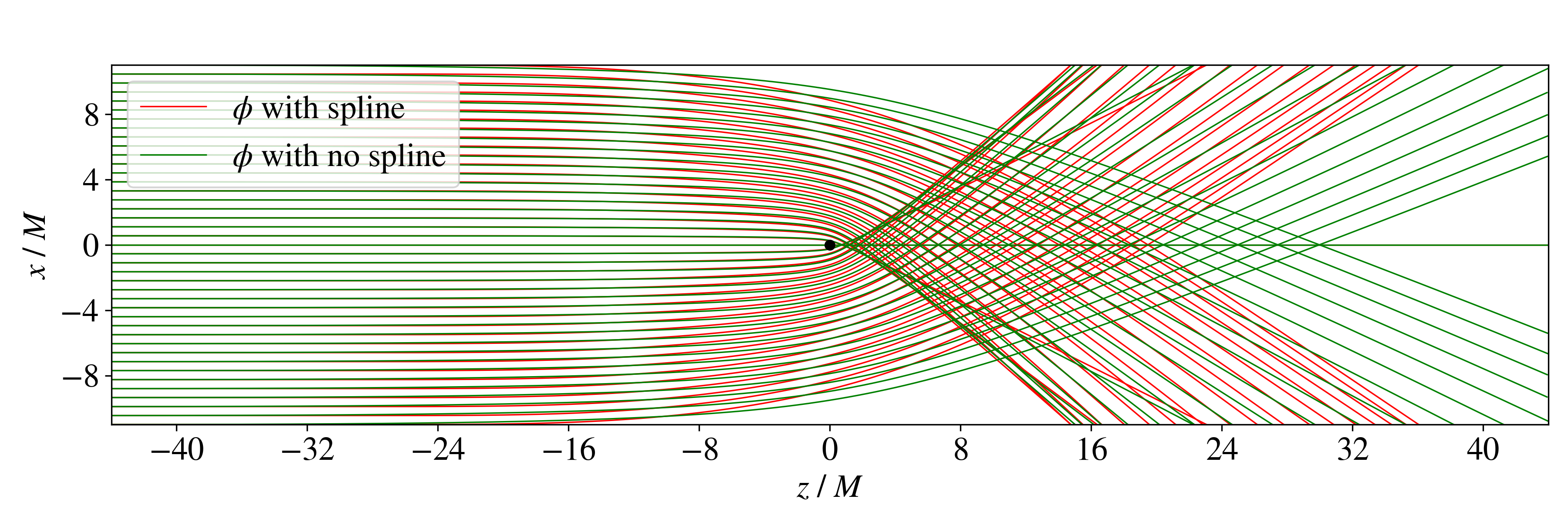}
   \caption{\small Null geodesics for NFW potential - strong gravity case}    
   \label{subfig:geodesic_SG_nfw}
   \end{subfigure}
       \hfill
   \begin{subfigure}[b]{0.475\textwidth}
   \centering
   \includegraphics[width=8cm]{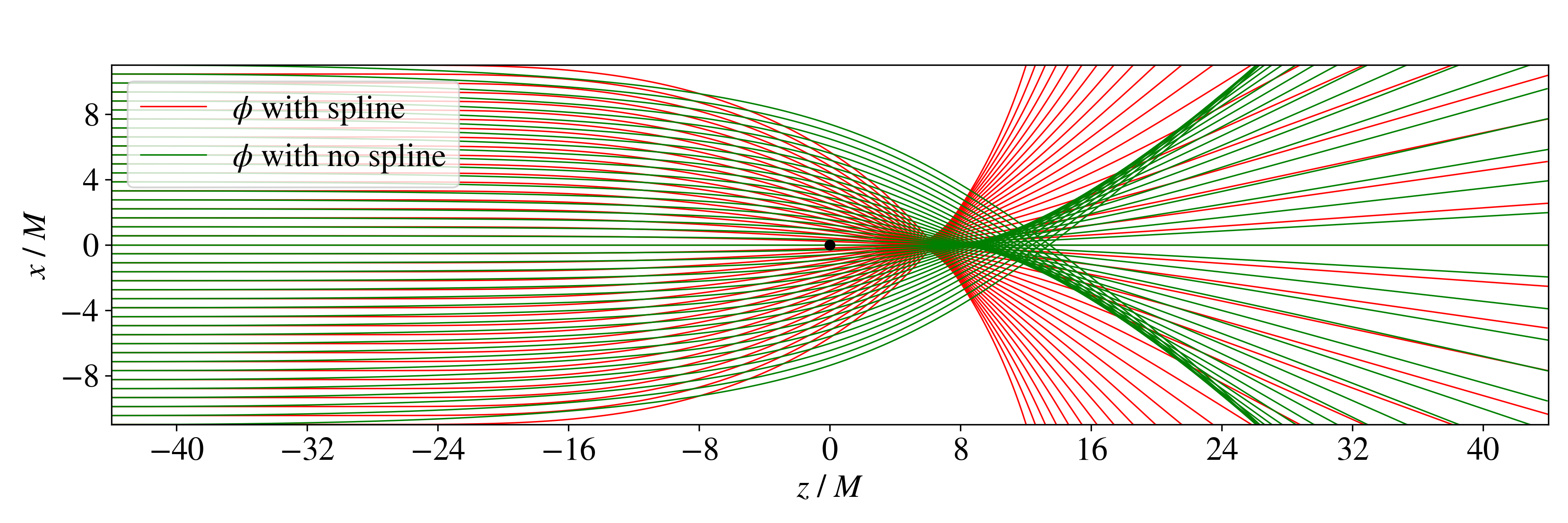}
   \caption{\small Null geodesics for Burkert potential - strong gravity case}    
   \label{subfig:geodesic_SG_bk}
   \end{subfigure}

\caption{Null geodesic trajectories of a plane wavefront for three gravitational potentials for the Gaussian, NFW, and Burkert profiles with $\phi_{\text{max}} = -0.49$ corresponding to the strong gravity case. The black dot denotes the peak curvature point. We compare the geodesics of the spacetime with and without the splines, showing that they are qualitatively similar and differ only by focusing distance and sharpness of focus. }
\label{fig:geodesics2}
\end{figure}

In this appendix we review the geodesic equations for our background spacetime and explain how these were solved to obtain the geodesic overlays presented in the figures in the main text. Further, in figures \ref{fig:geodesics1} and \ref{fig:geodesics2} we plot null geodesics for the spacetimes both with and without the spline approximation to demonstrate that the introduction of the spline and corresponding negative density only has minimal impact on the results.
\newline
\newline
The geodesic equations corresponding to the line element in equation \eqref{metric_exp} are
\begin{equation}
\begin{split}
\frac{d^2t}{d\lambda^2} &+ 2\frac{dt}{d\lambda}\left(\phi_x\frac{dx}{d\lambda}
  + \phi_y\frac{dy}{d\lambda} + \phi_z\frac{dz}{d\lambda}\right) = 0, \label{eq:geo_t}\\
\end{split}
\end{equation}
\begin{equation}
\begin{split}
\frac{d^2x}{d\lambda^2} &= \phi_x\left[-e^{4\phi}\left(\frac{dt}{d\lambda}\right)^2
  + \left(\frac{dx}{d\lambda}\right)^2 - \left(\frac{dy}{d\lambda}\right)^2
  - \left(\frac{dz}{d\lambda}\right)^2\right]
  \\ &+ 2\frac{dx}{d\lambda}\left(\phi_y\frac{dy}{d\lambda} + \phi_z\frac{dz}{d\lambda}\right), \label{eq:geo_x}
\end{split}
\end{equation}
\begin{equation}
\begin{split}
\frac{d^2y}{d\lambda^2} &= \phi_y\left[-e^{4\phi}\left(\frac{dt}{d\lambda}\right)^2
  - \left(\frac{dx}{d\lambda}\right)^2 + \left(\frac{dy}{d\lambda}\right)^2
  - \left(\frac{dz}{d\lambda}\right)^2\right]
  \\&+ 2\frac{dy}{d\lambda}\left(\phi_x\frac{dx}{d\lambda} + \phi_z\frac{dz}{d\lambda}\right), \label{eq:geo_y}
\end{split}
\end{equation}
\begin{equation}
\begin{split}
\frac{d^2z}{d\lambda^2} &= \phi_z\left[-e^{4\phi}\left(\frac{dt}{d\lambda}\right)^2
  - \left(\frac{dx}{d\lambda}\right)^2 - \left(\frac{dy}{d\lambda}\right)^2
  + \left(\frac{dz}{d\lambda}\right)^2\right]
  \\ &+ 2\frac{dz}{d\lambda}\left(\phi_x\frac{dx}{d\lambda} + \phi_y\frac{dy}{d\lambda}\right).
\end{split}
\end{equation}
where and $\lambda$ is an affine parameter along the ray. When solving these equations for our spacetime setups, we assume that both the initial positions and tangent vectors lie in the equatorial plane and we consider $xz$-plane only. For an incident plane wave, the corresponding ray bundles are therefore initially parallel to one another along the $x$-axis. These rays correspond to wavefronts propagating from left to right across the simulation domain. Figures \ref{fig:geodesics1} and \ref{fig:geodesics2} show six representative configurations for the Gaussian, NFW, and Burkert densities in both the weak and strong gravity regimes. In each plot in fig. \ref{fig:geodesics1} and \ref{fig:geodesics2}, the black dot marks the centre of the lens/density distribution. In these figures we show lines of null geodesics for the gravitational potentials with and without the spline approximation plotted on top of each other. From this, we see that the the introduction of the spline only affects the focusing distance of the GWs for the Gaussian and NFW potential in both weak and strong gravity. For the Burkert potential, the spline affects focusing distance and as well as how  tightly the rays are focused. Despite these differences, qualitatively, the behaviour of spline vs. no-spline geodesics are very similar. 
\newline \indent
Since the propagation direction is given by the normal to the wavefront, the null geodesics provide a useful geometric characterization of the GW trajectories. Although we do not explicitly construct the normals to the simulated wavefronts, we compare the geodesics with the observed propagation direction of the waves on a visual basis to assess whether the rays are roughly orthogonal to the wave fronts.

The plots in figures \ref{fig:geodesics1} and \ref{fig:geodesics2} show that the different potential configurations considered produce qualitatively different focusing behaviour. In the weak gravity regime, the rays are only moderately deflected and converge at a finite focal point determined by the lens profile. In the strong gravity regime, however, the deflection near the lens is much stronger and the rays no longer form a finite focal point of the type seen in the weak gravity case. Instead, the ray structure becomes more distorted in the vicinity of the central potential.
\newline\indent
The regions where neighbouring rays converge and cross, mark where the ray bundle is most strongly focused and hence where the wave amplitude is
expected to be largest. In wave optics terms, the contributions from these neighbouring rays arrive nearly in phase and add constructively. Thus, the null geodesics provide a useful geometric guide for interpreting the main features observed in the full wave simulations.

\section{Numerical convergence of constraints}\label{app:convergence}
\begin{figure}[htbp]
    \centering
    \includegraphics[width=8cm]{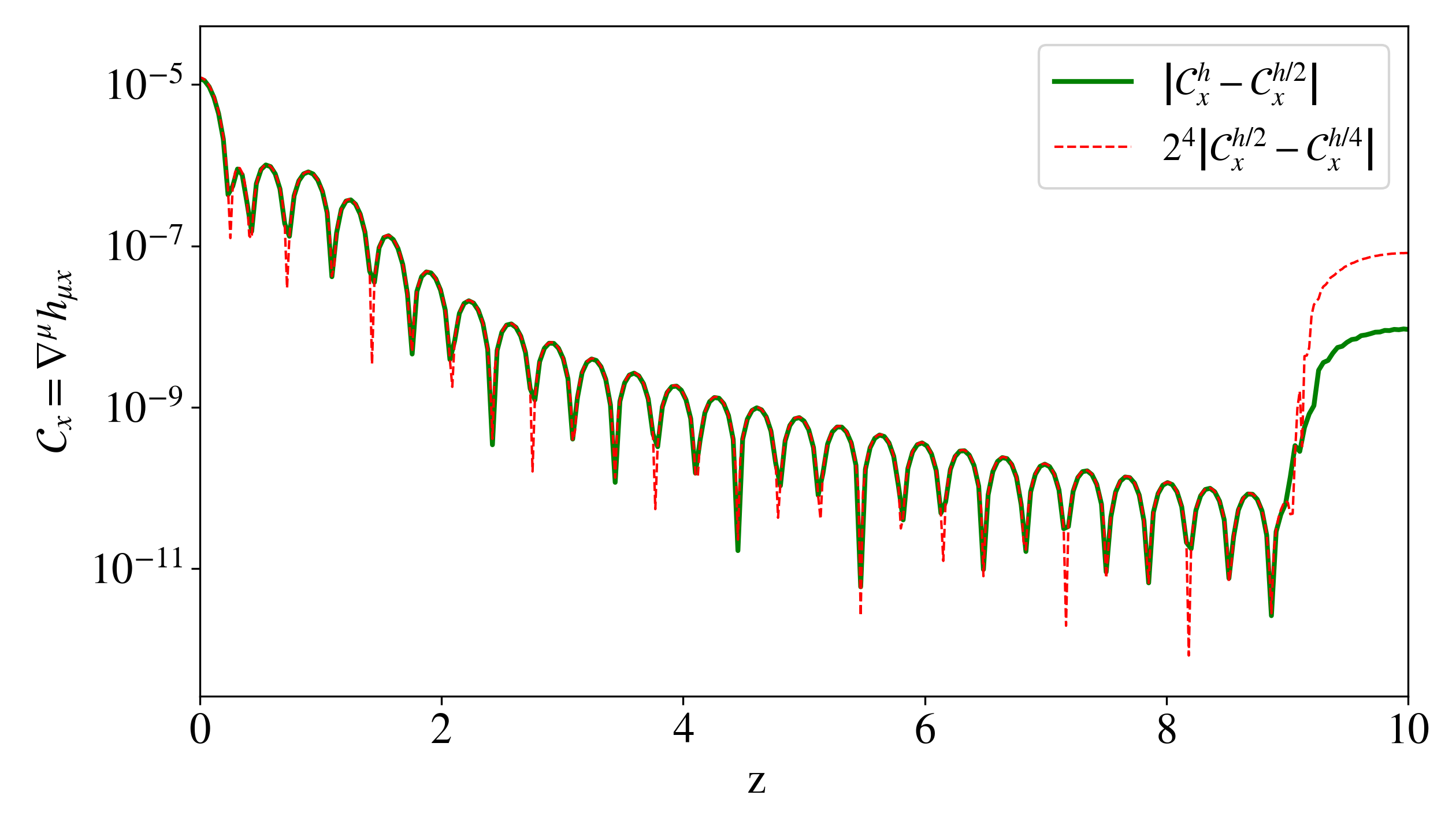}
    \includegraphics[width=8cm]{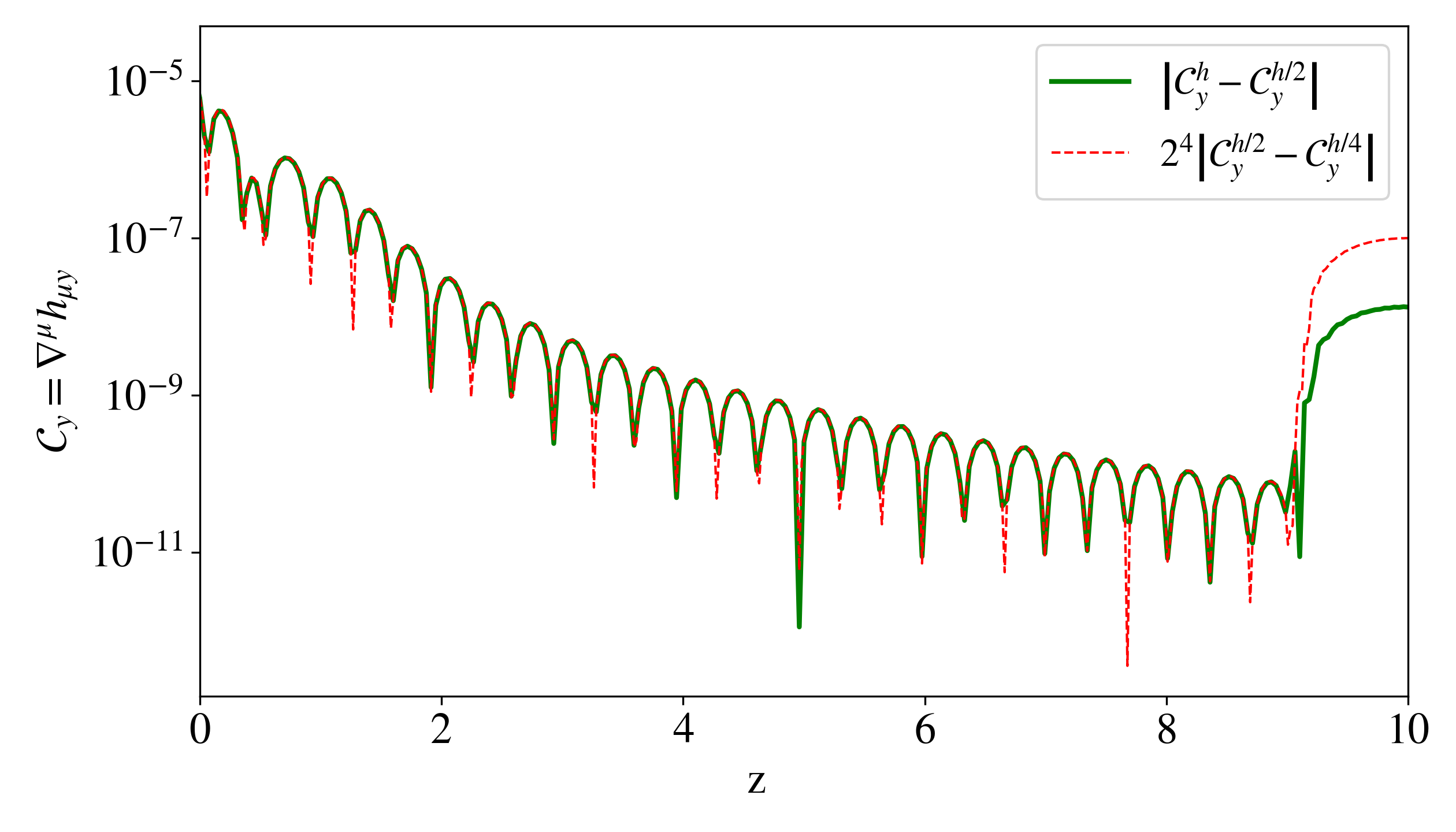}
    \caption{Convergence plots for the constraints $\mathcal{C}_x$ and $\mathcal{C}_y$ at time $t=0.879$ for three different resolutions denoted by $h$, $h/2$ and $h/4$, respectively. The data was sampled along a line from $(0,0,0)$ to $(0,0,10)$. For fourth order convergence, we expect the high resolution error to be $16$ times lower than the low resolution error. Hence, we multiply the high resolution error by a factor of $2^4$ to show numerical convergence.}
    \label{convergence_plot}
\end{figure}
To show that the numerical simulations presented in this work are valid and preserve the Lorenz gauge, we here present a convergence test using the constraint violation to demonstrate that the constraints converge. We perform three simulations (with no AMR) with grid resolutions $128 \times 128 \times 512$, $256 \times 256 \times 1024$ and $512 \times 512 \times 2048$ denoted by $h$, $h/2$ and $h/4$ respectively. We set the domains for $\{x,y,z\}$ as $\{(-4<x<4),(-4<y<4),(-10<z<10)\}$. We initialize plane wave $\bar h_{+}$ and $\bar h_{\times}$ as
\begin{equation}
    \begin{split}
        \bar h_+ & = \cos[ 2 \pi (k_zz-\omega t)] \\
        \bar h_\times & = \sin[ 2 \pi (k_zz-\omega t)], \\
    \end{split}
\end{equation}\label{eq:planewave_conv}
with $k_z \simeq 1.47$ in a Gaussian background (with weak gravity setting $\phi_{\text{max}} = -0.1$). The wave is initialized with the same non-vanishing amplitude over the entire domain, ensuring an initial constraint violation. Although it is unphysical (and does not satisfy Lorenz gauge) to initialize plane waves as in eqn. \eqref{eq:planewave_conv} in curved spacetime, we choose these initial conditions for the convergence test in order to demonstrate that the constraint damping method efficiently suppresses constraint violations.
\newline\indent
We initialize $\partial_t \bar h$ such that the wave propagates along the z-axis. We use reflecting boundary conditions in all three spatial dimensions and use constraint damping as specified in section \ref{subsec:constraint_damp}.
\newline\newline
We compute the absolute difference between the constraint violation $\mathcal{C}_v$ defined by eqn. \eqref{cv}, i.e. we calculate $\left| \mathcal{C}_v ^{ h} - \mathcal{C}_v ^{ h/2} \right|$ and $ \left| \mathcal{C}_v ^{ h/2} - \mathcal{C}_v ^{ h/4} \right|$ at three time instances. We verify the order of convergence, $p$, which should be $p=4$ in our case, i.e.,
\begin{equation}\label{conv_eqn}
\begin{split}
    \left| \frac{\mathcal{C}_v ^{ h} - \mathcal{C}_v ^{ h/2}}{\mathcal{C}_v ^{ h/2} - \mathcal{C}_v ^{ h/4}} \right|  &= 2^p + O(h).\\ 
    \text{or, } |\mathcal{C}_v ^{ h} - \mathcal{C}_v ^{ h/2}| & \simeq 2^4  |\mathcal{C}_v ^{ h/2} - \mathcal{C}_v ^{ h/4}| .
\end{split}
\end{equation}
Figure \ref{convergence_plot} compares the errors of the simulations with the three different resolutions. The errors seen in the subfigures agree with equation \eqref{conv_eqn}, demonstrating that the constraints converge. The subfigures also demonstrate that with increasing simulation resolution, the constraint violation goes toward zero. This shows that the constraint violations developed during the evolution are due to the numerical discretisation error and not physical constraint violations, and are efficiently damped by the constraint damping scheme.

\section{Plots of density, derivatives of $\phi$ and negative energy density contributions} \label{app:dens_negative_energy}

\begin{figure*}
\centering
    \begin{subfigure}[b]{0.95\textwidth}
    \centering
    \includegraphics[width=16cm]{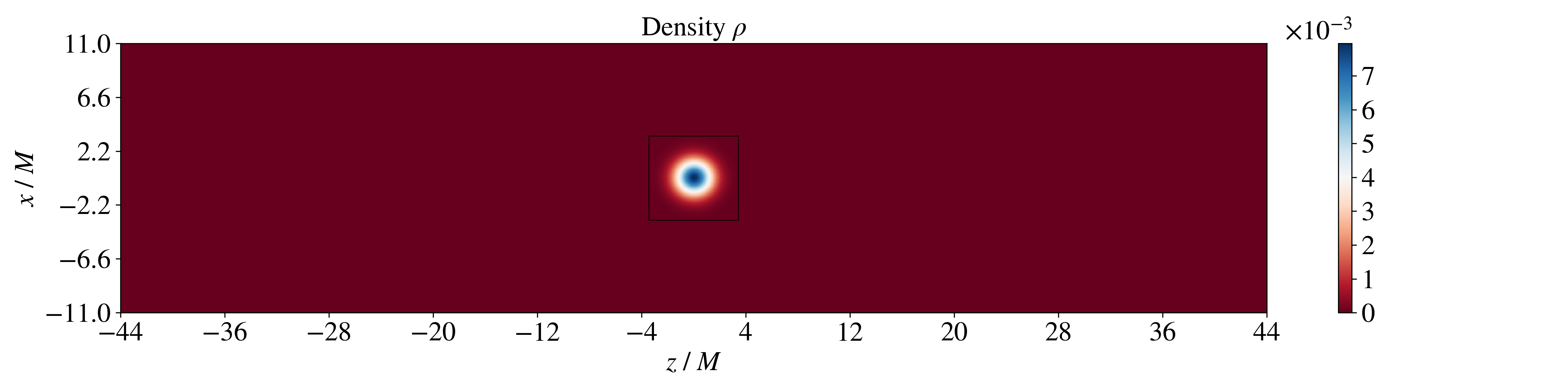}
    \caption{Gaussian density}    
    \end{subfigure}
        \vskip\baselineskip
    \begin{subfigure}[b]{0.95\textwidth}
    \centering
    \includegraphics[width=16cm]{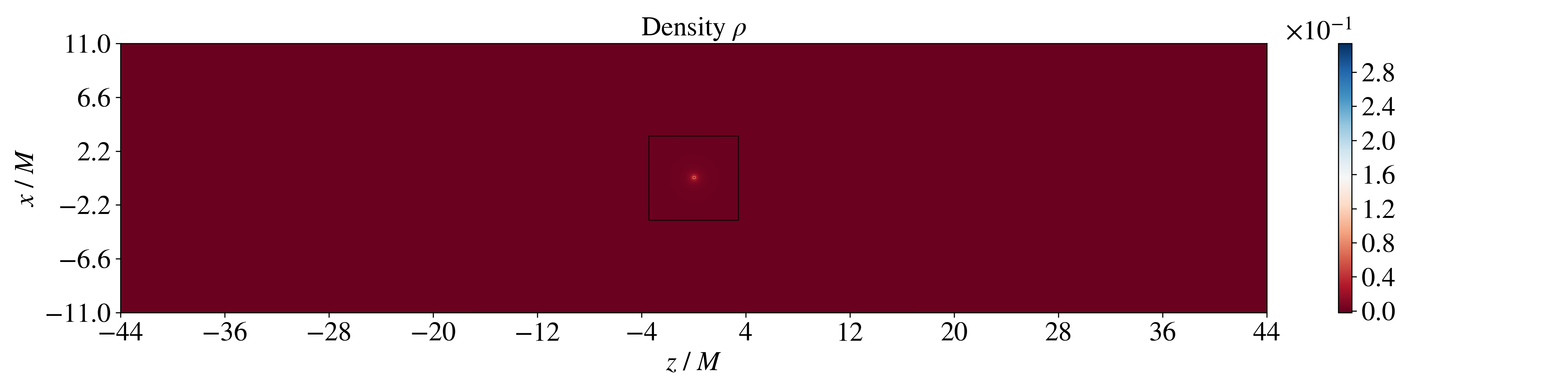}
    \caption{Dark matter halo density - NFW profile} 
    \end{subfigure}
        \vskip\baselineskip
    \begin{subfigure}[b]{0.95\textwidth}
    \centering
    \includegraphics[width=16cm]{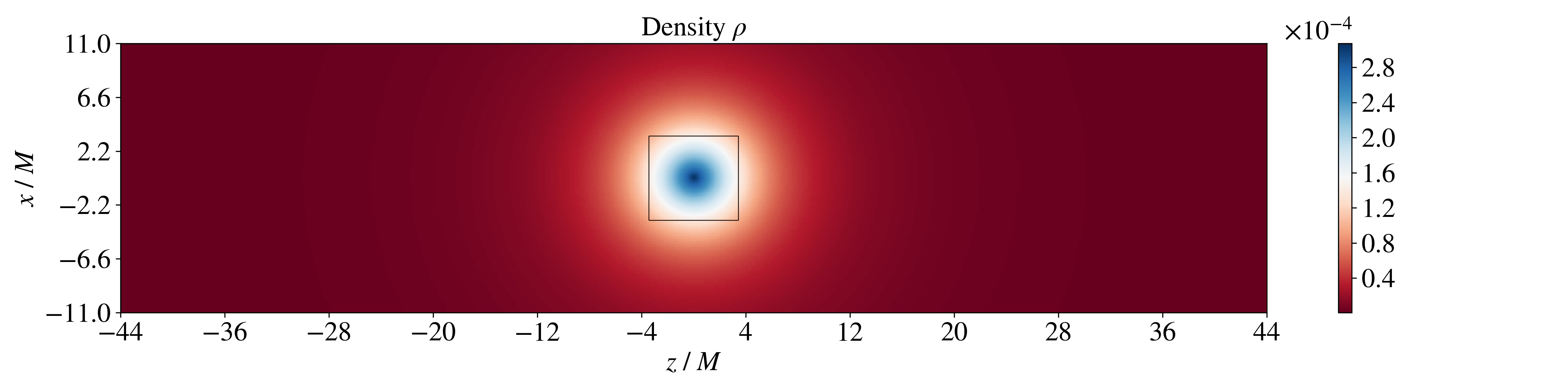}
    \caption{Dark matter halo density - Burkert profile}    
    \end{subfigure}

\caption{2D slice of density plots for Gaussian, NFW and Burkert profiles (weak gravity case with $\phi_{\text{max}}=0.1$). The black outlined boxes inside the simulation domain are where the AMR patch exists and has twice the resolution of the base grid.} 
\label{fig:densities_weak}
\end{figure*}

\begin{figure*}
\centering
    \begin{subfigure}[b]{0.95\textwidth}
    \centering
    \includegraphics[width=16cm]{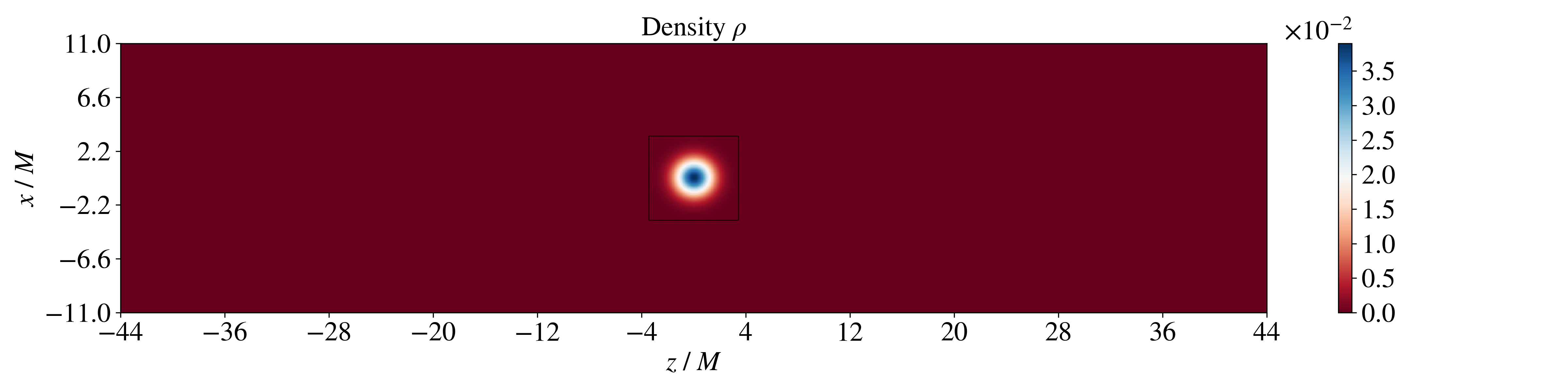}
    \caption{Gaussian density}    
    \end{subfigure}
        \vskip\baselineskip
    \begin{subfigure}[b]{0.95\textwidth}
    \centering
    \includegraphics[width=16cm]{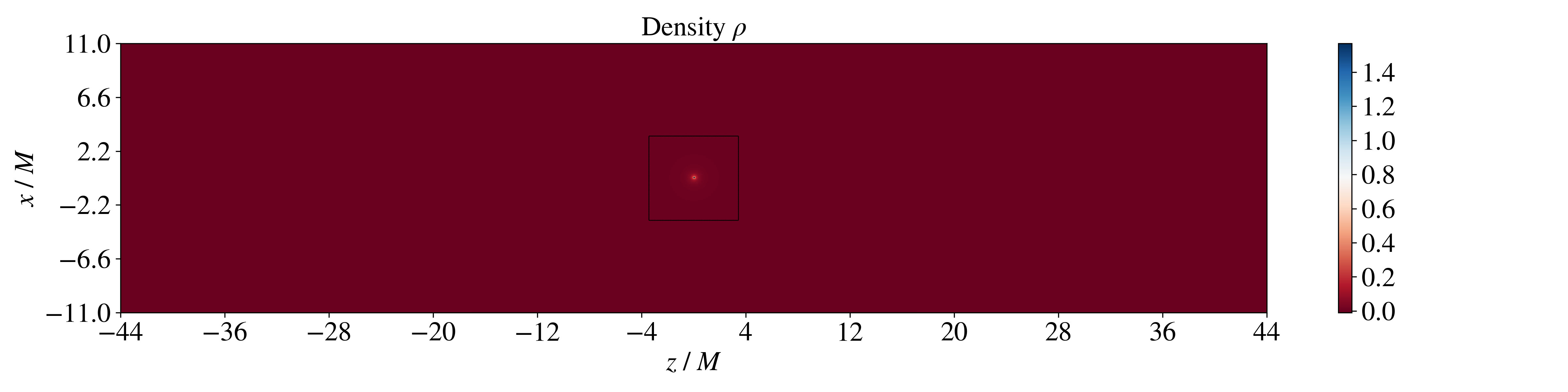}
    \caption{ Dark matter halo density - NFW profile}    
    \end{subfigure}
        \vskip\baselineskip
    \begin{subfigure}[b]{0.95\textwidth}
    \centering
    \includegraphics[width=16cm]{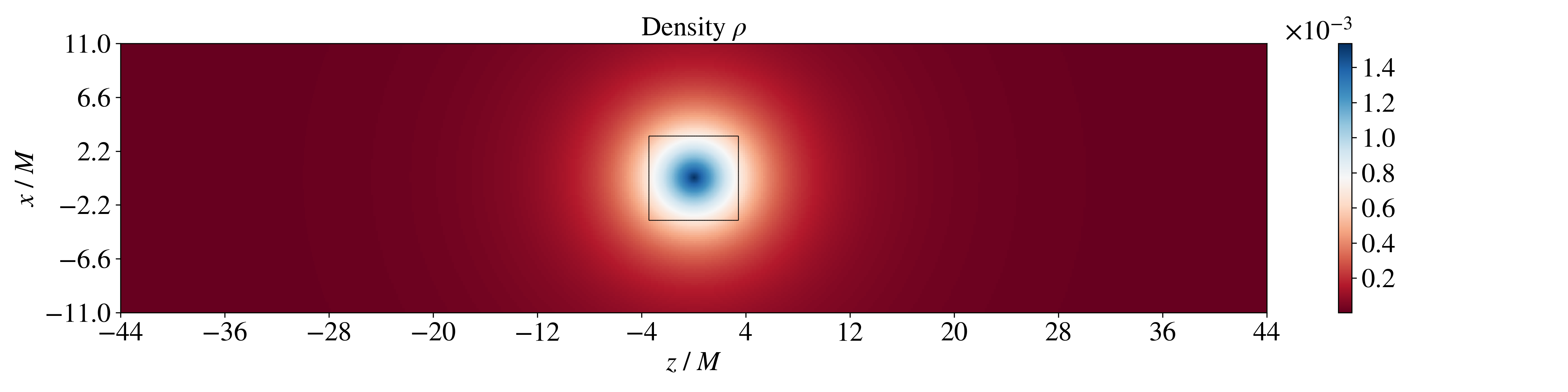}
    \caption{ Dark matter halo density - Burkert profile}    
    \end{subfigure}

\caption{2D slice of density plots for Gaussian, NFW and Burkert profiles (strong gravity case with $\phi_{\text{max}}=0.49$). The black outlined boxes inside the simulation domain is where the AMR patch exists and has twice the resolution of the base grid.} 
\label{fig:densities_strong}
\end{figure*}

\begin{figure}
\centering
    \begin{subfigure}[b]{0.475\textwidth}
    \centering
    \includegraphics[width=8cm]{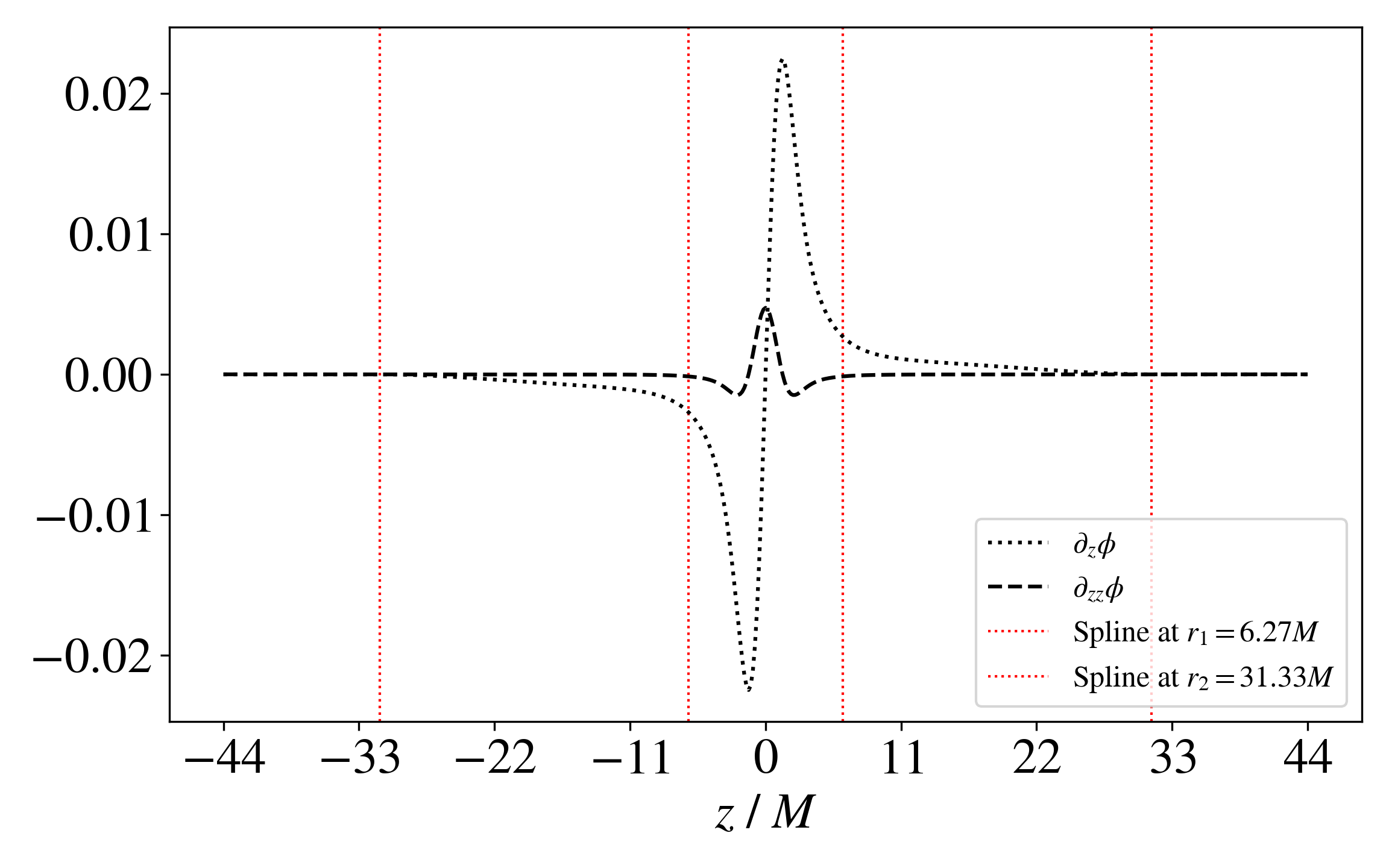}
    \caption{\small Derivatives of Gaussian potential}    
    \end{subfigure}
        \vskip\baselineskip
    \begin{subfigure}[b]{0.475\textwidth}
    \centering
    \includegraphics[width=8cm]{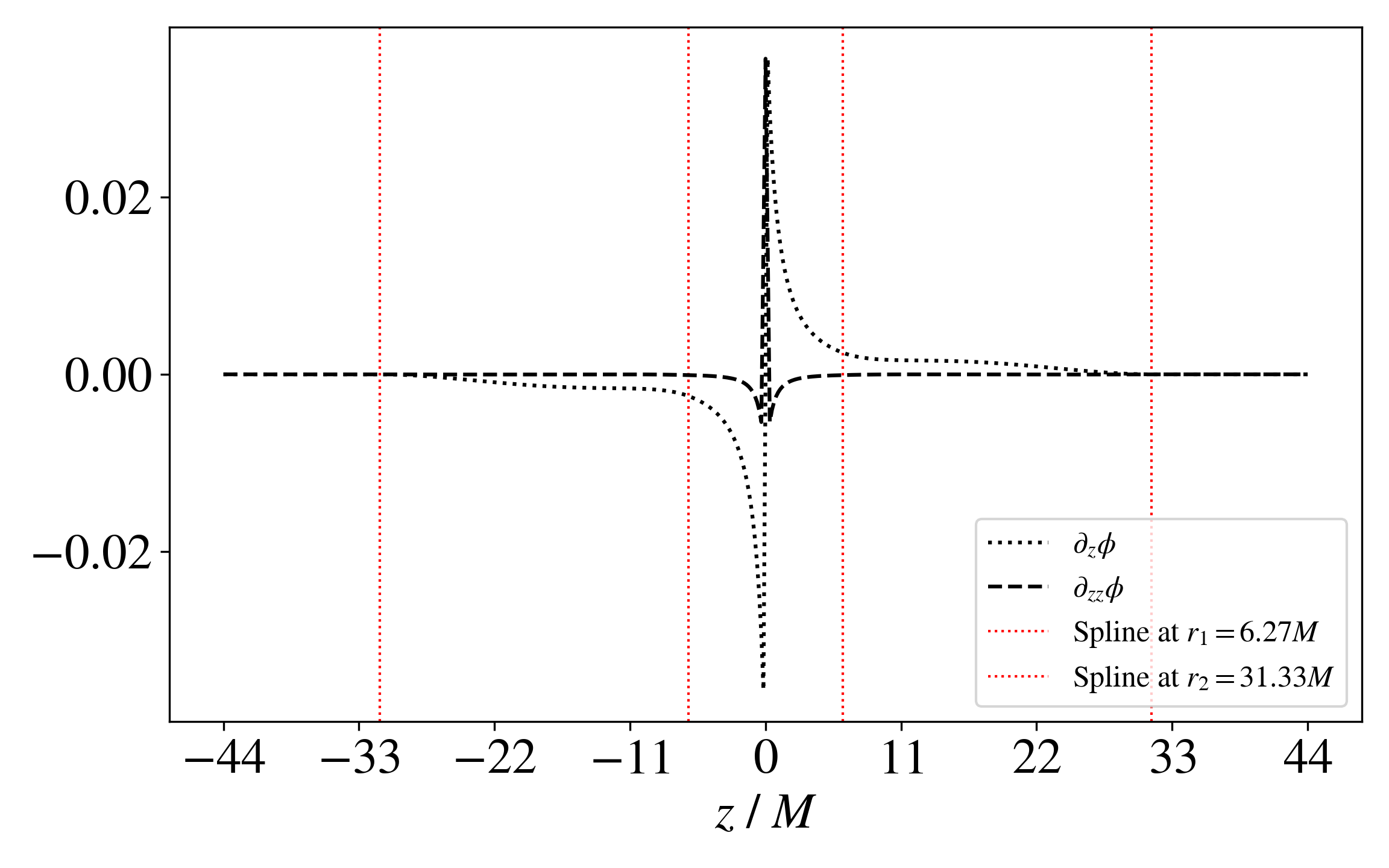}
    \caption{\small Derivatives of NFW potential}    
    \end{subfigure}
        \vskip\baselineskip
    \begin{subfigure}[b]{0.475\textwidth}
    \centering
    \includegraphics[width=8cm]{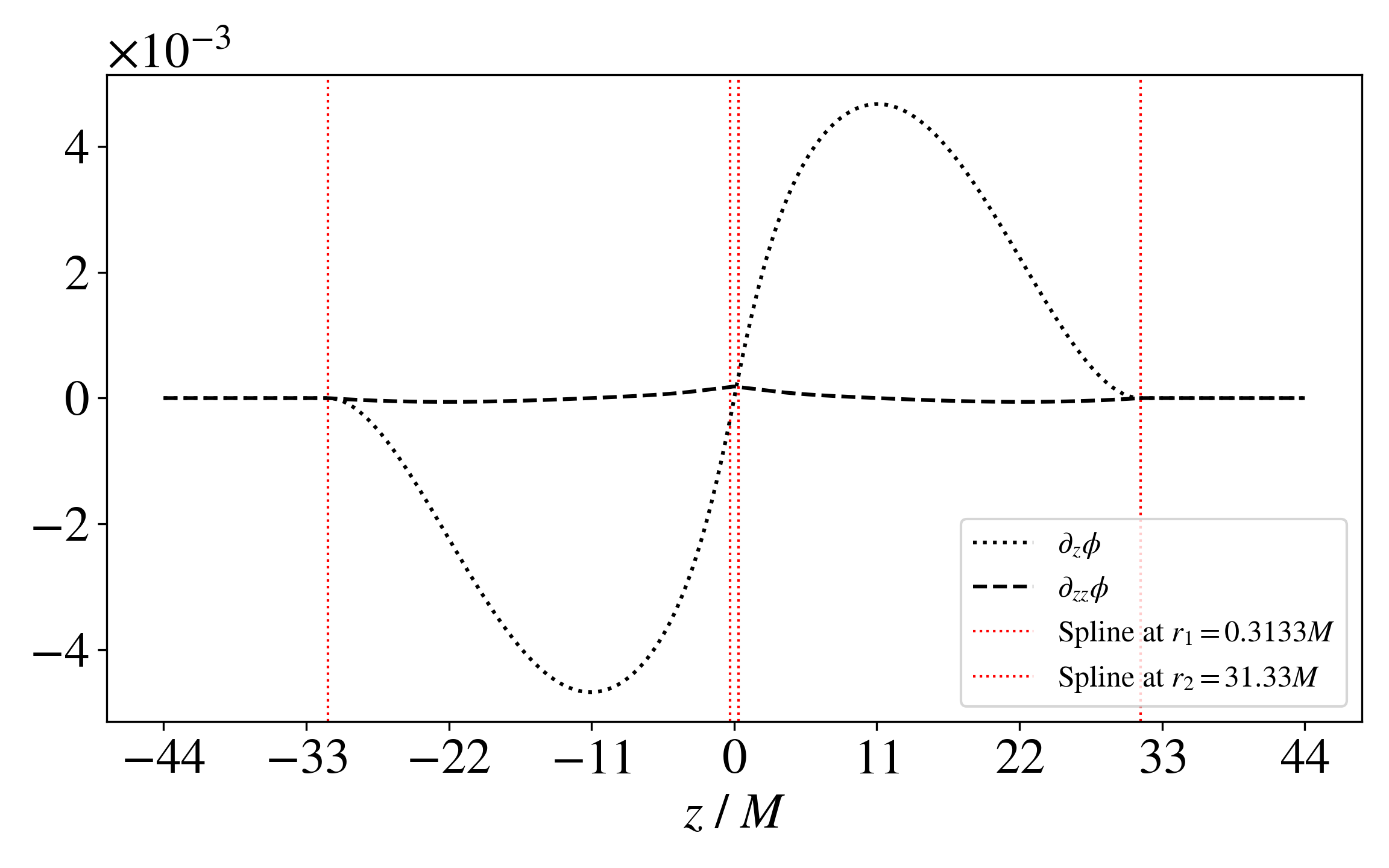}
    \caption{\small Derivatives of Burkert potential}    
    \end{subfigure}
\caption{The first and second derivatives of $\phi$ along the $z$-axis for three different potentials. From top to bottom: Gaussian, NFW and Burkert profiles, all corresponding to the strong gravity case with $\phi_{\text{max}}=0.49$.  }\label{fig:d_phi}
\end{figure}

\begin{figure}
    \centering
    \begin{subfigure}[b]{0.475\textwidth}
    \includegraphics[width=7cm]{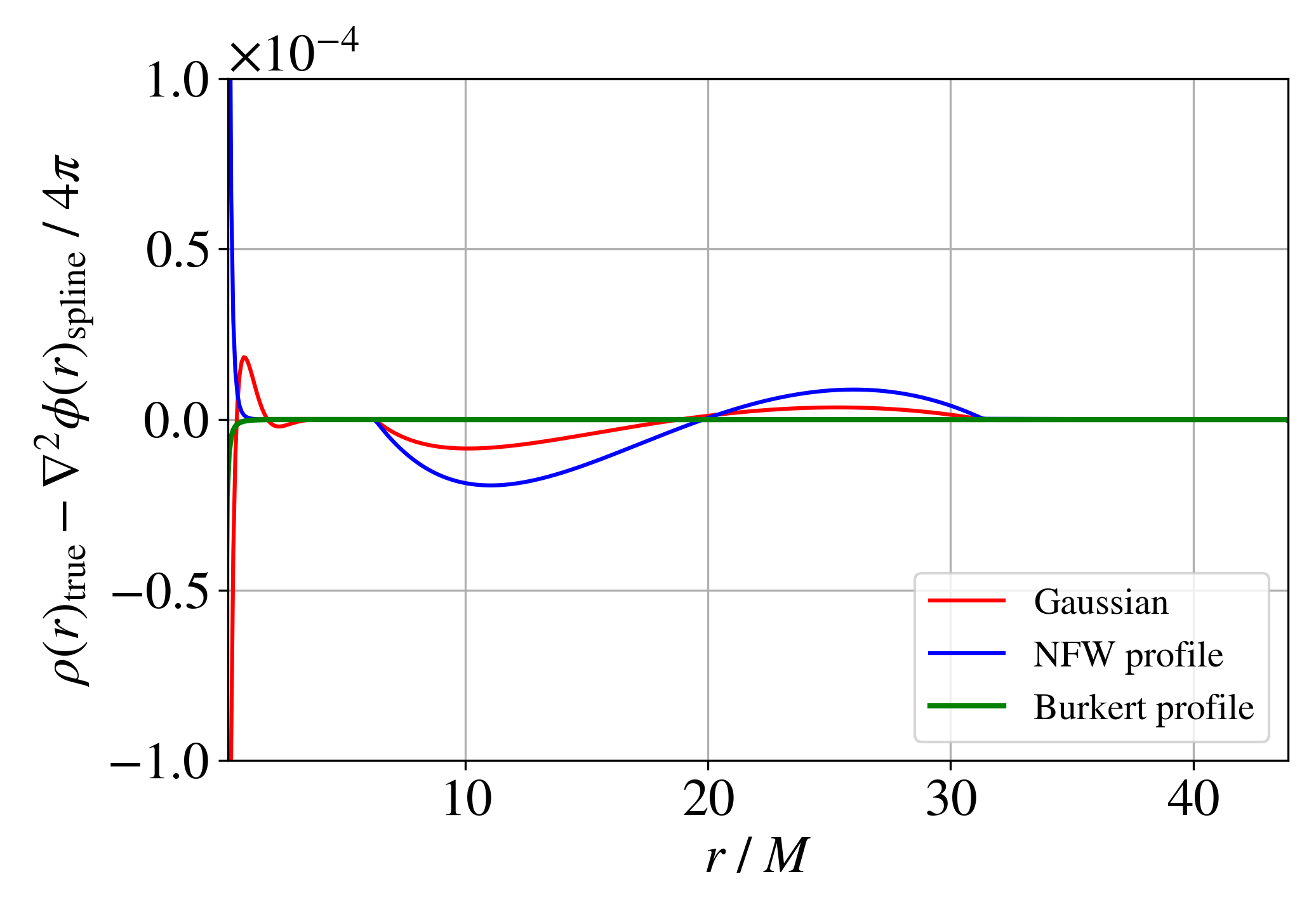}
    \caption{Deviation of $\rho(r)$ from its true form for the weak gravity case}
    \end{subfigure}
    \hfill
    \begin{subfigure}[b]{0.475\textwidth}
    \includegraphics[width=7cm]{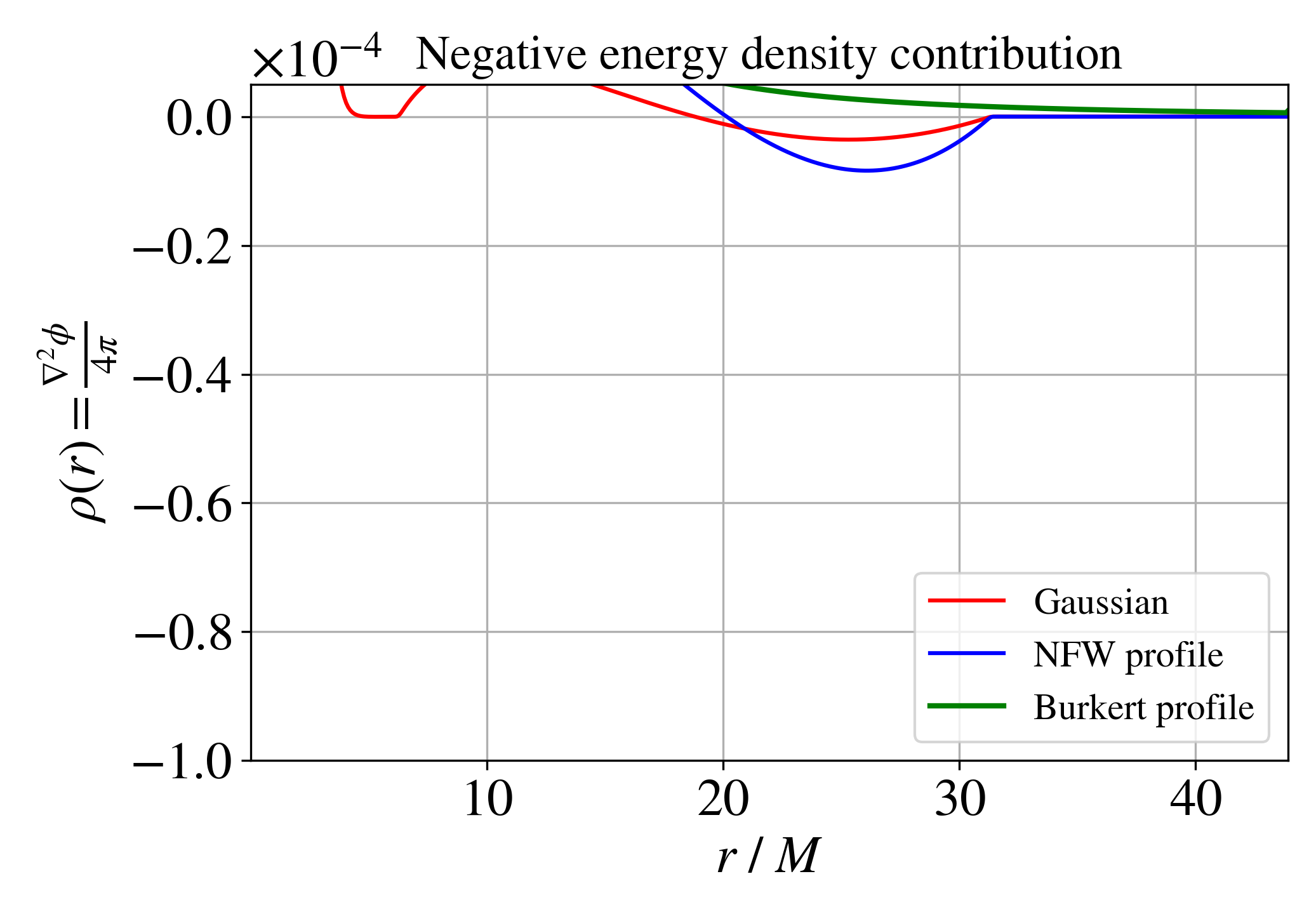}
    \caption{The negative energy density contribution for the weak gravity case}
    \end{subfigure}
    \hfill
    \begin{subfigure}[b]{0.475\textwidth}
    \includegraphics[width=7cm]{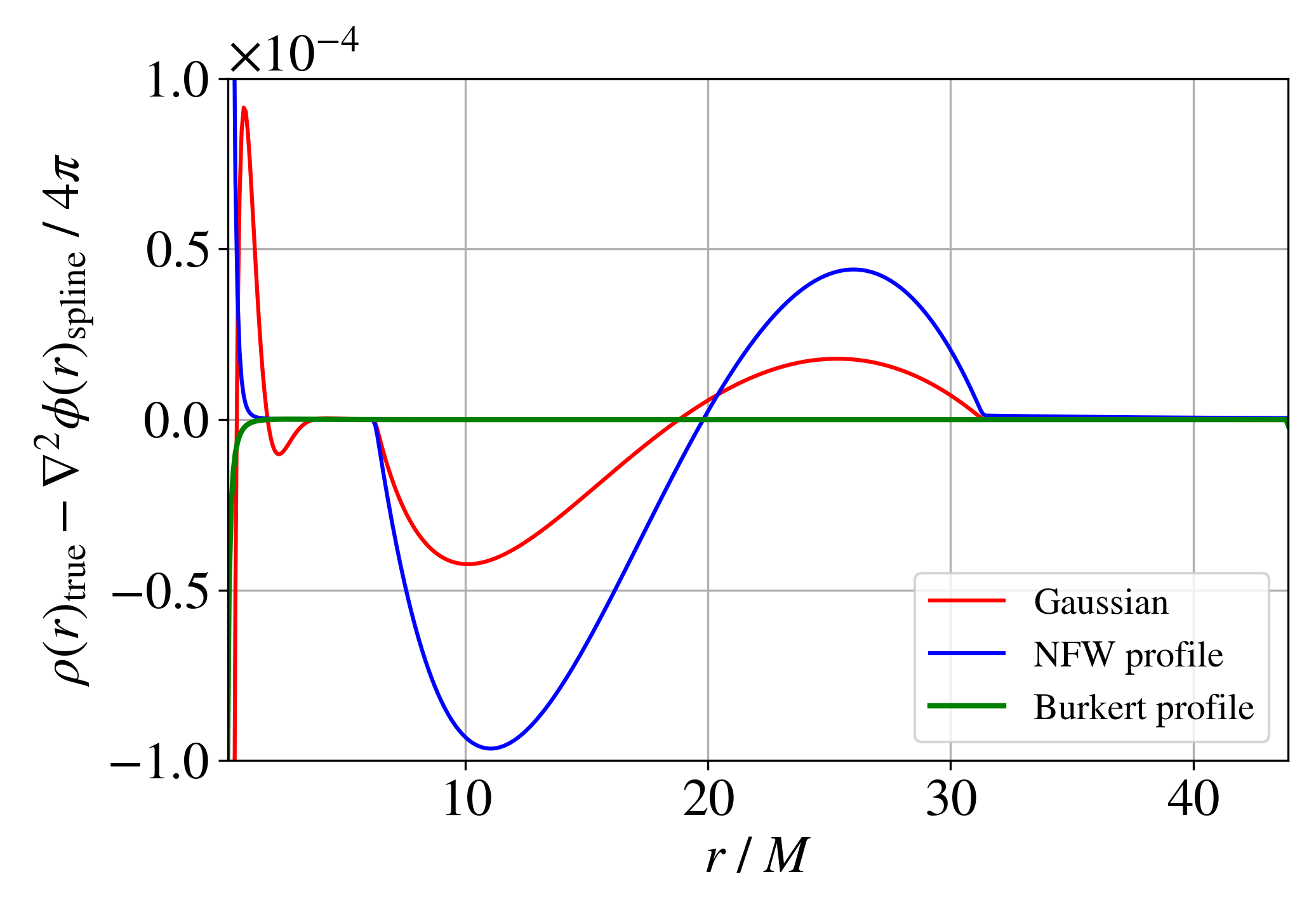}
    \caption{Deviation of $\rho(r)$ from its true form for the strong gravity case}
    \end{subfigure}
    \hfill
    \begin{subfigure}[b]{0.475\textwidth}
    \includegraphics[width=7cm]{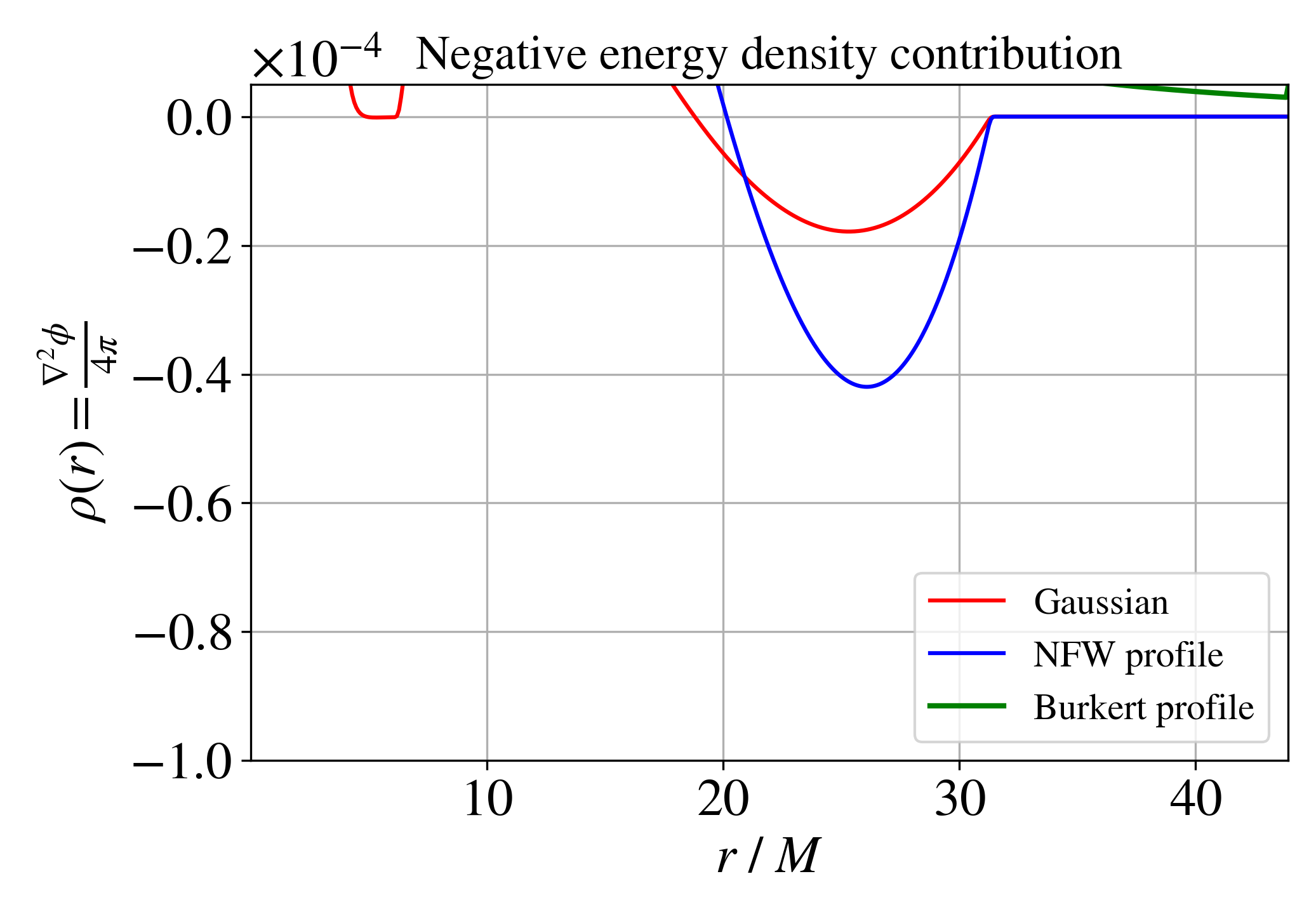}
    \caption{The negative energy density contribution for the strong gravity case}
    \end{subfigure}
    \caption{Figures shows the deviation of the density function $\rho(r)$ due to the spline modification to $\phi(r)$ including the negative energy density contribution for all three density configurations in both the weak and strong gravity cases.}
    \label{negative_dens_plots}
\end{figure}

This appendix contains the plots of density for Gaussian, NFW and Burkert profile for strong (Fig. \ref{fig:densities_strong}) and weak (Fig. \ref{fig:densities_weak}) gravity configurations, derivatives of $\phi$ for all three profiles (Fig. \ref{fig:d_phi}) and plots showing negative energy contribution due to the spline implementation (Fig. \ref{negative_dens_plots}). This appendix also includes table of values (Table \ref{tab:spline}) for $r_1$ and $r_2$ that implements the spline for potential $\phi$.

\begin{figure*}
\centering
    \begin{subfigure}[b]{0.475\textwidth}
    \centering
    \includegraphics[width=8.5cm]{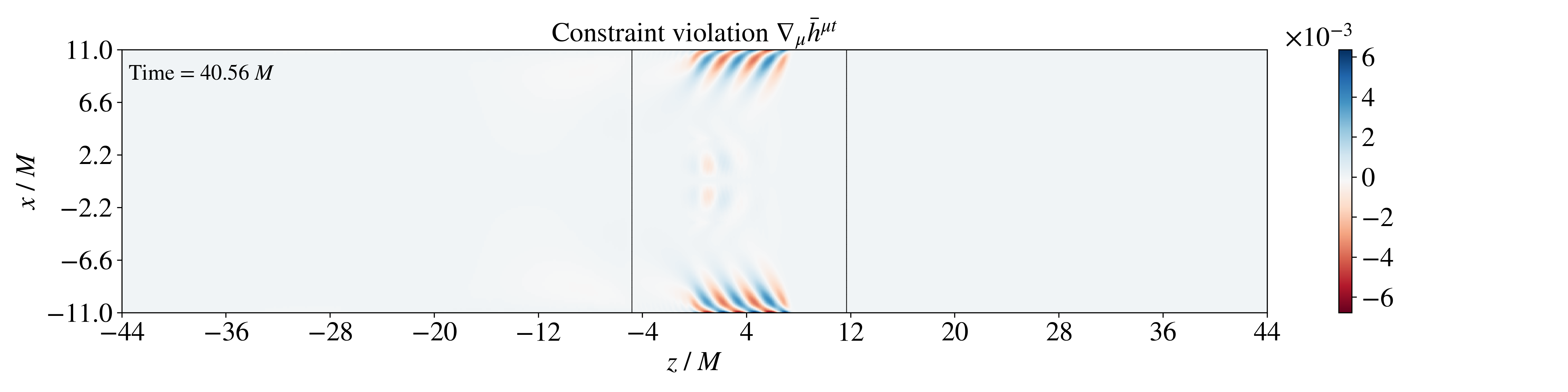}
    \caption{\small Lorenz gauge violation $C_t$}    
    \label{subfig:gauss_wg_cvt}
    \end{subfigure}
        \hfill
    \begin{subfigure}[b]{0.475\textwidth}
    \centering
    \includegraphics[width=8.5cm]{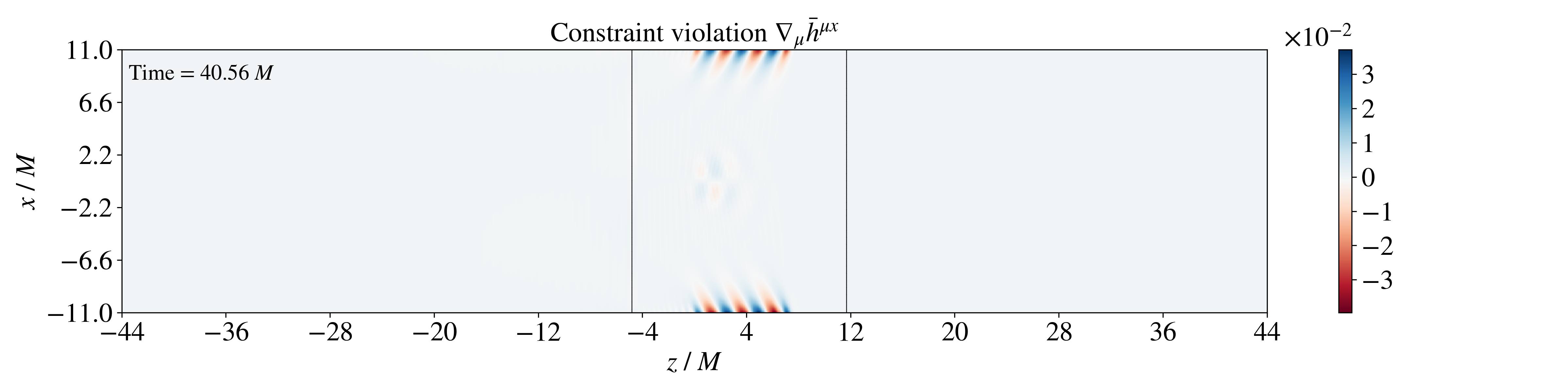}
    \caption{\small Lorenz gauge violation $C_x$}    
    \label{subfig:gauss_wg_cvx}
    \end{subfigure}
        \vskip\baselineskip
    \begin{subfigure}[b]{0.475\textwidth}
    \centering
    \includegraphics[width=8.5cm]{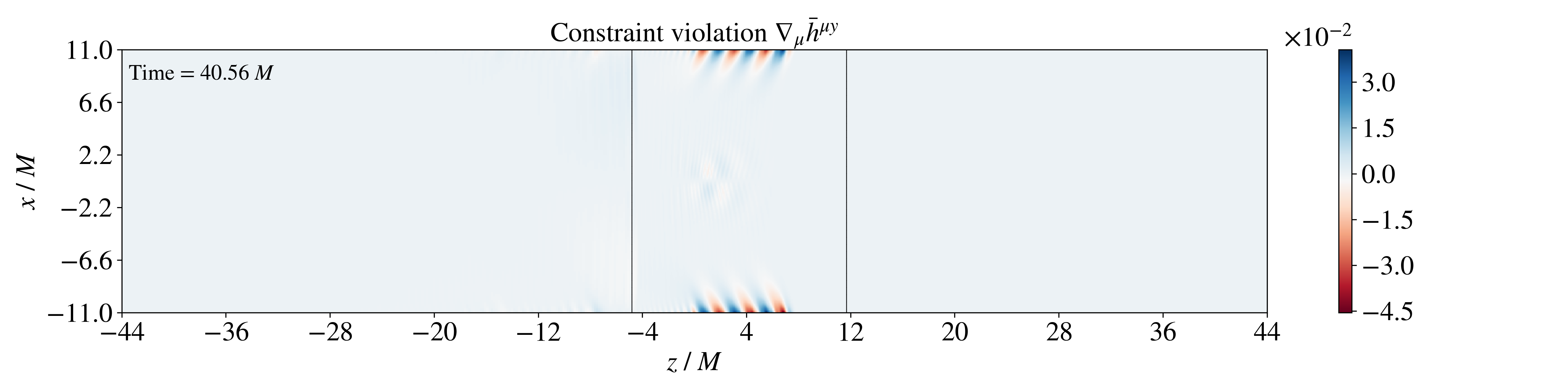}
    \caption{\small Lorenz gauge violation $C_y$}    
    \label{subfig:gauss_wg_cvy}
    \end{subfigure}
        \hfill
    \begin{subfigure}[b]{0.475\textwidth}
    \centering
    \includegraphics[width=8.5cm]{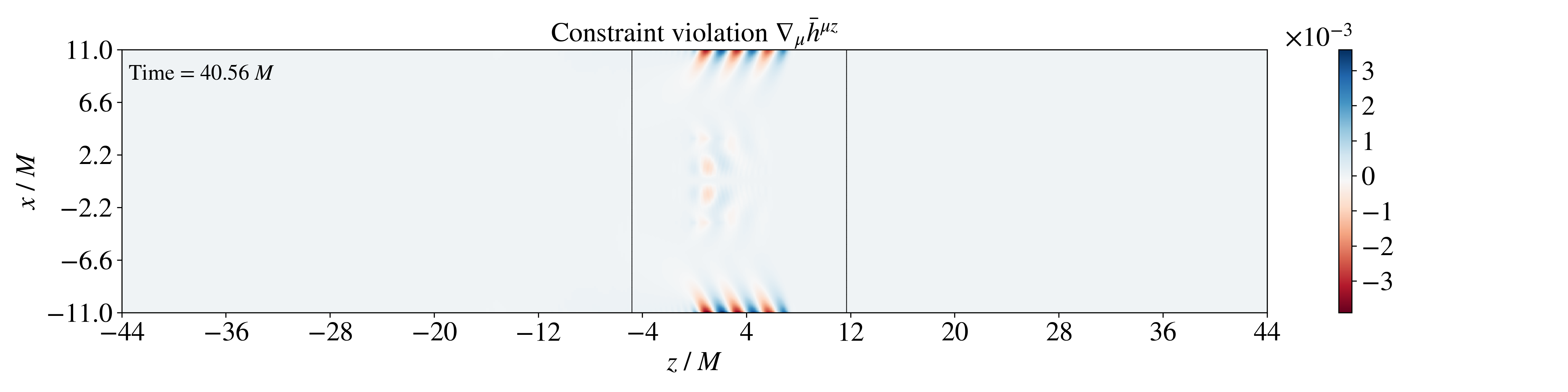}
    \caption{\small Lorenz gauge violation $C_z$}    
    \label{subfig:gauss_wg_cvz}
    \end{subfigure}
\caption{2D slice of constraint violation plots at the instance when the GW is propagating inside the Gaussian density (weak gravity case with $\phi_{\text{max}}=0.1$). We note that at the considered simulation resolution, the constraint violation is well below the order of GW amplitude. The black outlined boxes inside the simulation domain is where the AMR patch exists and has twice the resolution of the base grid.} 
\label{fig:gauss_wg_CV}
\end{figure*}

\begin{figure*}
\centering
    \begin{subfigure}[b]{0.475\textwidth}
    \centering
    \includegraphics[width=8.5cm]{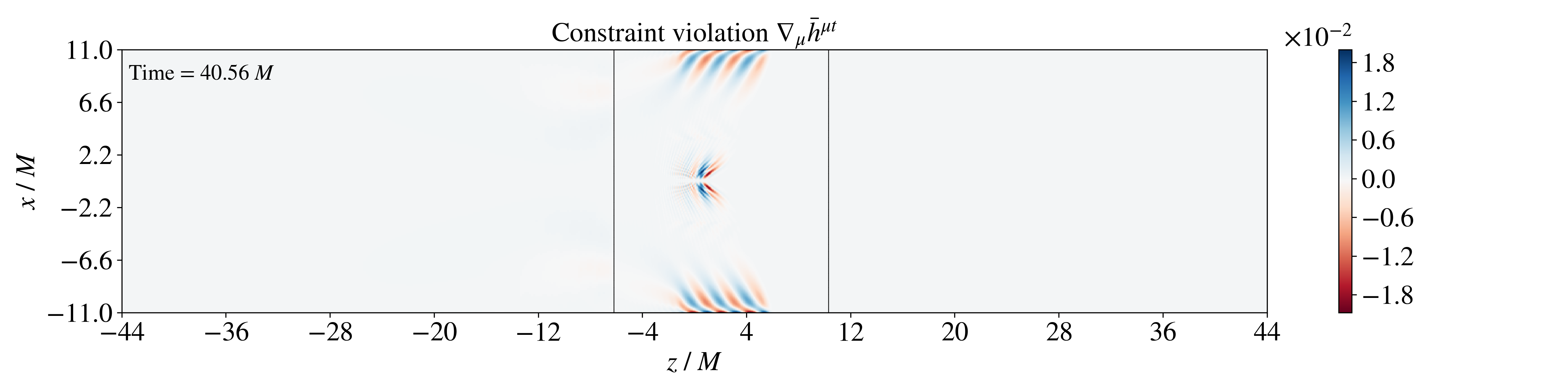}
    \caption{\small Lorenz gauge violation $C_t$}    
    \label{subfig:gauss_sg_cvt}
    \end{subfigure}
        \hfill
    \begin{subfigure}[b]{0.475\textwidth}
    \centering
    \includegraphics[width=8.5cm]{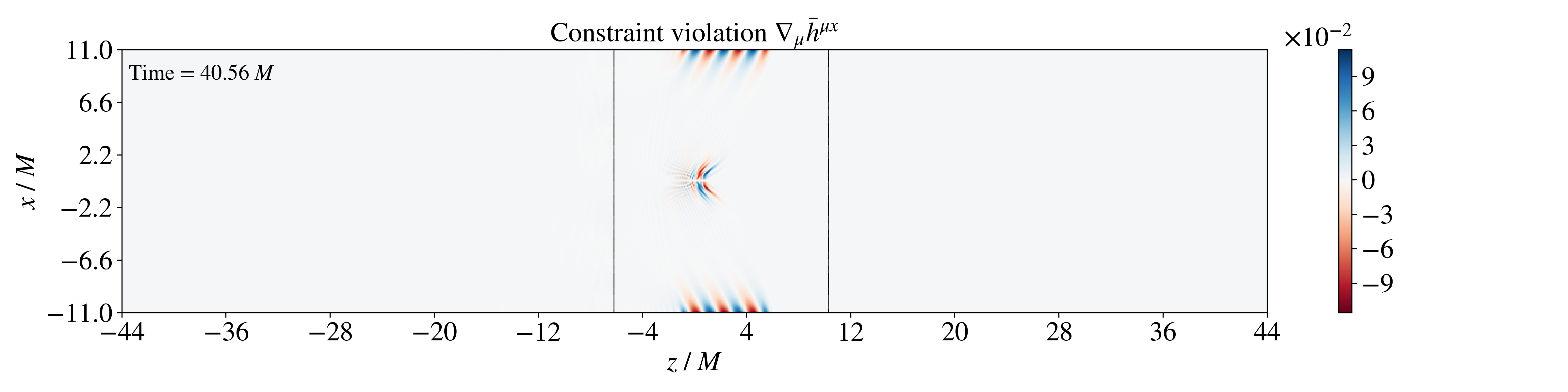}
    \caption{\small Lorenz gauge violation $C_x$}    
    \label{subfig:gauss_sg_cvx}
    \end{subfigure}
        \vskip\baselineskip
    \begin{subfigure}[b]{0.475\textwidth}
    \centering
    \includegraphics[width=8.5cm]{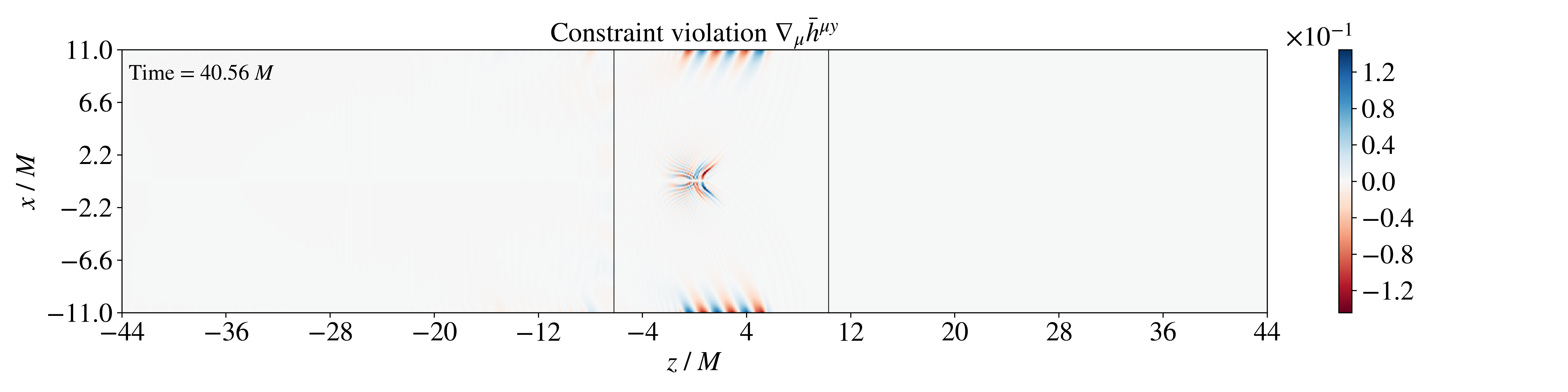}
    \caption{\small Lorenz gauge violation $C_y$}    
    \label{subfig:gauss_sg_cvy}
    \end{subfigure}
        \hfill
    \begin{subfigure}[b]{0.475\textwidth}
    \centering
    \includegraphics[width=8.5cm]{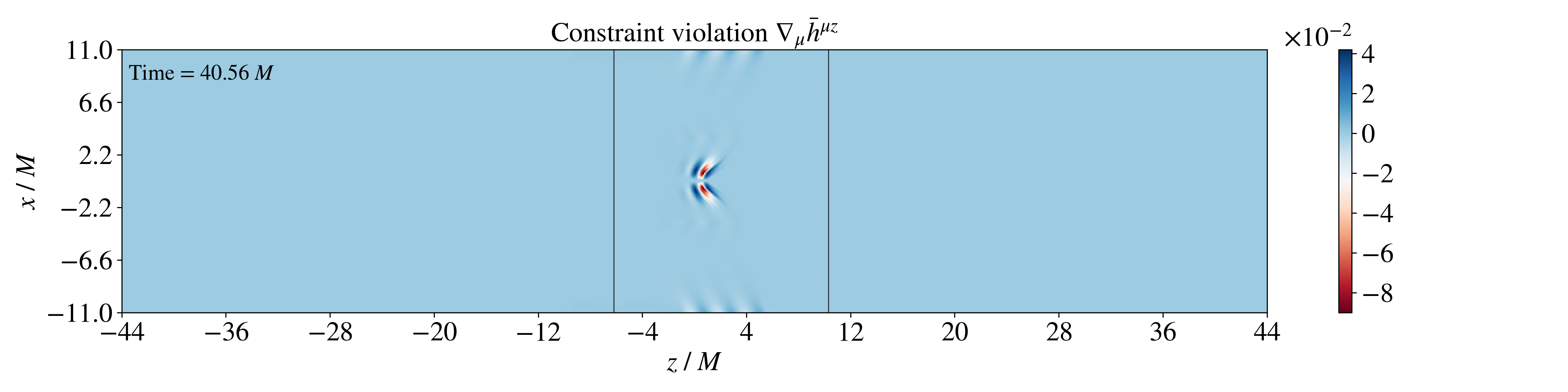}
    \caption{\small Lorenz gauge violation $C_z$}    
    \label{subfig:gauss_sg_cvz}
    \end{subfigure}
\caption{2D slice of constraint violation at the instance when the GW is propagating inside the Gaussian density (strong lensing case with $\phi_{\text{max}}=0.49$). At the considered simulation resolution, the constraint violation is well below the order of the GW amplitude. The black outlined boxes inside the simulation domain is where the AMR patch exists and has twice the resolution of the base grid.} 
\label{fig:gauss_sg_CV}
\end{figure*}

\begin{figure*}
\centering
    \begin{subfigure}[b]{0.475\textwidth}
    \centering
    \includegraphics[width=8.5cm]{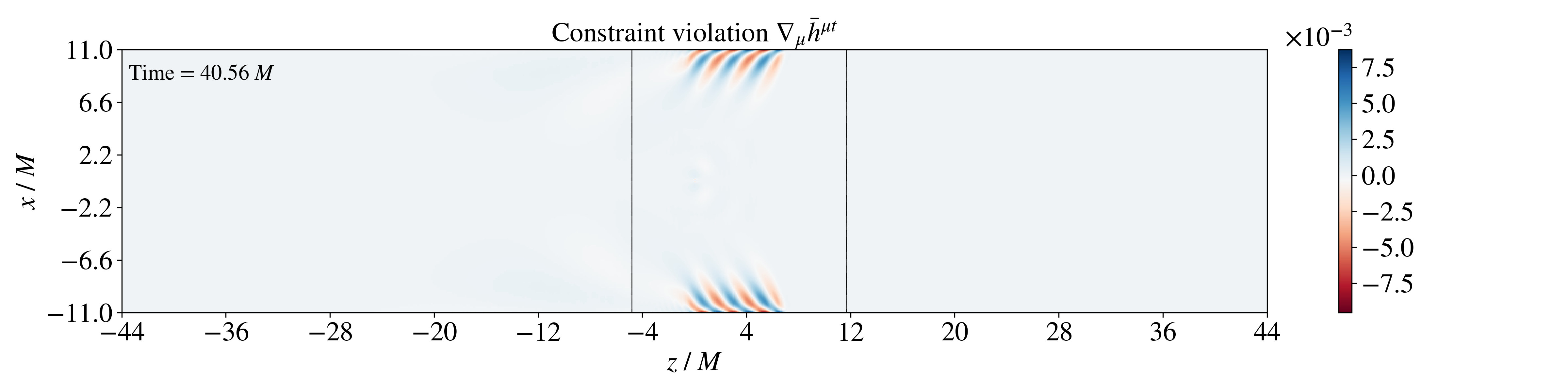}
    \caption{\small Lorenz gauge violation $C_t$}    
    \end{subfigure}
        \hfill
    \begin{subfigure}[b]{0.475\textwidth}
    \centering
    \includegraphics[width=8.5cm]{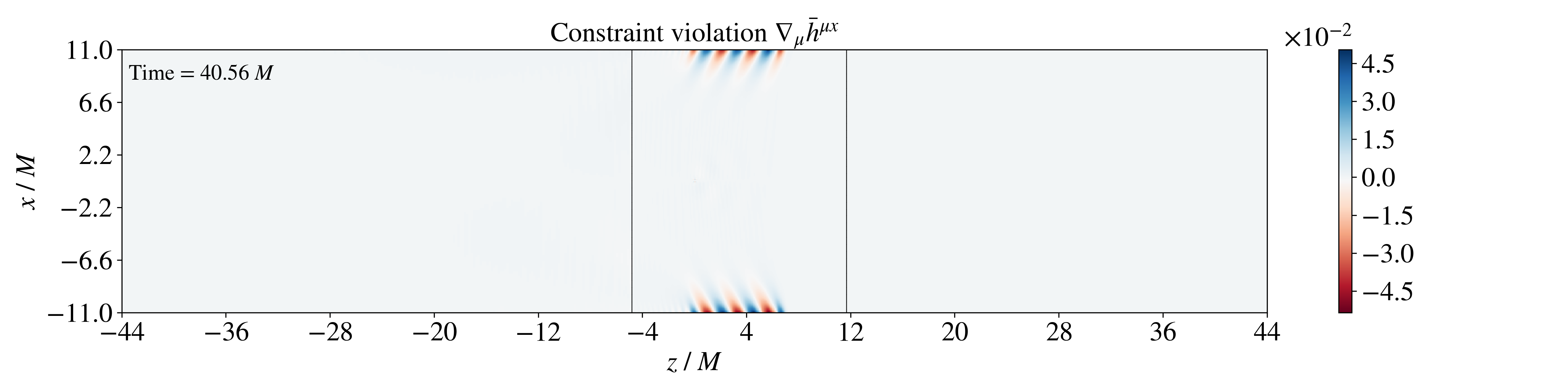}
    \caption{\small Lorenz gauge violation $C_x$}    
    \end{subfigure}
        \vskip\baselineskip
    \begin{subfigure}[b]{0.475\textwidth}
    \centering
    \includegraphics[width=8.5cm]{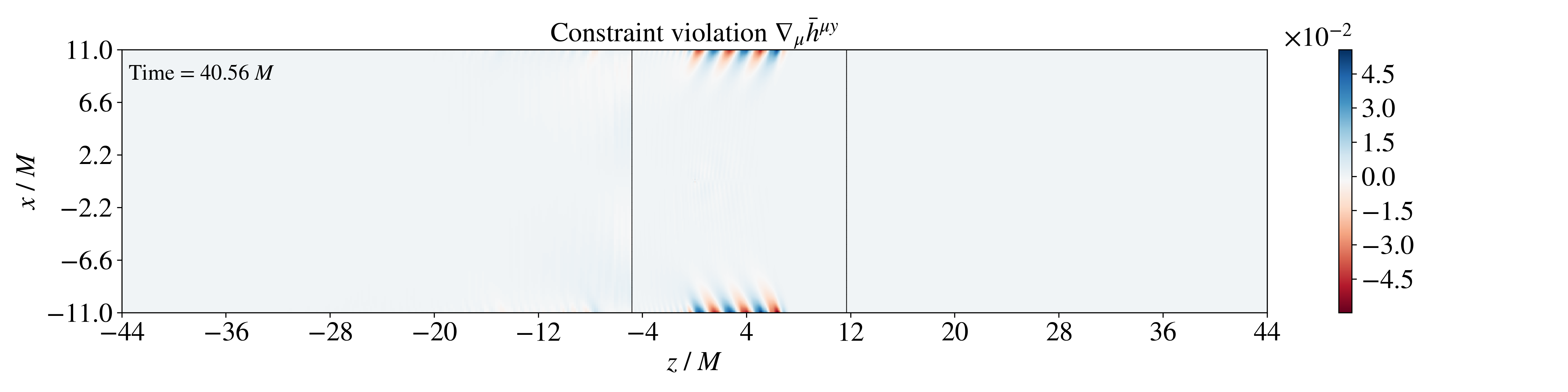}
    \caption{\small Lorenz gauge violation $C_y$}    
    \end{subfigure}
        \hfill
    \begin{subfigure}[b]{0.475\textwidth}
    \centering
    \includegraphics[width=8.5cm]{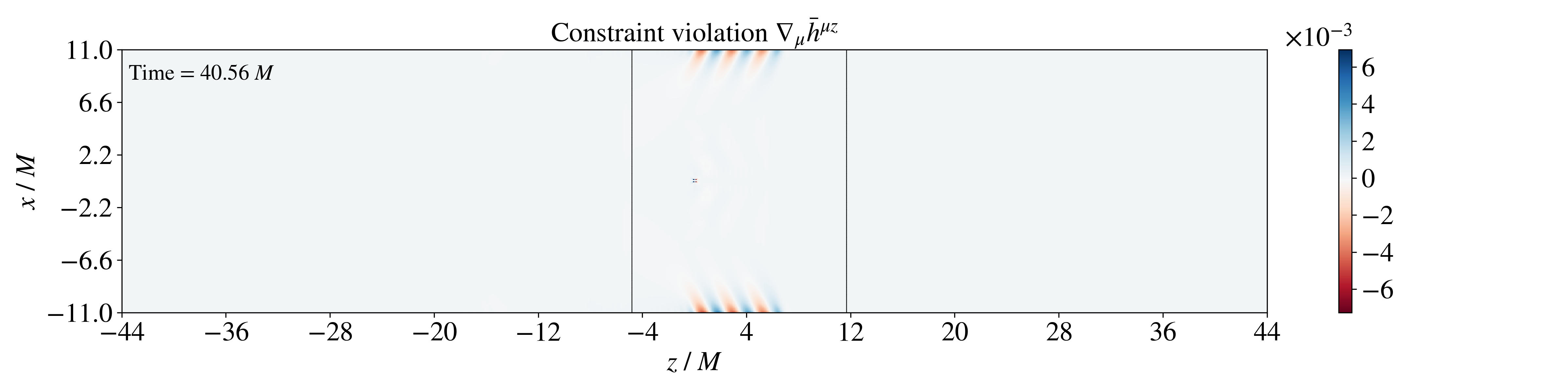}
    \caption{\small Lorenz gauge violation $C_z$}    
    \end{subfigure}
\caption{2D slice of constraint violation at the instance when the GW is propagating inside the NFW potential (weak lensing case with $\phi_{\text{max}}=0.1$). At the considered simulation resolution, the constraint violation is well below the order of the GW amplitude. The black outlined boxes inside the simulation domain is where the AMR patch exists and has twice the resolution of the base grid.} 
\label{fig:NFW_wg_CV}
\end{figure*}

\begin{figure*}
\centering
    \begin{subfigure}[b]{0.475\textwidth}
    \centering
    \includegraphics[width=8.5cm]{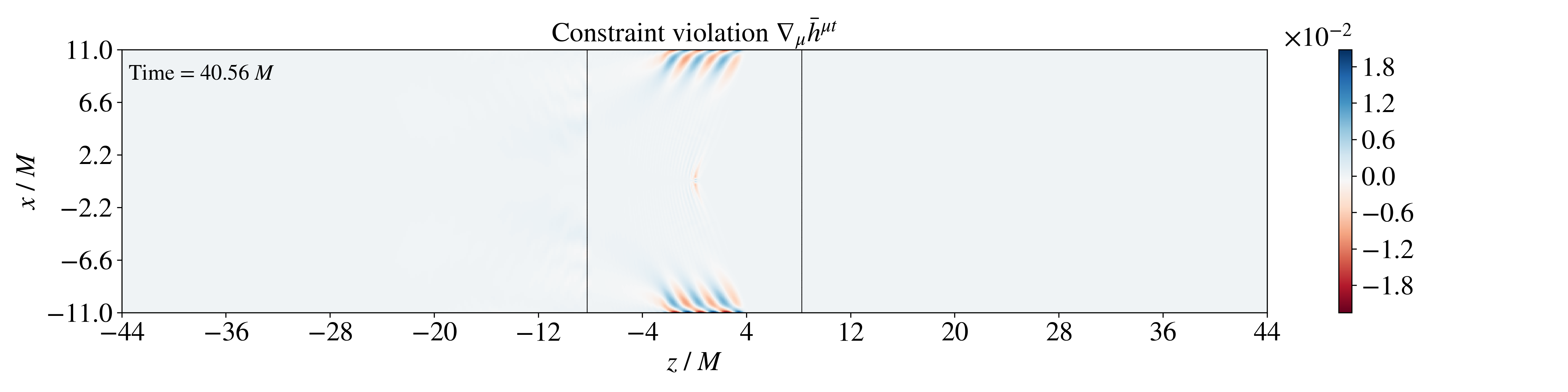}
    \caption{\small Lorenz gauge violation $C_t$}    
    \end{subfigure}
        \hfill
    \begin{subfigure}[b]{0.475\textwidth}
    \centering
    \includegraphics[width=8.5cm]{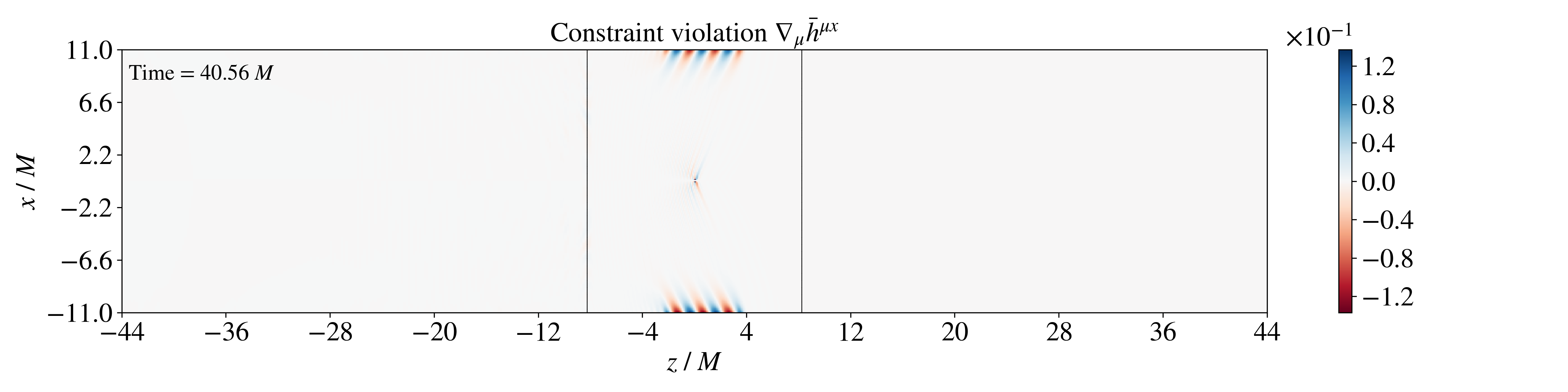}
    \caption{\small Lorenz gauge violation $C_x$}    
    \end{subfigure}
        \vskip\baselineskip
    \begin{subfigure}[b]{0.475\textwidth}
    \centering
    \includegraphics[width=8.5cm]{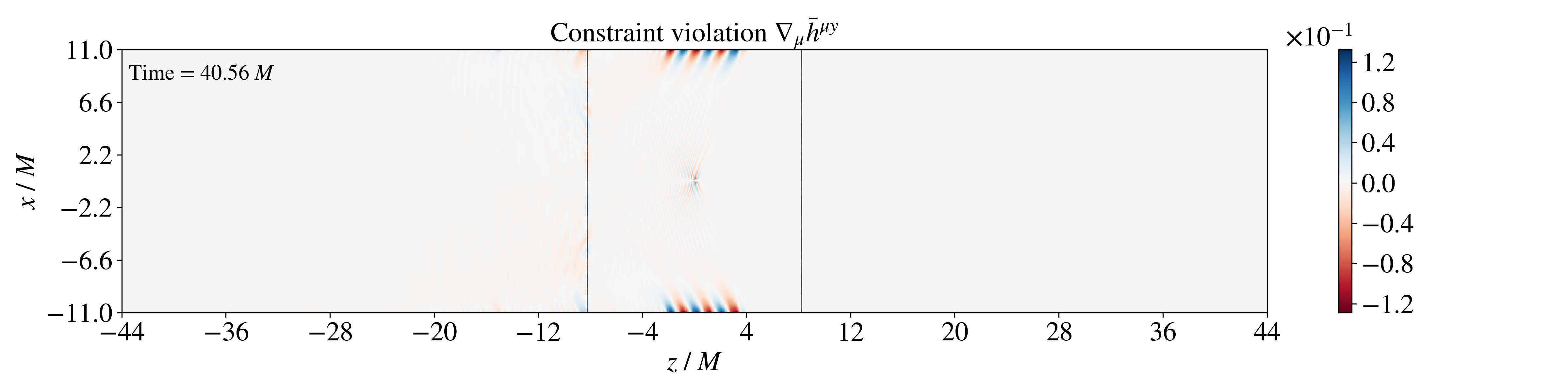}
    \caption{\small Lorenz gauge violation $C_y$}    
    \end{subfigure}
        \hfill
    \begin{subfigure}[b]{0.475\textwidth}
    \centering
    \includegraphics[width=8.5cm]{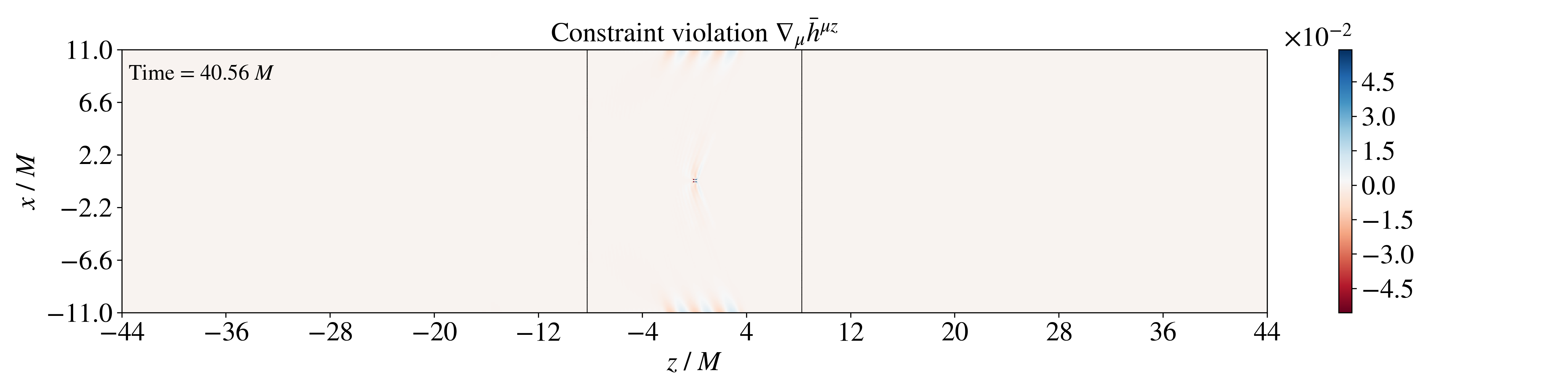}
    \caption{\small Lorenz gauge violation $C_z$}    
    \end{subfigure}
\caption{2D slice of constraint violation plots at the instance when the GW is propagating inside the NFW potential (strong gravity case with $\phi_{\text{max}}=0.49$). At the considered simulation resolution, the constraint violation is well below the order of the GW amplitude. The black outlined boxes inside the simulation domain is where the AMR patch exists and has twice the resolution of the base grid.} 
\label{fig:NFW_sg_CV}
\end{figure*}

\begin{figure*}
\centering
    \begin{subfigure}[b]{0.475\textwidth}
    \centering
    \includegraphics[width=8.5cm]{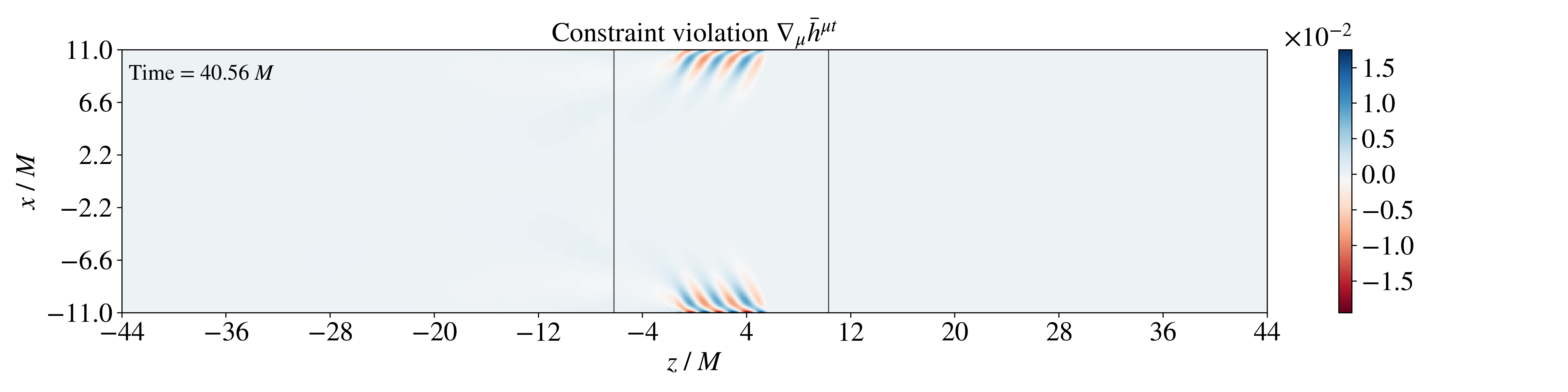}
    \caption{\small Lorenz gauge violation $C_t$}    
    \end{subfigure}
        \hfill
    \begin{subfigure}[b]{0.475\textwidth}
    \centering
    \includegraphics[width=8.5cm]{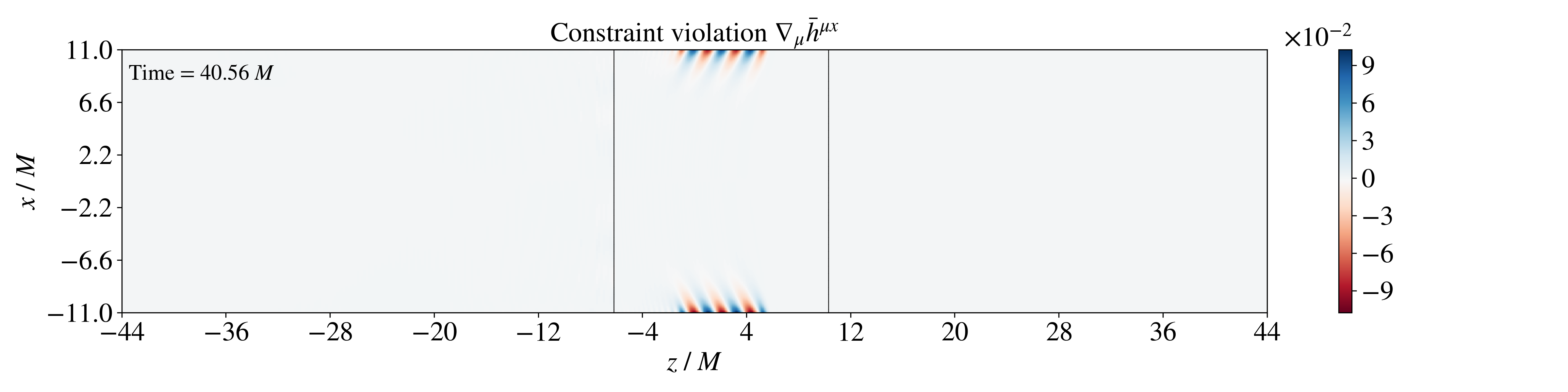}
    \caption{\small Lorenz gauge violation $C_x$}    
    \end{subfigure}
        \vskip\baselineskip
    \begin{subfigure}[b]{0.475\textwidth}
    \centering
    \includegraphics[width=8.5cm]{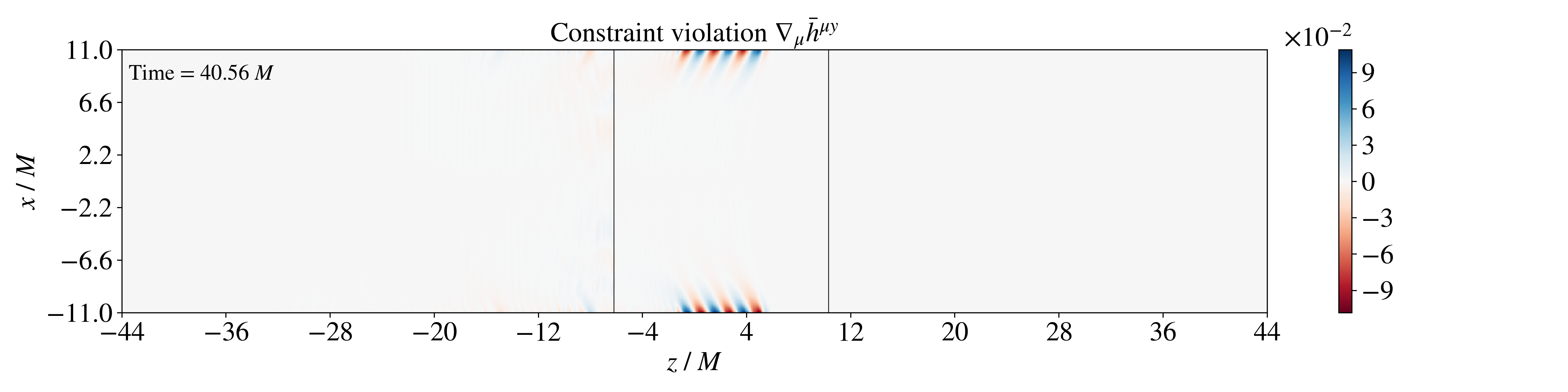}
    \caption{\small Lorenz gauge violation $C_y$}    
    \end{subfigure}
        \hfill
    \begin{subfigure}[b]{0.475\textwidth}
    \centering
    \includegraphics[width=8.5cm]{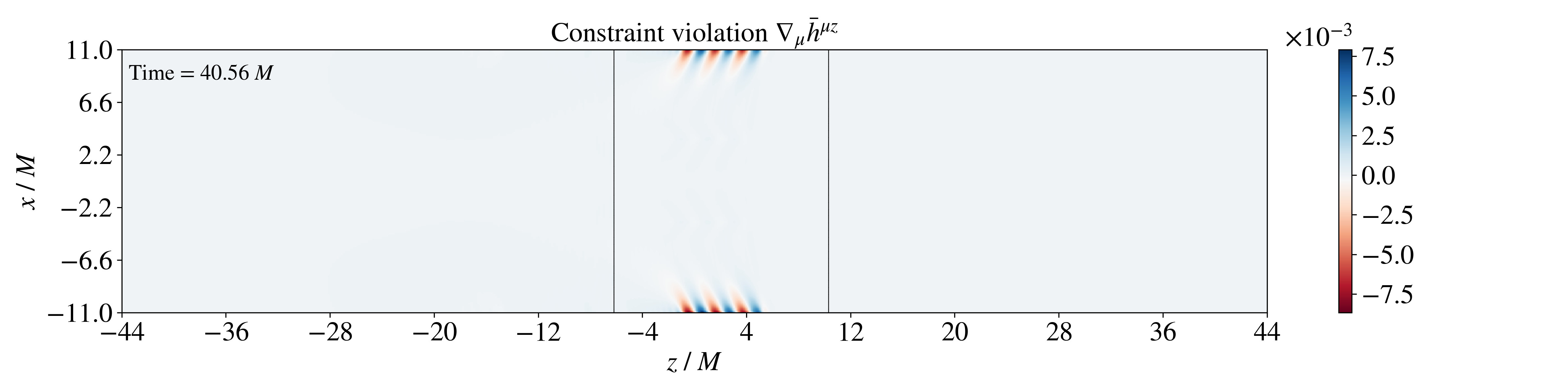}
    \caption{\small Lorenz gauge violation $C_z$}    
    \end{subfigure}
\caption{2D slice of constraint violation at the instance when the GW is propagating inside the Burkert potential (weak gravity case with $\phi_{\text{max}}=0.1$). At the considered simulation resolution, the constraint violation is well below the order of the GW amplitude. The black outlined boxes inside the simulation domain are where the AMR patch exists and has twice the resolution of the base grid.} 
\label{fig:bk_wg_CV}
\end{figure*}

\begin{figure*}
\centering
    \begin{subfigure}[b]{0.475\textwidth}
    \centering
    \includegraphics[width=8.5cm]{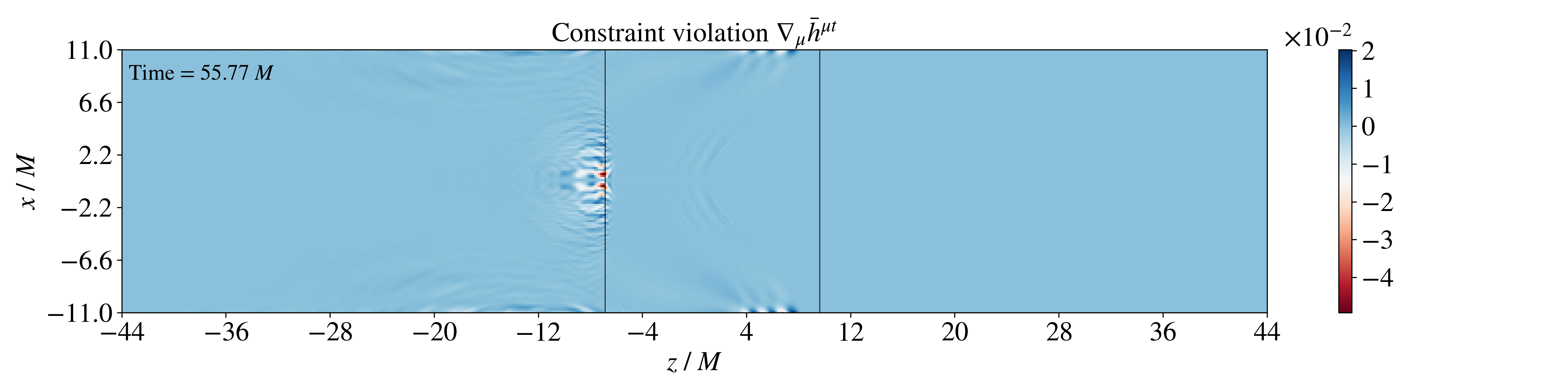}
    \caption{\small Lorenz gauge violation $C_t$}    
    \end{subfigure}
        \hfill
    \begin{subfigure}[b]{0.475\textwidth}
    \centering
    \includegraphics[width=8.5cm]{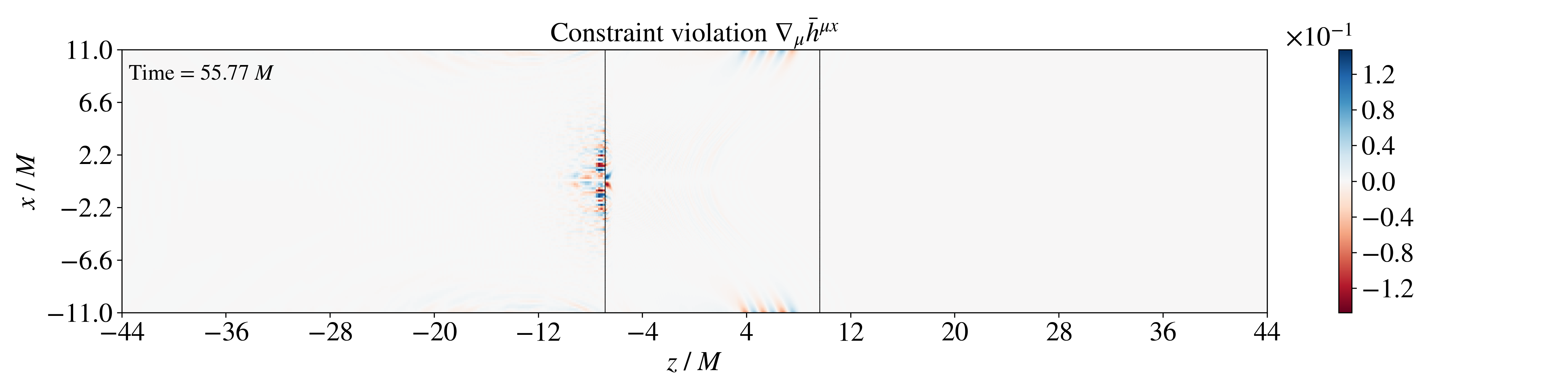}
    \caption{\small Lorenz gauge violation $C_x$}    
    \end{subfigure}
        \vskip\baselineskip
    \begin{subfigure}[b]{0.475\textwidth}
    \centering
    \includegraphics[width=8.5cm]{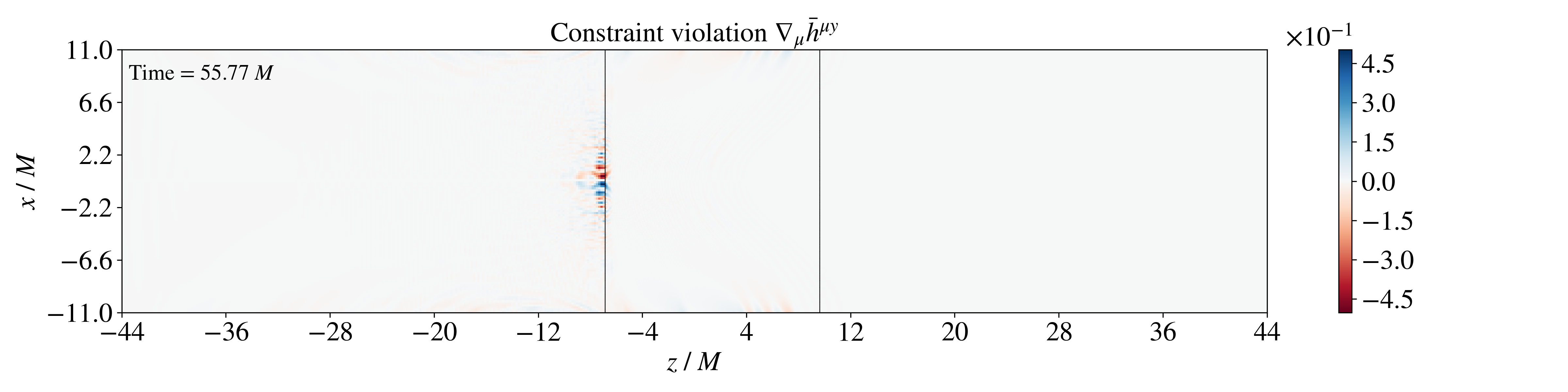}
    \caption{\small Lorenz gauge violation $C_y$}    
    \end{subfigure}
        \hfill
    \begin{subfigure}[b]{0.475\textwidth}
    \centering
    \includegraphics[width=8.5cm]{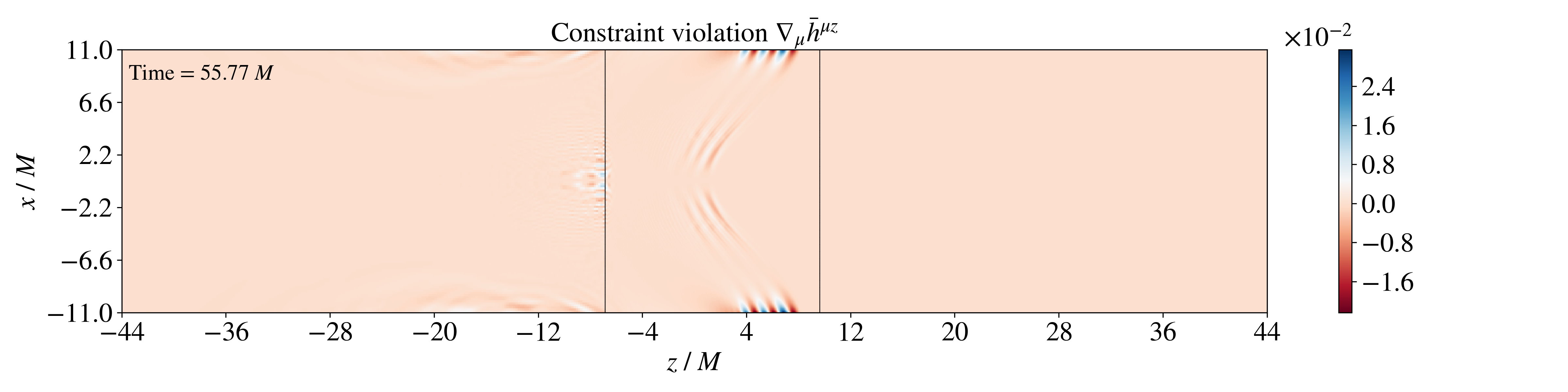}
    \caption{\small Lorenz gauge violation $C_z$}    
    \end{subfigure}
\caption{2D slice of constraint violation at the instance when the GW is propagating inside the Burkert potential (strong gravity case with $\phi_{\text{max}}=0.49$). At the considered simulation resolution, the constraint violation is well below the order of the GW amplitude. The black outlined boxes inside the simulation domain are where the AMR patch exists and has twice the resolution of the base grid.} 
\label{fig:bk_sg_CV}
\end{figure*}

\begin{table}
    \centering
    \begin{tabular}{|c|c|c|}\hline
         Spline distance from $\phi_{\text{max}}$ &  $r_1$& $r_2$\\\hline
         Gaussian potential&  $6.27 \ M$& $31.33 \ M$\\\hline
         NFW Potential&  $6.27 \ M$& $31.33 \ M$\\\hline
         Burkert Potential&  $0.3133 \ M$& $31.33 \ M$\\ \hline
    \end{tabular}
    \caption{Table of values for points $r_1$ and $r_2$  where spline is implemented.  $r_1$ and $r_2$ are the distances from the $\phi_{\text{max}}$ where spline is setup. The values were chosen empirically by ensuring that the chosen values ensure that both $\phi$ and its derivatives do not have any abnormal jumps or discontinuities due to the spline. These values also guarantee that the negative energy density contributions remain well below the numerical finite difference stencil order used.}
    \label{tab:spline}
\end{table}

\clearpage

\bibliographystyle{apsrev4-2}  
\bibliography{refs}

\end{document}